# Energy-efficient switching of nanomagnets for computing: Straintronics and other methodologies

Noel D'Souza[1], Ayan Biswas[2], Hasnain Ahmad[2], Mohammad Salehi Fashami[1], Md Mamun Al-Rashid[1,2], Vimal Sampath[1], Dhritiman Bhattacharya[1], Md Ahsanul Abeed[2], Jayasimha Atulasimha[1,2,*] and Supriyo Bandyopadhyay[2, *]

[1]Department of Mechanical and Nuclear Engineering,

[2]Department of Electrical and Computer Engineering,

Virginia Commonwealth University, Richmond VA 23284, USA.

**Abstract**

The need for increasingly powerful computing hardware has spawned many ideas stipulating, primarily, the replacement of traditional transistors with alternate "switches" that dissipate miniscule amounts of energy when they switch and provide additional functionality that are beneficial for information processing. An interesting idea that has emerged recently is the notion of using two-phase (piezoelectric/magnetostrictive) multiferroic nanomagnets with bistable (or multi-stable) magnetization states to encode digital information (bits), and switching the magnetization between these states with small voltages (that strain the nanomagnets) to carry out digital information processing. The switching delay is ~ 1 ns and the energy dissipated in the switching operation can be few to tens of aJ, which is comparable to, or smaller than, the energy dissipated in switching a modern-day transistor. Unlike a transistor, a nanomagnet is "non-volatile", so a nanomagnetic processing unit can store the result of a computation *locally* without refresh cycles, thereby allowing it to double as both logic and memory. These dual-role elements promise new, robust, energy-efficient, high-speed computing and signal processing architectures (usually non-Boolean and often non-von-Neumann) that can be more powerful, architecturally superior (fewer circuit elements needed to implement a given function) and sometimes faster than their traditional transistor-based counterparts. This topical review covers the important advances in computing and information processing with nanomagnets with emphasis on strain-switched multiferroic nanomagnets acting as non-volatile and energy-efficient switches – a field known as "straintronics". It also outlines key challenges in straintronics.

[*] Corresponding authors' e-mails: jatulasimha@vcu.edu, sbandy@vcu.edu

# 1. Introduction

The rapid development of computing technology in the latter half of the 20$^{th}$ century has had a dramatic impact on human life and society. In the early part of the 20$^{th}$ century, the famed slide rule was perhaps the most commonly available sophisticated computing tool to the public. The need for a more powerful computing tool for code breaking during World War II led to the development of the machine called Colossus that was demonstrated at Bletchley Park in 1943 [1]. Shortly thereafter, the computer known as ENIAC was demonstrated at the Univ. of Pennsylvania in 1946 [2]. However, these machines employed vacuum tubes and were extremely unwieldy and cumbersome as they occupied a large volume of space. Additionally, they were not user-friendly and difficult to operate. The development of the transistor by Shockley, Bardeen, and Brattain in 1947 [3] led to a revolution in miniaturizing computing devices, enabling pocket sized calculators and personal computers that became commonplace as the size of the transistor and the cost of manufacturing decreased dramatically.

The continuous downscaling of the semiconductor transistor that made this electronics revolution possible was anticipated by Gordon Moore, a co-founder of the Intel Corporation in the early 1960s. In his famed "Moore's law" [4], he predicted that the number of transistors in a chip will continue to double roughly every 18 months and this will sustain the electronics revolution, ushering in increasingly powerful computers with time. This prediction held during the 20$^{th}$ century, resulting in rapid advances in computing hardware and falling cost. Keeping up with Moore's law has been the mantra of the electronics industry and has enabled a plethora of lightweight compact devices in the late 20$^{th}$ century such as laptops, mobile phones and implantable medical devices that continue to impact our everyday lives.

However, not all is well with Moore's law. A realization has been dawning since the late 20$^{th}$ century that the relentless miniaturization may eventually come to an end, not so much because the laws of classical physics will no longer hold beyond a certain device size and stop miniaturization dead in its tracks, but more because the energy dissipation density in electronic chips will balloon to unmanageable magnitudes when more and more devices are packed within a given area. If the heat that is dissipated when transistors switch from the "on" to the "off" state, or vice versa, remains constant and does not decrease, then the amount of heat generated per unit area in a chip will increase in proportion to the number of devices per unit area. If the generated heat cannot be removed from the chip quickly and efficiently enough, then the chip will inevitably fail owing to excessive temperature rise. Since thermal management technology has its own stringent limitations, this could hinder further miniaturization and not allow any increase in the device packing density. The resulting stagnation will impede further improvements of computers and ubiquitous computing devices, such as wearable electronics and embedded processors. While pundits may not agree on the best strategies to overcome this impasse, there is reasonable consensus among them that this is a serious roadblock that has to be addressed. Reducing energy dissipation in the switch during the switching process is certainly one way to address the roadblock. Moreover, if the energy dissipation can be reduced substantially, it may enable certain types of embedded processors that consume so little power that they can run entirely on energy harvested from the environment and never need a battery!

## 1.1 Other needs for energy efficient computing devices

Let us start with a simple calculation to estimate how much energy is dissipated in a modern-day transistor switch in the manner of ref. [5]. Consider, the Intel® Core™ i7-6700K processor, built with 14-nm FINFET technology, and released in 2015. It contains ~1.75 billion transistors and dissipates 91 Watts while operating at a clock frequency of 4 GHz [6]. The power dissipated in the chip can be expressed as $P_d = NE_d f$ where $N$ is the number of transistors switching at any given instant, $E_d$ is the average energy dissipated by a transistor during switching, and $f$ is the clock frequency. Assuming an activity level of 10%, i.e., one in ten transistors switches at any given time on the average, we get $N = 1.75 \times 10^9 \times 0.1 = 1.75 \times 10^8$ in the Intel processor. Therefore, $E_d = \frac{P_d}{Nf} = \frac{91}{1.75 \times 10^8 \times 4 \times 10^9} = 130$ aJ, which means that a transistor of circa 2015 dissipates ~130 aJ to switch. The actual dissipation may be a little less since we have ignored dissipation in the wiring. It could also be a little more (less) if the activity level is lower (higher) than what we assumed. Nonetheless, a good round number to keep in mind is ~100 aJ per transistor per switching event.

Now assume that the energy dissipation per transistor remains the same, but transistor density increases owing to Moore's law of miniaturization, so that a future chip has $10^{10}$ transistors/cm$^2$. At the same time, let us assume that the clock frequency has gone up to 5 GHz. In that case (if we again assume 10% activity level), the power dissipated will approach 1 kW/cm$^2$! Although there are reports of handling this level of power dissipation [7], it nonetheless challenges heat-sinking technology.

Let us take this one step further and examine a different scenario. Consider a future application-specific-integrated-circuit (ASIC) that has $10^8$ transistors, an activity level of 10%, and a clock speed of 10 MHz (certain applications do not need high speed). The power required to run this processor (implemented with 14-nm Intel FINFET technology) will be ~12.5 mW and that could eventually drain the battery in hand-held or medically implanted devices that might use such ASICs. This motivates the search for an alternate to the transistor that will dissipate a mere ~ $10^{-17}$ Joules (10 aJ) per bit operation. An ASIC built with such an alternate device will dissipate only ~ 1.25 mW power under identical circumstances and hence can be run with energy harvesting devices alone [8] without needing a battery. Such "battery-less processors" could lead to hitherto unimaginable applications.

Medically implanted ASICs of this type could monitor and process brain signal patterns of epileptic patients to warn of impending seizures [9] while being powered by the patient's head movements alone and not require a battery. This could prevent frequent medical procedures on the patient to replace/recharge the dead battery.

There are other similar applications that can be enabled by extremely low power processors. Buoy-mounted low-power processors that monitor ship and submarine movements in the oceans using inputs from a network of acoustic sensors can be powered solely by energy harvested from the rocking motion of the buoy induced by sea waves and not need a separate power source. Similarly, processors mounted on tall buildings and bridges [10] to monitor their health could be powered exclusively by energy captured from vibrations of the structure due to wind or passing traffic. This would eliminate the need to replace batteries in these processors, which is inconvenient, not to mention risky when the processors are placed in not easily accessible locations. All these applications could become a reality if energy-efficient ASICs, built with low power devices that dissipate ~10 aJ to switch, come about.

The burning question then is what is the *alternate* to the transistor that will dissipate ~ 10 aJ of energy to switch and can enable such gadgets? There are many likely answers to this question and opinion is divided, but in this review, we will approach this question from a fundamental perspective and then venture into specifics.

## 1.2    Energy dissipation in a transistor

A transistor switch, used in digital electronics, is a *charge-based* device which has two conductance states ("on" and "off") that encode the binary bits 0 and 1. A good example of that is an enhancement mode *n*-channel metal-oxide-semiconductor-field effect transistor (MOSFET) shown in Fig. 1(a). A positive gate voltage draws electrons into the channel from the source via Coulomb attraction and turns the transistor "on". A negative gate voltage pushes electrons out of the channel and turns the transistor off. Switching between the two states is therefore accomplished by moving charge in and out of the MOSFET's channel.

All charge-based switches work within this paradigm of moving charge in and out of the device. After all, charge is a *scalar* quantity that has a magnitude and no other attribute. Therefore, two states encoding the two binary bits '0' and '1' *must* be encoded in two different *amounts* (or magnitudes) of charge. More charge present in the device will represent one state or bit, and less charge the other state or bit.

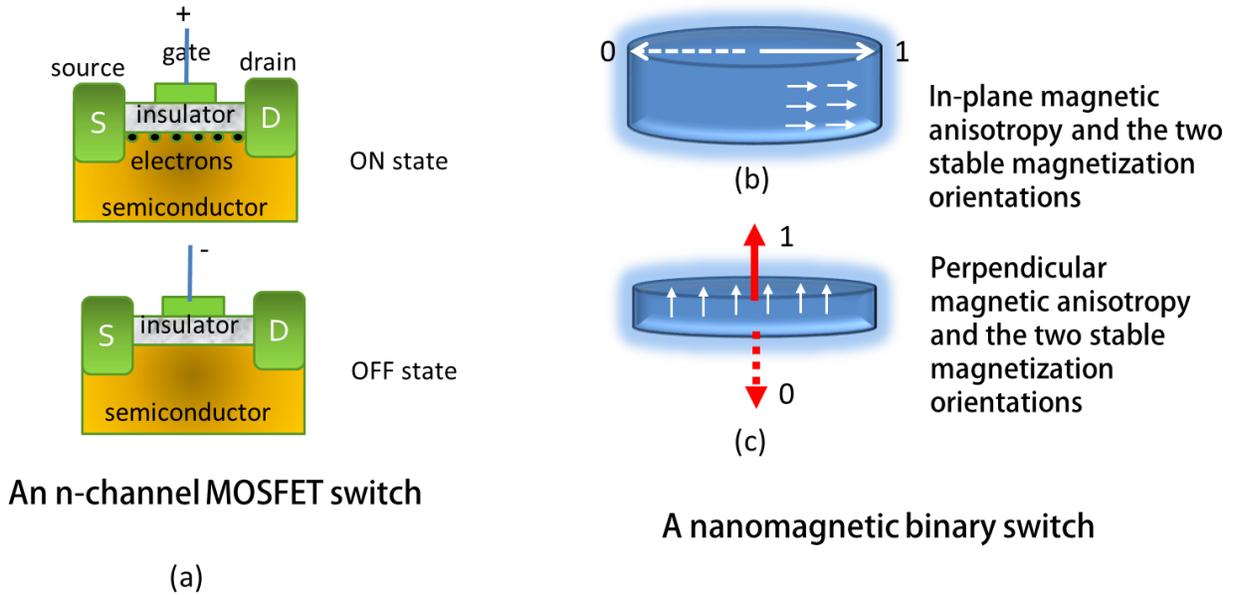

**Fig. 1**: (a) An n-channel MOSFET is switched by moving charges in and out of the channel with two different polarities of the gate voltage. (b) A nanomagnet shaped like an elliptical disk has two stable in-plane magnetization orientations along the major axis which can encode the bits 0 and 1. Switching between the two involves reversing the magnetization (equivalent to flipping the spins within the nanomagnet). No charge is moved within the nanomagnet, although charge may have to be moved in an external circuit to reverse the magnetization. (c) Nanomagnet shaped as a circular disk with perpendicular magnetic anisotropy (PMA). Here, the magnetization has two stable states perpendicular to the nanomagnet's plane ("up" and "down") that can encode the bits 0 and 1. PMA nanomagnets are often preferred over nanomagnets with in-plane magnetic anisotropy since they can be scaled down to smaller lateral sizes, which increases device density on a chip. However, in this review, we mostly use nanomagnets with in-plane anisotropy to illustrate straintronic switches.

Every time the device is switched, the amount of charge in the device must be changed from $Q_1$ to $Q_2$, or vice versa. If the switching duration is $\Delta t$, then the motion of charges will result in a current $I = |Q_1 - Q_2|/\Delta t = \Delta Q/\Delta t$ to flow [5, 11]. The associated energy dissipation is given by

$$E_d = I^2 R \Delta t = (\Delta Q/\Delta t) IR\Delta t = \Delta Q IR = \Delta Q \Delta V \ , \tag{1.1}$$

where $R$ is the resistance in the path of the current and $\Delta V = IR$. Basically, $\Delta V$ is the voltage required to move $\Delta Q$ amount of charge in or out of the device.

We can estimate the quantity $\Delta Q$ in the 14-nm FINFET transistor used in the Intel® Core™ i7-6700K processor. The power supply voltage of this processor is 1.2 V [6], which means $\Delta V = 1.2$ V. Since $E_d = 130$ aJ, we estimate that $\Delta Q = 673$ electrons. In other words, fewer than 700 electrons are moved to switch a state-of-the-art FINFET transistor today.

Unfortunately, we cannot arbitrarily decrease $\Delta Q$ to reduce energy dissipation. The *minimum* $\Delta Q$ that is acceptable is determined by the acceptable noise margin. If the device is operated in a noisier environment, then maintaining sufficient distinction between the bits mandates a larger difference between the amounts of charge $Q_1$ and $Q_2$ that encode the bits, which means $\Delta Q$ will have to be larger, resulting in more energy dissipation. This tells us that there is a tradeoff between energy dissipation and error-resilience.

The quantity $\Delta V = IR = (\Delta Q/\Delta t) R$, on the other hand, is obviously not independent of the switching speed. For a fixed $\Delta Q$ and $R$, $\Delta V$ is inversely proportional to the switching time $\Delta t$. If the transistor is switched faster $(\Delta t$ smaller$)$, $\Delta V$ will increase and therefore the energy dissipation $E_d = \Delta Q \Delta V$ increases [11]. Hence, there is a tradeoff between energy dissipation and speed as well.

While the above argument is at fixed "$R$", one could try to decrease $R$ to the extent possible so as to decrease $\Delta t$ while keeping $\Delta V$ constant. This is reflected in the expression for the energy delay product:

$$E_d \Delta t = (I\Delta t)^2 R = (\Delta Q)^2 R \ . \tag{1.2}$$

Because of the dependence on $\Delta Q$, the minimum energy-delay product in a *charge-based device* will depend on what is the minimum $\Delta Q$ we can live with, which, in turn, depends on *how much error-probability we can tolerate*. Therefore, there is a tradeoff between energy-delay product and error-resilience. If we try to reduce the energy-delay product by decreasing $\Delta Q$, then it will decrease the logic separation between bits and the error probability will increase. Therefore, the price for a lower energy-delay product is lower error-resilience. We cannot overemphasize the fact that the energy-delay product is not divorced from the error probability and the two are inexorably linked. When benchmarking of devices is based solely on their energy-delay product [12], it does not include this link. Perhaps a more comprehensive benchmarking metric is the product of the energy, delay and error probability.

Finally, it is perhaps worthwhile to ponder the question of how low the energy dissipation in a transistor (built with current technology, e.g. the 14 nm FINFET) can become if it were to operate at the current clock speed of 4 GHz and at the same time maintain reasonable reliability. In order to remain reasonably resilient against background charge fluctuations, let us assume that the minimum $\Delta Q$ we can live with is $\Delta Q|_{min} \sim 10 \times \Delta Q|_{thermal}$ where $\Delta Q|_{thermal}$ is the background charge fluctuation due to thermal noise. In a MOSFET type device, the latter quantity will be given by $\Delta Q|_{thermal} = C_g V_{thermal}$ where $C_g$ is the gate capacitance (including parasitics) and $V_{thermal}$ is the thermal noise voltage in the gate terminal. The quantity $V_{thermal} = \sqrt{\frac{kT}{C_g}}$ and hence $\Delta Q|_{thermal} = \sqrt{C_g kT}$. We can reasonably assume that $C_g \sim 1$ fF, which will make $\Delta Q|_{thermal} = 2 \times 10^{-18}$ Coulombs, or about 12 electrons at room temperature. Thus, a good estimate for $\Delta Q|_{min}$ is 100 electrons, which is roughly an order of magnitude higher than the background charge fluctuation.

Since in the i7-6700K chip the voltage needed to move 673 electrons is 1.2 V, we can extrapolate (assuming linear scaling) that the voltage needed to move 100 electrons in the same amount of time will be ~180 mV. Therefore, the energy dissipation would be $\Delta Q \Delta V = 1.61 \times 10^{-17} \times 0.18 = 2.8$ aJ if we maintain the same clock speed of 4 GHz. This number will go up or down if we increase or decrease the clock speed. The corresponding energy-delay product (for 4 GHz clock rate) will be $7 \times 10^{-28}$ J-sec. The purpose of this analysis was to show that mainstream transistors like the FINFET are unlikely to achieve much lower switching energy dissipation than ~1 aJ.

The next question is whether energy dissipation lower than ~1 aJ per switching event is even meaningful or beneficial from a circuits perspective. Line losses and other circuit overheads per device may exceed 1 aJ/bit (unless the latter can also be reduced with development of new interconnect materials and improved circuit architectures), which makes the energy dissipation in the device per se a moot point when we approach dissipation figures of ~1 aJ/bit. In this article, we discuss nanomagnetic switches which offer a tantalizing alternative to the transistor, but in the end, even if they can become more energy-efficient than transistors and dissipate less than 1 aJ/bit, they may still not make circuits any more energy-efficient since the bulk of the dissipation may not occur in the device but occur in the peripherals. While we recognize this, we point out that nanomagnetic devices have a key advantage that charge based devices do not have. Charges leak and that makes charge-based devices *volatile*. Magnets do not leak in that fashion and information encoded in the magnetization state of a nanomagnet can be retained for centuries, even after all power is switched off, making them *non-volatile*. The non-volatility can be exploited in non-volatile memory that do not need refresh cycles, non-von-Neumann circuit architectures, as well as certain types of circuits that exploit the non-volatility to reduce device count, improve overall energy-delay product, and perform certain types of functions that cannot be realized with transistors. At this time, this may be the primary motivation for research in nanomagnetic devices.

## 1.3      Nanomagnetic switch

The simplest nanomagnetic switch is a single nanomagnet, small enough to have a single ferromagnetic domain, shaped like an elliptical disk as shown in Fig. 1(b). This geometry ensures that the nanomagnet's magnetization has two stable orientations directed in opposite directions along the ellipse's major axis. These two stable orientations can encode the binary bits 0 and 1. Alternatively, a nanomagnet with circular disk geometry with perpendicular magnetic anisotropy (PMA), whose easy axis of magnetization is perpendicular to plane (see Fig 1 (c)), can be employed and the two stable "up" and "down" states can encode the bits 0 and 1. PMA nanomagnets are preferred in most applications since they can be scaled to smaller lateral sizes because of the higher energy density of PMA and they are relatively immune to random variations in lateral dimensions caused by lithographic imperfections. However, in this review, for the sake of simplicity, we will illustrate most of the straintronic devices with in-plane nanomagnets.

How the bits encoded in the two magnetization orientations or "states" can be read electrically, or how one can switch from one to the other (write data), will be discussed later, but it is obvious that switching requires simply reversing the magnetization, or, equivalently, flipping every spin within the nanomagnet *without moving the spin carrying electrons in space and causing a current flow*. Some current, however, may flow in an external circuit to make the magnetization (and the spins) flip and this would cause an *external* $\Delta Q \Delta V$ dissipation, but there need not be any internal $\Delta Q \Delta V$ dissipation since ideally $\Delta Q = 0$.

Both the transistor and the nanomagnet experience external and internal energy dissipation during switching. The external dissipation is the $\Delta Q \Delta V$ dissipation associated with charge motion in an external circuit that causes switching to occur. In the case of the MOSFET transistor, the internal dissipation is $\int_0^{\Delta t} I_{DS}(t) V_{DS}(t) dt$, where $I_{DS}$ is the source-to-drain is current, $V_{DS}$ is the source-to-drain voltage and $\Delta t$ is the time it takes for the transistor to turn fully on or off. In the case of the nanomagnet, the internal dissipation is the Gilbert damping loss associated with damped magnetization rotation (which is material dependent among other things) plus some additional loss that depends on how the magnetization is rotated. If the magnetization is rotated by straining a magnetostrictive nanomagnet (which is the subject of this review), then this additional loss is $\text{stress} \times \text{strain} \times \text{nanomagnet volume}$. The total energy dissipation during switching is

$$E_d^{total} = \Delta Q \Delta V \big|_{external} + E_d^{internal} \qquad (1.3)$$

The nanomagnet will be less dissipative than the transistor only if $E_d^{total}\big|_{nanomagnet} < E_d^{total}\big|_{transistor}$.

*Coherent switching:* One of the reasons for focusing on single domain nanomagnets is that they have an intriguing feature which, if taken advantage of, could lead to low $E_d^{total}$. Because of exchange interaction between them, all the spins in a single domain nanomagnet rotate *simultaneously in unison* when the magnetization reverses [13]. This is called *coherent switching*, which does not always happen (it will not happen in larger nanomagnets that are multi-domain), but if it happens then we can view all the numerous spins in the nanomagnet acting together as just *one single giant classical spin* [13]. As a result, there is only one effective information carrier (the giant single spin) in a single domain nanomagnet, whereas in a transistor, there are multiple information carriers since the different charges (electrons) in the channel act independently and their motions are uncorrelated. Switching is always incoherent in a transistor. It has been shown from fundamental arguments that the minimum energy that can be dissipated in non-adiabatic (and hence relatively fast) switching process in a switch is *NkTln(1/p)* where *N* is the number of information carriers (or degrees of freedom), *kT* is the thermal energy at a temperature *T* and *p* is the switching error probability [14]. For a single domain nanomagnet, regardless of how many spins are within the sample, *N* = 1 if all the spins in the nanomagnet switch coherently (i.e. in perfect unison and they remain parallel to each other at all times). However, if the switching is incoherent, then the effective $N > 1$. In that case, it may be of the order of 10. For a transistor with 673 electrons as the 14-nm FINFET we discussed, $N = 673$. Thus, the nanomagnet still has an inherent advantage since $N\sim 10$. However, as the number of electrons in a CMOS device scales to $N \sim 100$, this advantage will reduce to a factor of ~10.

Furthermore, the possibility of partially (or even fully) coherent switching does not guarantee that switching a nanomagnetic switch will be more energy-efficient than switching a transistor since it may not be possible to exploit this inherent advantage easily. No nanomagnet switching scheme that is presently extant is able to exploit the full benefit of reduced $N$ ($N \sim 10$) since the energy dissipation is dominated by that in the switching circuitry and not the energy dissipation in the nanomagnet. Therefore, at this time, the reduced $N$ is somewhat immaterial and nanomagnetic switches may not exhibit an energy advantage over transistors in terms of total $E_d$.

*Non-volatility:* There is, however, a different reason why nanomagnets are favored as a potential replacement for a transistor. It has nothing to do with lower dissipation during switching but has to do with the fact that a nanomagnet can store information for a very long time without requiring energy to retain the information. We call this property "non-volatility" and a transistor does not possess this property. Let us examine this further.

A nanomagnet whose shape is anisotropic (or one that has magneto-crystalline anisotropy associated with the crystalline structure) has one or more "easy axis" (or preferred axis) of magnetization, meaning that magnetization orientations collinear with these axes are stable. In an elliptical disk that has no magneto-crystalline anisotropy, the easy axis coincides with the major axis of the ellipse (ideally). The two mutually anti-parallel directions collinear with the easy axis are the two stable orientations of the magnetization (other orientations are unstable). The potential energy profile of the nanomagnet as a function of the magnetization orientation in the plane of an elliptical nanomagnet is shown in Fig. 2. Note that there are two energy minima corresponding to the two orientations along the major axis (easy axis). These are obviously the two stable orientations. They are separated by an in-plane energy barrier $E_b$. The material composition (saturation magnetization and magnetocrystalline anisotropy in case of single crystal or textured materials), shape and volume of the nanomagnet determine the height of the potential energy barrier $E_b$. By making the energy barrier $E_b \sim 60$ $kT$ at room temperature (or 1.56 eV) by appropriately choosing the material, shape and volume, one can make either stable direction (or, equivalently, the bit encoded in a stable direction) extremely "stable" in the presence of thermal noise. The stability is measured by the mean time that elapses before the bit (or magnetization direction) may flip spontaneously from one direction to the other owing to thermal agitation. This time is given by $\tau =$ $\tau_a e^{E_b/kT}$ [15], where $\tau_a$ is the so-called inverse attempt frequency of demagnetization due to noise, which is typically $10^{-12}$ - $10^{-9}$ seconds at room temperature in most magnetic materials [16]. Using the lower limit of $\tau_a$, we get that the storage time exceeds ~36,000 centuries if $E_b = 60\ kT$. This long storage time makes either stable state extremely stable. Therefore, the nanomagnet becomes a "non-volatile" entity that can retain information almost indefinitely without being fed external energy. Non-volatility is the most important advantage that a nanomagnetic switch has over a transistor switch which is "volatile" since charges leak off when the device is powered off. Refresh cycles are needed to retain information in a volatile device and typically consume more than 20% of the energy budget [17]. Therefore, regardless of whether a nanomagnetic switch is more energy-efficient than a transistor switch, it can certainly lead to more energy efficient circuit *architectures* by eliminating the need for refresh cycles.

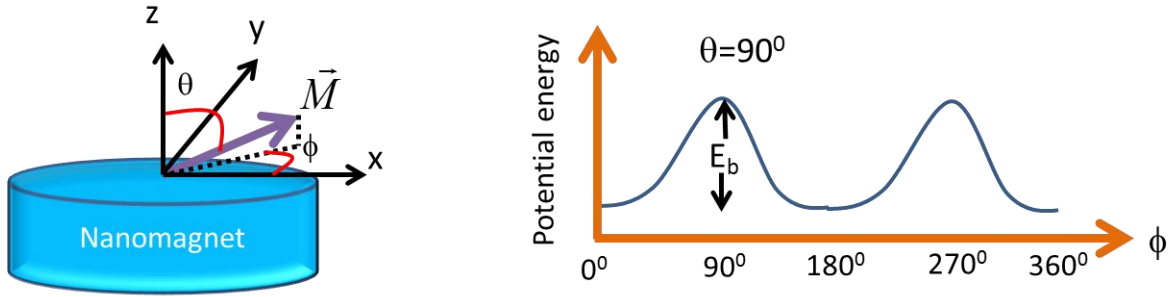

**Fig. 2**: The potential energy profile of an isolated single domain nanomagnet shaped like an elliptical disk (in the plane of the nanomagnet).

There are other advantages of non-volatility. Logic gates, fashioned out of nanomagnets, are non-volatile computing units unlike those fashioned out of transistors. They can perform a computation and then retain/store the output (i.e. the result of the computation) internally in the magnetization state(s) of the nanomagnetic logic elements without the need to store it in an external memory unit. This allows the same piece of hardware to double as both logic and memory. One immediate outcome of this feature is that nanomagnetic logic can implement non-von-Neumann computer architectures with no physical partition between processor and memory. The processor also acts as the memory. Instruction sets for running a program do not have to be fetched from a remote memory into a processor since they are stored in situ, cutting down on the time and improving reliability of the computation. This can lead to computers with zero boot delay as well as certain other types of computer architectures that can operate more efficiently than their traditional counterparts built with transistors [18-20].

## 1.4 In-plane and perpendicular magnetic anisotropy

Nanomagnets are sometimes classified into two types based on the orientation of the easy axis of magnetization (which is the direction of the two stable magnetization states): nanomagnets with *in-plane anisotropy* (IPA) and nanomagnets with *perpendicular magnetic anisotropy* (PMA). In the former, the preferred direction of magnetization (or the "easy axis") is determined primarily by the lateral shape (e.g. if the lateral shape is elliptical, then the shape anisotropy energy will tend to make the easy axis coincide with the major axis of the ellipse). However, when the nanomagnet is very thin, the surface anisotropy may override the shape anisotropy and make the easy axis perpendicular to the nanomagnet's plane [21-23]. This results in PMA. The difference between the two is shown in Fig. 3.

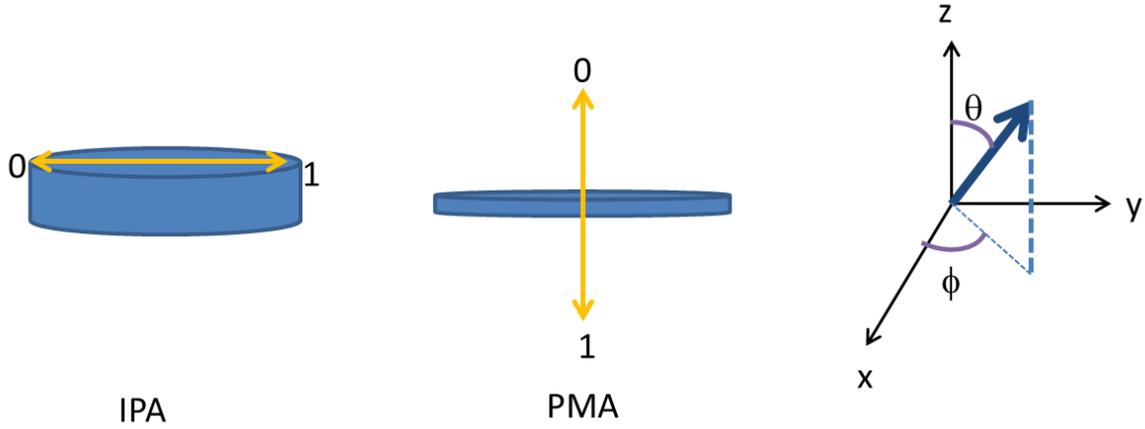

**Fig. 3**: Two elliptical nanomagnets with in-plane and perpendicular magnetic anisotropy.

If the angle subtended between the magnetization of a nanomagnet and the normal to its plane is $\theta$, then the total anisotropy energy is expressed as

$$E_{\text{anisotropy}} = -K\Omega\cos^2\theta, \quad (1.4)$$

where $K$ is the anisotropy energy density and $\Omega$ is the nanomagnet volume. The anisotropy energy has three contributions – due to volume (including any magnetocrystalline anisotropy), due to surface and due to shape. The last term depends on the orientation of the magnetization and is given by $E_{\text{shape}} = (1/2)\mu_0 M_s^2 \left[ N_{d-xx}\sin^2\theta\cos^2\phi + N_{d-yy}\sin^2\theta\sin^2\phi + N_{d-zz}\cos^2\theta \right]$ [24] where $N_{d\text{-}mm}$ is the demagnetization factor along the *m*-th coordinate axis and $\phi$ is the azimuthal angle of the magnetization as defined in Fig. 3. The demagnetization factors depend on the ellipse's major and minor axis dimensions, as well as the thickness, and obey the relation $N_{d-xx} + N_{d-yy} + N_{d-zz} = 1$. In the event that the thickness of the ellipse is much smaller than the major or minor axis, $N_{d-zz} \gg N_{d-xx}, N_{d-yy}$, in which case $N_{d-zz} \approx 1; N_{d-xx}, N_{d-yy} \approx 0$ [25] and we get

$$K \approx K_{vol} + \underbrace{(2K_s/t)}_{\text{surface ansisotropy}} \underbrace{-(1/2)\mu_0 M_s^2}_{\text{shape anistropy}}, \tag{1.5}$$

where $K_{vol}$ is the volume contribution, $K_s$ is the surface contribution, $t$ is the film thickness and the last term is the shape contribution where $\mu_0$ is the permeability of free space and $M_s$ is the nanomagnet's saturation magnetization. If the thickness $t$ is large (though smaller than the lateral dimensions), then the last term dominates and $K$ is negative. In that case, $E_{\text{anisotropy}}$ is minimized when $\theta = 90^0$, i.e. the easy axis is in-plane. However, when $t$ is very small, the second term dominates and $K$ is positive. In that case, $E_{\text{anisotropy}}$ is minimized when $\theta = 0^0$ and the easy axis will be perpendicular to the plane.

Clearly, whether PMA is exhibited or IPA depends on whether $K > 0$ or $K < 0$. Based on this simple consideration, we can find an expression for the critical thickness below which the nanomagnet will exhibit PMA and above which IPA. That expression is obtained from Equation (1.5) and is given by [26]

$$t_{cr} = \frac{2K_s}{K_{vol} - (1/2)\mu_0 M_s^2}. \tag{1.6}$$

PMA nanomagnets have the advantage that the easy axis is always perpendicular to the plane even if the ellipse is not a perfect ellipse. IPA nanomagnets have the disadvantage that the easy axis direction is a little uncertain if the ellipse is not a perfect ellipse. PMA nanomagnets are therefore more tolerant of lithographic imperfections. PMA nanomagnets can also be made smaller than IPA nanomagnets without losing the anisotropy that makes the magnetization direction bistable. Therefore, PMA nanomagnets are more *scalable*. These advantages have made PMA nanomagnets the preferred choice for implementing non-volatile memory with MTJs.

## 1.5  Overview of this topical review

That a nanomagnetic switch is superior to a transistor switch because of non-volatility is an incontrovertible fact. What is less clear, however, is whether and when a nanomagnetic switch also has an energy advantage over a transistor. Ultimately, whether switching a nanomagnet dissipates less energy than switching a modern-day transistor, i.e. whether $E_d^{total}\big|_{nanomagnet} < E_d^{total}\big|_{transistor}$, depends on *how the magnet is switched*. In the next section, we discuss a number of methodologies for rotating the magnetization of a nanomagnet (switching) that have been explored by different research groups. This discussion cannot be exhaustive since this is a very active field of research and new methodologies are being proposed, demonstrated and examined frequently.

Among nanomagnet switching methodologies, one that is among the most energy-efficient is "straintronic switching". This involves a two phase (piezoelectric/magnetostrictive) multiferroic nanomagnet. A schematic is shown in Fig. 4. Consider an elliptical nanomagnet delineated on a poled piezoelectric substrate. A voltage is applied through the piezoelectric with electrically shorted gate pads on either side of the nanomagnet to generate a biaxial strain in the piezoelectric underneath the nanomagnet [27-29]. Some of this strain is transferred to the magnetostrictive nanomagnet and rotates its magnetization by ~$90^0$ by virtue of the Villari effect. If the strain pulse is timed such that the strain is relaxed as soon as the $90^0$ rotation is completed, then a residual torque acting on the magnetization due to the strain induced magnetization dynamics continues to rotate the magnetization after the strain is relaxed until a $180^0$ rotation is completed and the magnetization has flipped (switched) [30]. Our calculations [31, 32] and experiments [33-37] seem to show that strain mediated magnetization control in magnetostrictive nanomagnets (with appropriate scaling) is a very energy-efficient scheme to rotate the magnetization of a nanomagnet. The energy dissipation in a properly scaled multiferroic nanomagnet can be ~ 5 aJ, albeit the rotation speed is relatively slow (~ 1 ns), which makes the energy-delay product ~ 5× $10^{-27}$ J-sec. The rotation speed is similarly slow in other magnetization rotation schemes, except there is now some thought that anti-ferromagnets may switch faster than ferromagnets (the Néel vector can rotate in a fraction of a nanosecond) and if that can be harnessed to implement a switch by itself, then the switching speed can be dramatically increased. In this review, however, we will stay focused on discussion of ferromagnet-based switches. Note that the energy-delay product of a straintronic switch is about an order of magnitude larger than that of a transistor (because a nanomagnet switches much slower than a transistor). Thus, whether or not a straintronic switch (2-phase multiferroic) is more energy-efficient than a transistor, it is almost surely inferior in terms of energy-delay product. However, it is important to remember that a straintronic switch, like other nanomagnetic switches, is non-volatile (unlike a transistor) and therefore has all the advantages of non-volatility that we had discussed earlier. Consequently, it is quite possible that even though the nanomagnetic switch may have higher energy-delay product than a transistor, the overall circuit will have a lower energy-delay product (because of reduced device count, absence of refresh cycles, etc.) if it is built with nanomagnetic switches instead of transistor switches. Furthermore, slow switching of an isolated device is usually not a serious disadvantage since it can be mitigated with massive parallelism. Furthermore, in most embedded processor applications (wearable electronics, health monitoring devices), it is the energy dissipation that is important, not the speed. The slow speed can also be ameliorated by more efficient circuit architectures (enabled by non-volatility of the devices) that can speed up the operation of the circuit despite the use of more sluggish devices.

In this article, we emphasize the straintronic nanomagnet switching scheme primarily because it is an extremely energy-efficient scheme among nanomagnet switching schemes. This scheme is discussed in detail in Section 3, while other magnet switching methodologies are discussed briefly in Section 2.

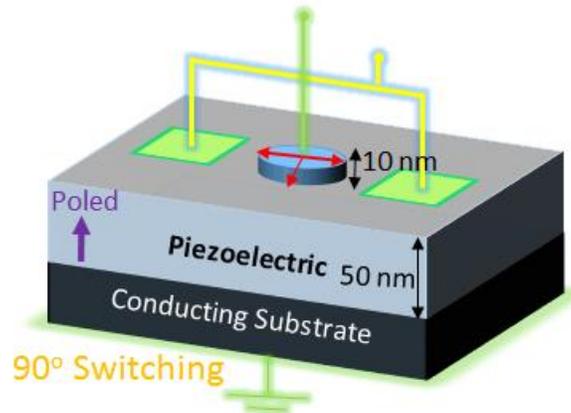

**Fig. 4**: Straintronic switching scheme.

Section 4 reviews some straintronic memory schemes and logic device designs along with some experimental results. Section 5 deals with straintronic Magnetic Tunnel Junctions (MTJs). Section 6 looks at the potential application of straintronics for non-Boolean computing and Section 7 concludes with a note summarizing possible future challenges and potential developments in the area of straintronics.

## 2. Nanomagnet switching schemes

### 2.1. Switching with a current-generated magnetic field

The most obvious method of switching or rotating the magnetization of a nanomagnet from one orientation to another is to use a magnetic field pointing in the direction of the desired orientation. For example, if we wish to flip the magnetization of the elliptical nanomagnet in Fig. 1(b) from the right-pointing stable direction to the left-pointing stable direction, we can apply a magnetic field in the left-pointing direction, wait until the magnetization has flipped, and then remove the magnetic field. The magnetic field can be generated by an on-chip current. There are, however, two major disadvantages with this approach. First, because it is very difficult to confine a magnetic field to a small region of space, it is difficult to selectively address an individual nanomagnet surrounded by other nanomagnets close by, using this approach. Therefore, nanomagnets have to be placed relatively far apart in a chip if this method is used, which reduces device density. Second, this method of switching a nanomagnet is energy-*inefficient*.

Let us make a rough estimate of the current that we will need to generate a magnetic field of sufficient strength to rotate a nanomagnet's magnetization through $180^0$ and flip the magnetization. To understand what this entails, refer to the potential profile in Fig. 2. Obviously, we have to transcend the energy barrier to take the magnetization state from one energy minimum to the other. Therefore, the magnetostatic energy generated by the magnetic field must equal or exceed the energy barrier $E_b$. In other words, the minimum magnetic field strength $H_{min}$ that is required will be given by $\mu_0 M_s H_{min} \Omega = E_b$, where $\Omega$ is the nanomagnet volume, $\mu_0$ is the permeability of free space and $M_s$ is the saturation magnetization of the nanomagnet. The loop current needed to generate this magnetic field is found from Ampere's law:

$$I = \int \vec{H}_{min} \bullet d\vec{l} = \int \frac{E_b}{\mu_0 M_s \Omega} dl \ . \tag{2.1}$$

Assuming $E_b$ = 60 kT, $M_s$ = $10^5$ A/m, $\Omega$ = 40,000 nm$^3$, and the loop length for the current is $2\pi \times 100$ nm, we get $I$ = 30 mA. Let us assume that the loop wire is made of silver which has the highest conductivity among normal metals (resistivity $\rho$ = 16 n-ohms-m). Assuming further that the wire diameter is 50 nm, the loop resistance turns out to be $R$ = 5 ohms and the power dissipation in the loop is $I^2 R = 4.5$ mW . If the time it takes to switch the magnetization is ~ 1 ns (a reasonable estimate), then the energy dissipated in the external circuitry to switch the nanomagnet is $4.5 \text{ mW} \times 1 \text{ ns} = 4.5$ pJ or $10^9$ kT at room temperature, which is excessive. Therefore, the use of a current generated magnetic field is *not* energy-efficient [24] and therefore not advisable.

## 2.2  Switching a nanomagnet with spin transfer torque (STT) generated by a spin-polarized current

One method frequently employed to rotate a nanomagnet's magnetization to a desired orientation is to pass a spin-polarized current through it, which allows selectively addressing a nanomagnet in close proximity to others (an advantage over the previous scheme). The spins of the electrons carrying the current are polarized in the direction of the intended magnetization. The spin polarized electrons transfer their spin angular momentum to the resident electron spins in the nanomagnet [38-43], thereby applying a spin-transfer-torque (STT) on the nanomagnet's magnetization and switching it to the desired orientation. This is shown in Fig. 5(a).

There are many ways of generating the spin-polarized current. The most common way is to use a magneto-tunneling junction (MTJ) shown in Fig. 5 (b). It consists of a "hard" nanomagnet (with stiff magnetization that is not easily rotated), an ultrathin spacer layer, and a "soft" nanomagnet whose magnetization can be rotated by a spin-polarized current passing through it and imparting spin angular momentum to the resident electrons. The hard layer is implemented with a synthetic anti-ferromagnet to reduce the dipole coupling between the hard and soft layers.

The MTJ is an iconic device used to *write, store and read* bits in spin-transfer-torque magnetic random access memory (STT-MRAM). Typically, when the magnetizations of the hard and soft layers are anti-parallel, the resistance is highest, and when they are parallel, the resistance is lowest. When the two magnetizations are at an angle $\theta$, the resistance of the MTJ, $R(\theta)$, obeys the relation [44]

$$\left[\frac{R(\theta)-R(0)}{R(\pi)-R(0)}\right]_{MTJ} = \frac{1-\cos\theta}{\chi(1+\cos\theta)+2}, \quad (2.2)$$

where $\chi = 2P_1P_2/(1-P_1P_2)$ and $P_1$, $P_2$ are the interfacial spin polarizations at the two ferromagnet/spacer interfaces. If we align the soft layer's magnetization parallel to that of the hard layer's, the MTJ will have a low resistance which can store and encode the binary bit 1 and when the two layers have anti-parallel magnetization, the resistance will be high and that could store and encode the binary bit 0. These are shown in Figs. 5 (c) and (d). In the end, the bit is stored in the MTJ's resistance state and therefore we can read it by measuring the resistance and determining whether it is high or low. The bit can be written into the MTJ by aligning the soft layer's magnetization parallel or anti-parallel to that of hard layer by rotating the soft layer's magnetization with a spin polarized current flowing through it. Note that the "parallel" and "anti-parallel" states are the only two stable states of the MTJ if the hard and soft layers are implemented with nanomagnets possessing PMA or IPA.

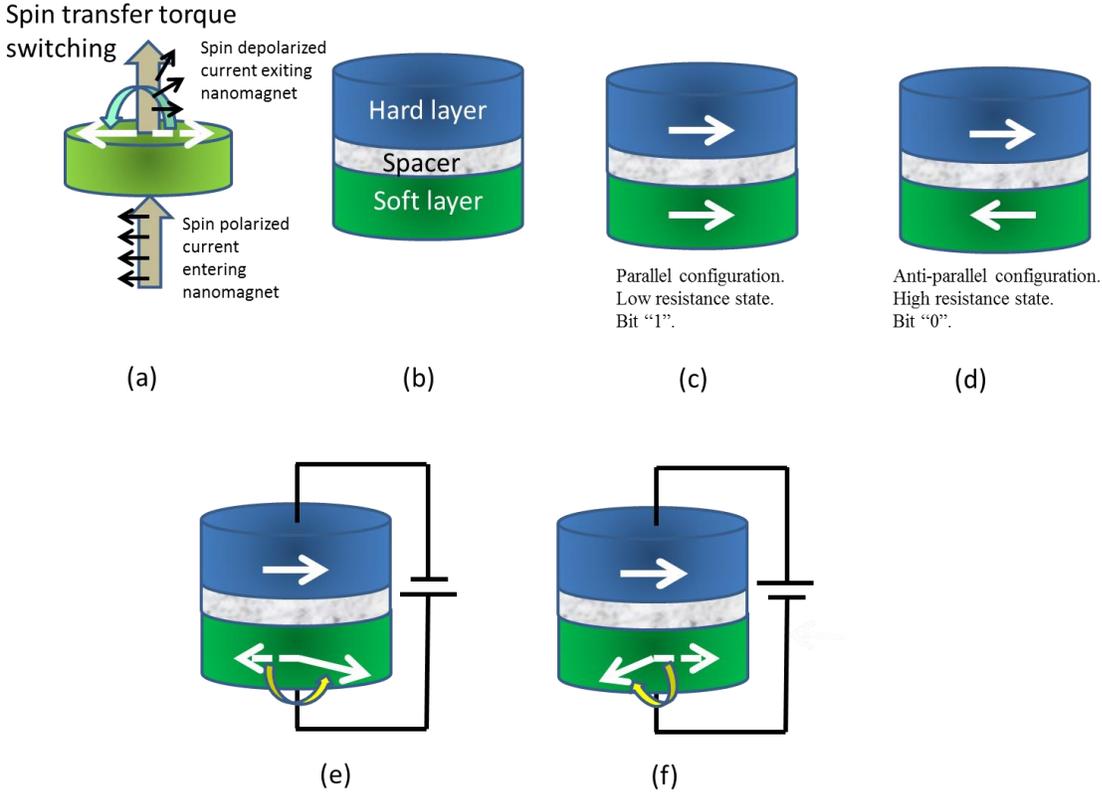

**Fig. 5**: (a) Spin transfer torque switching of a nanomagnet's magnetization from the right-direction (broken arrow) to the left-direction (solid arrow) with a spin polarized current in which the spins of electrons are oriented in the left-direction. (b) A magneto-tunneling junction (MTJ) structure. (c) When the hard and soft layers' magnetizations are mutually parallel, the MTJ resistance measured between the hard and soft layers is low. (d) When the hard and soft layers' magnetizations are mutually anti-parallel, the MTJ resistance is high. (e) Switching the soft layer from the anti-parallel configuration (broken arrow) to the parallel configuration (solid arrow) with the polarity of the battery shown. The hard layer acts as a spin polarizer. (f) Switching the soft layer from the parallel configuration (broken arrow) to the anti-parallel configuration (solid arrow) with the polarity of the battery reversed. The hard layer acts as a spin analyzer.

In Fig. 5, we show an MTJ built with hard and soft layers possessing IPA. The spin-polarized current flows in a direction perpendicular to the heterointerfaces by tunneling through the spacer and it is generated in the following way. If we connect the negative terminal of the battery to the hard layer and the positive terminal to the soft layer, then the hard layer will inject its majority spin electrons (spins polarized parallel to the hard layer's magnetization) into the soft layer. This constitutes a spin-polarized current injected into the soft layer. The injected spins will transfer their momenta to the electrons in the soft layer and ultimately align the latter's spins in the direction of the hard layer's magnetization, thereby making the two magnetizations mutually parallel (Fig. 5 (e)), writing the bit 1. The hard layer acts as the spin polarizer and generates the spin polarized current that switches the soft layer.

If we reverse the polarity of the battery, the soft layer will try to inject electrons into the hard layer and generate a spin-polarized current. Electrons whose spins are aligned parallel to the hard layer's magnetization will be preferentially transmitted by the hard layer which acts as a spin analyzer or filter. Therefore, the soft layer will more successfully inject those spins that are parallel to the hard layer's magnetization. Continued injection depletes the population of these spins in the soft layer, so that ultimately spins that are anti-parallel to the hard layer's magnetization become majority spins in the soft layer. This makes the soft layer's magnetization anti-parallel to that of the hard layer's (Fig. 5 (f)) and writes the bit 0. Thus, we can write either bit 0 or bit 1 into the memory by choosing the polarity of the battery.

This method of switching magnetization with a spin polarized current is not particularly energy-efficient either and is likely to dissipate about $10^7$ $kT$ of energy (~1.6 pJ) to switch a single-domain nanomagnet in ~ 1 ns, even when the energy barrier $E_b$ within the magnet is only few tens of $kT$ [45]. More recent estimates bring this number down to ~100 fJ [46], which is still excessive. The key advantage of this method is that it can be used to address individual nanomagnets among an assembly of many, unlike a magnetic field which cannot be confined to a single nanomagnet easily. The magnetic field approach is useful when a large number of densely packed nanomagnets have to be switched in the same direction, e.g. in "initialization" steps. One would not need to access every nanomagnet with electrical contacts and that reduces the lithography overhead. But whenever different nanomagnets have to be switched in different directions, because they will encode different bits, we cannot use the magnetic field approach and must use a method such as spin-transfer-torque (STT) induced by spin polarized currents injected into nanomagnets individually. This requires contacting every nanomagnet electrically, but this is not difficult to do with crossbar architectures.

In spin transfer torque (STT) switching, there is a minimum amount of current needed to switch the magnetization of a nanomagnet and that is called the "critical current" $I^{cr}$. It depends on the energy barrier between the two stable magnetization states of the nanomagnet, the degree of spin polarization of the current, and a few other factors. This critical current density can be quite high (~ 1MA/cm$^2$) and that is the primary reason for the high energy dissipation in this mode of switching. The external energy dissipation is $I^2 R \Delta t$ where $I$ is the current through the MTJ, $R$ is the resistance of the MTJ (quite high because of the spacer layer) and $\Delta t$ is the time taken to rotate the soft layer's magnetization. Since $\Delta t$ is a function of $I$, there is some optimization involved in choosing the right amount of current to minimize the energy dissipation.

The use of heat assisted switching to lower the critical current has been investigated theoretically [47] while thermally assisted switching of the soft layer of a magnetic tunnel junction that is exchange biased has been experimentally demonstrated [48]. It was shown that a current pulse can raise the temperature of the soft layer above its blocking temperature without significantly affecting the hard layer and that makes it easier to rotate the former's magnetization with a lower current. The design challenges associated with these strategies are discussed in Ref [49]. Application of a short-duration large-amplitude pulse and long-duration small-amplitude pulse lead to two different switching regimes: one dominated by the angular momentum of the current pulse and the other a spin transfer assisted thermal activation over the energy barrier separating the two easy directions [50]. In the end, however, the switching current is not reduced enough to make STT current based switching mechanism competitive in terms of energy dissipation. STT may be better than using an on-chip current-generated magnetic field to switch nanomagnets, but there are switching strategies that are potentially more energy efficient than STT.

**2.2.1  STT where the spin polarized current is generated via the giant Spin Hall Effect without a spin polarizing magnet**

A recent idea to reduce the threshold current needed to switch a nanomagnet with STT incorporates the giant Spin Hall Effect [51-53] which is elucidated in Fig. 6.

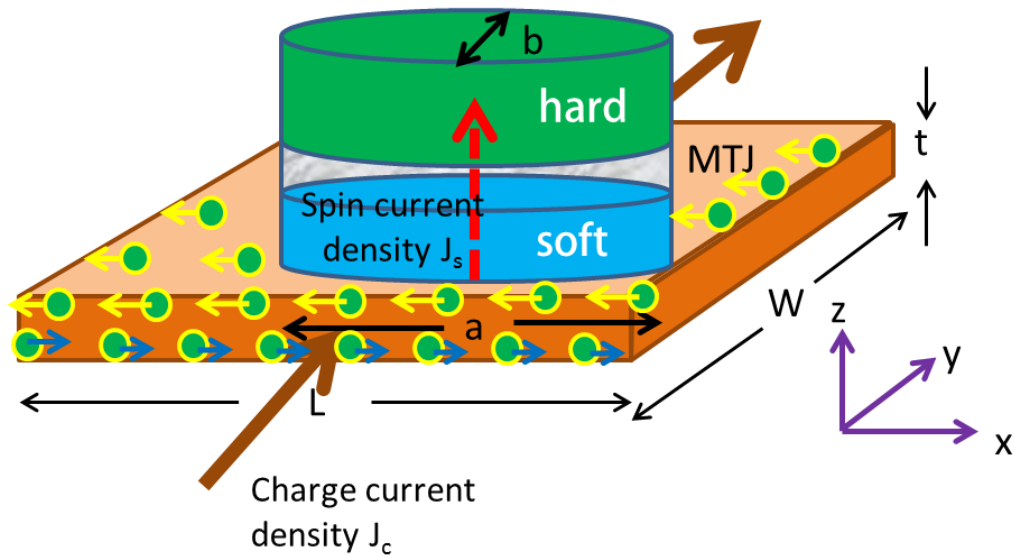

**Fig. 6**: Switching the magnetization of an elliptical nanomagnet with spin transfer torque via the giant Spin Hall Effect.

Consider a two-dimensional slab of material shown in Fig. 6 that has strong spin-orbit interaction. A "charge" current of density $J_c$ is injected into it in the *y*-direction and has no net spin polarization. The electrons in the injected current suffer spin-dependent scattering as they travel through the slab (because of the spin-orbit interaction), as a result of which, +*x*-polarized spins are deflected to the bottom edge of the slab and –*x*-polarized spins are deflected to the top edge as shown by the green circles in Fig. 6. This builds up a spin imbalance (preponderance of +*x*-polarized spins in the bottom surface of the slab and preponderance of –*x*-polarized spins in the top surface) that drives a spin current of density $J_s$ in the *z*-direction. This spin (diffusion) current flows through the MTJ and rotates the magnetization of the soft layer by delivering a spin-transfer torque. In Fig. 6, the magnetization of the soft layer will turn to the –*x* direction if it was originally pointing in the +*x*-direction, and it will remain unaffected if it was originally pointing in the –*x*-direction. For the reverse process of switching the magnetization of the soft layer from the –*x*-direction to the +*x*-direction, we simply have to reverse the polarity of the charge current, which will reverse the spin orientations at the two edges.

There may be other sources for the spin torque in the device in Fig. 6 such as the Rashba-Edelstein effect in the ferromagnet itself [54, 55]. In a ferromagnet with Rashba spin orbit interaction [56], passage of a current can cause spin polarization in a particular direction and switch the magnet's magnetization to that direction. Alternately, spin-orbit interaction acts like an effective magnetic field [57] and that field can switch the magnetization of the magnet by delivering a torque. There is no consensus yet about the actual mechanism for producing the torque, but there is plenty of experimental evidence that a torque is produced and that it can switch the magnet [55]. Here, we will assume, for simplicity, that the Spin Hall Effect is the source of the torque and the spin polarized current caused by the charge current is responsible for switching.

The ratio of the two current densities – spin current density and charge current density – is called the "spin Hall angle" $\theta_{SH}$:

$$\theta_{SH} = \frac{J_s}{J_c}. \tag{2.3}$$

A more accurate expression relating the charge current density to the spin current density is [58]

$$\frac{J_s}{J_c} = \theta_{SH}\left(1 - \text{sech}\frac{t}{L_s}\right), \tag{2.4}$$

where $t$ is the slab thickness and $L_s$ is the spin diffusion length. The spin Hall angle is usually quite small in most materials, but in certain materials it can be large. It is reported to be 0.15 in β-Ta [51], 0.3 in β-W [52] and 0.24 in CuBi alloys [53]. These materials are therefore said to exhibit the *giant* Spin Hall Effect. The spin current $I_s$ can be used to deliver a spin transfer torque on a soft magnet and rotate its magnetization.

Note that the spin current does not dissipate any power since the scalar product $\vec{J}_s \bullet \vec{\mathcal{E}} = 0$, where $\vec{\mathcal{E}}$ is the electric field driving the charge current and it is collinear with $\vec{J}_c$ which is perpendicular to $\vec{J}_s$. Any power dissipation is due to the charge current. In our MTJ of elliptical cross-section with major axis = $a$ and minor axis = $b$, the minimum power dissipation can be approximately written as

$$P_d = \left(I_s^{cr}\right)^2 \left(\frac{4}{\pi \theta_{SH}}\right)^2 \rho \frac{t}{ab} \tag{2.5}$$

Clearly, there are two ways to make the power (and energy) dissipation small: first by using a material with large spin Hall angle, and second by using a slab with very small thickness $t$ [5, 46]. The energy dissipation can be reduced to ~$10^4$ kT (1.6 fJ) by using this approach, and perhaps even lower [59].

Note also from Equation (2.3) that the ratio of the spin current to charge current is

$$\beta = \frac{I_s}{I_c} = \theta_{SH} \frac{(\pi/4)ab}{Lt} = \theta_{SH} \frac{A_{MTJ}}{A_{slab}}, \tag{2.6}$$

where $A_{MTJ}$ is the cross-sectional area of the MTJ through which the spin current flows and $A_{slab}$ is the cross-sectional area of slab through which the charge current flows. Further, if we assume $L \sim a$, then the ratio of the spin current to charge current is

$$\beta = \frac{I_s}{I_c} = \theta_{SH} \frac{(\pi/4)b}{t} \approx \theta_{SH} \frac{b}{t} \tag{2.7}$$

The quantity $\beta$ acts like a "gain" [5] and its value can be made much larger than unity by ensuring that $A_{slab}/A_{MTJ} << \theta_{SH}$, or specifically by ensuring that the ratio of the nanomagnet lateral dimension to slab thickness $t/b << \theta_{SH}$. A scaled nanomagnet will have ~ 50 nm lateral dimension and the slab thickness can be ~ 2nm, resulting in $t/b$ ~ 0.04 << $\theta_{SH} = 0.3$ in β-W [50]. Thus, a gain of $\beta$ ~ 10 is quite feasible.

We can explain the role of the giant Spin Hall Effect in reducing energy dissipation in STT switching by invoking Equations (1.1) and (1.2). We follow the arguments presented in [5]. Recall from Equation (1.2) that the energy-delay product is proportional to $(\Delta Q)^2$, where $\Delta Q$ is the amount of charge that must move through the device. In STT-switching, we have to rotate the spins of a certain number electrons in the nanomagnet to switch. Let that number be $N$. Since the electrons passing through the nanomagnet impart their momenta to the spins for reorientation, the number of electrons needed to reorient $N$ spins must exceed $2N$ and therefore $\Delta Q \geq 2qN$, where $q$ is the electronic charge.

The quantity $N$, the number of spins in a nanomagnet of volume $\Omega$, is given by [5]

$$N = \frac{M_s \Omega}{\mu_B} , \qquad (2.8)$$

where $M_s$ is the saturation magnetization of the nanomagnet and $\mu_B$ is the Bohr magneton. The energy barrier $E_b$ that keeps the two magnetization orientations encoding logic bits 0 and 1 well separated in energy is proportional to the volume $\Omega$. Therefore, $N$ depends on $E_b$. Now, recall that the static error probability is related to $E_b$ as

$$p_{static} = e^{-E_b/kT} . \qquad (2.9)$$

$$\Rightarrow E_b = kT \ln(1/p_{static}) \qquad (2.10)$$

Hence $N$ (and consequently the energy-delay product) depends on the built-in error resilience. Once again, the energy-delay product depends on how much error probability we are able to tolerate.

In the case of giant Spin Hall Effect (GSHE), the constraint $\Delta Q \geq 2qN$ no longer applies since the charge current can be $\beta$ times less than the spin current. Therefore, the correct constraint will be

$$\Delta Q \geq 2qN/\beta \qquad (2.11)$$

Using Equations (2.7) - (2.11), we can recast the last condition as

$$\Delta Q|_{GSHE} \geq \frac{2q}{\beta}\left(\frac{M_s \Omega_{magnet}}{\mu_B}\right) = \frac{2q}{\beta}\left(\frac{M_s}{\mu_B}\right)\frac{E_b}{K_u} \qquad (2.12)$$

where $K_u$ is the magnetic anisotropy energy density, $\Omega_{magnet}$ is the volume of the nanomagnet (such that $K_u \Omega_{magnet} = E_b$) and other terms have been defined earlier. Assuming $M_s = 10^5$ A/m, $E_b = 1$ eV, $K_u = 10^6$ J/m$^3$ and $\beta \sim 10$, one gets $\Delta Q \approx 345$ electrons.

Compare this number with $\Delta Q \approx 673$ electrons for a modern day 14-nm FINFET. The GSHE-enhanced STT nanomagnetic switch is comparable with a scaled FINFET in terms of the magnitude of $\Delta Q$. However, the slab in Fig. 6 can be made of heavy metals which have lower resistances than semiconductor structures, so the energy delay product $(\Delta Q)^2 R$ could potentially be smaller in GSHE-enhanced STT-switched nanomagnets than in transistors.

It is clear that the energy-delay product, which is proportional to $(\Delta Q)^2$, will scale with $1/\beta^2$. Thus, the Giant Spin Hall Effect not only provides gain, but it also reduces the energy-delay product considerably. The factor $\beta$ can be of the order of 10; hence, it is possible to reduce energy-delay product in STT-switching by two orders of magnitude by utilizing the Giant Spin Hall Effect. Experiments have shown that the energy dissipation associated with switching via the Giant Spin Hall Effect in low loss magnetic materials like CoFeB is ~ $10^4$ kT or 1.6 fJ and with further scaling can be below 100 aJ [59].

We will show later that there are other switching methodologies where the constraint $\Delta Q \geq 2qN$ does not apply. These methodologies are also worthy of investigation since they provide a clear pathway to reducing the energy-delay product.

We conclude this section with a few words about a recent approach to make the quantity $(J_s/\theta_{SH} J_c) \gg 1$. This requires inserting a ferromagnetic insulator (e.g. yttrium iron garnet) supporting pure spin current via magnon diffusion between the giant spin Hall slab and the soft layer of the magnetotunneling junction [60]. The role of the insulator is to funnel spin current from the large area of the giant spin Hall material into the small area of the soft layer. As the cross-sectional areas of magneto-tunneling junctions are scaled down, this will be increasingly important.

### 2.2.2 STT generated by a topological insulator

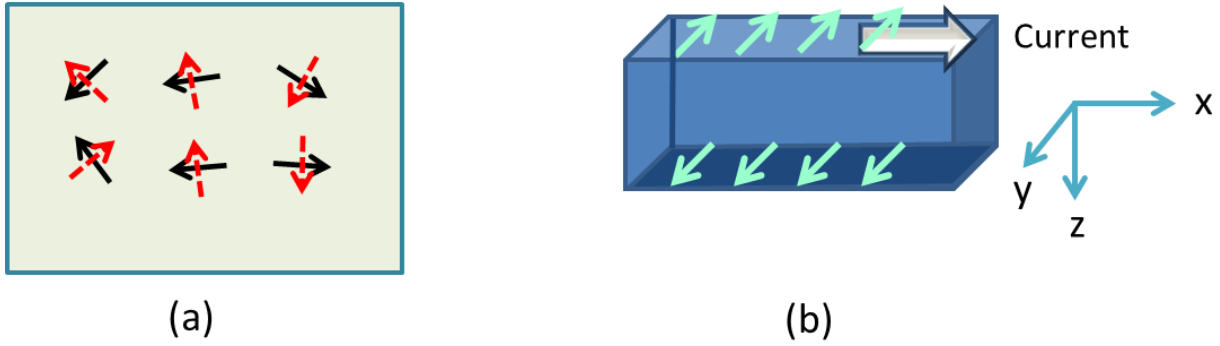

**Fig. 7**: (a) The spin orientation of an electron on the surface of a topological insulator (shown by the broken red arrow) is uniquely determined by the direction of its velocity (shown by the solid black arrow). (b) When a current flows along the surface of a topological insulator, there is a net velocity in the direction of current flow. Hence there is a preponderance of spins associated with that velocity, resulting in a spin-polarized surface (spins shown by the thin arrows).

Topological insulators (TI) are materials with many interesting properties [61, 62], but here we will be concerned with one special property that is of particular relevance. An ideal three-dimensional topological insulator will not allow current conduction in the bulk (hence an "insulator"), but current can flow in any direction on the surface. The direction in which the current carrying electron is moving uniquely determines its spin orientation and vice versa as shown in Fig. 7. This is an intriguing feature of topological insulators and is sometimes referred to as "spin-momentum-locking". In equilibrium, no current flows and hence there is no net velocity of the electrons in the surface. However, when a current does flow, the net electron velocity becomes non-zero and hence the net spin becomes non-zero in the surface of a TI, resulting in a spin-polarized surface.

Consider the topological insulator (TI) shown in Fig. 7 (b). A current flows on the top surface along the +x-direction and hence the surface develops a net spin polarization in the -y-direction.

The Hamiltonian describing the surface state of the TI is

$$H_{\vec{k}} = v_F \left( \hat{z} \times \vec{\sigma} \right) \bullet \vec{k} - \varepsilon ,  \qquad (2.13)$$

where $\vec{k}$ is the electron's wavevector, $v_F$ is the Fermi velocity on the surface, $\vec{\sigma}$ is the Pauli spin matrix operator, $\varepsilon$ is a constant and $\hat{z}$ is the unit vector normal to the surface. The velocity operator $v = \vec{\nabla}_{\vec{k}} H_{\vec{k}} = 2 v_F \left( \hat{z} \times \vec{S} \right) / \hbar$, where $\vec{S}$ is the spin operator given by $\vec{S} = (\hbar/2) \vec{\sigma}$. A current density $J_x = nev_x$ in the $x$-direction ($n$ = electron concentration and $e$ = electronic charge) will result in a net spin in the $–y$-direction given by

$$\langle S_y \rangle = -\frac{\hbar}{2ev_F} J_x \qquad (2.14)$$

Consider a system shown in Fig. 8 where a nanomagnet is placed on top of the TI shown in Fig. 7(b). In addition to the spin accumulation described above, there is another possible source of spin accumulation if the TI happens to be certain materials like $Bi_2Se_3$ [63]. The interface between the nanomagnet and the TI contains a two-dimensional electron gas with Rashba spin-orbit interaction [64] which leads to a spin accumulation at the interface given by [65-67].

$$\langle S_y \rangle \big|_{Rashba} = \frac{\hbar}{2e} \frac{m^* \alpha_R J_x}{2 E_F}, \qquad (2.15)$$

where $E_F$ is the Fermi energy, $\alpha_R$ is the Rashba constant and $m^*$ is the electron's effective mass at the interface. Therefore, the total spin accumulation is the sum of the terms in Equations (2.14) and (2.15).

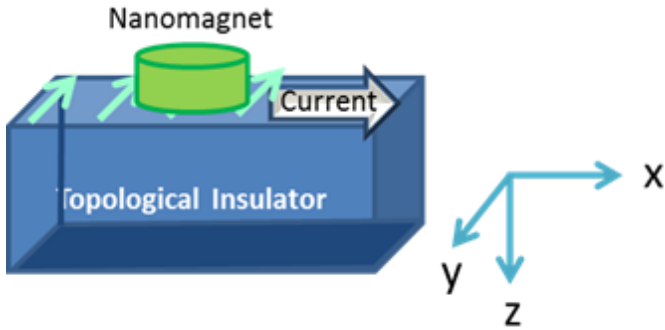

**Fig. 8**: A nanomagnet whose spins are coupled to the surface spins of the topological insulator underneath.

Because of exchange coupling between the spins on the surface of the TI and the spins in the nanomagnet at the interface with the TI, the spins in the nanomagnet at the interface *align parallel* to the spins on the TI surface, so the interface of the nanomagnet has a net spin accumulation. This spin accumulation produces a diffusion of spins in the direction normal to the magnet-TI interface ($z$-direction) in the nanomagnet, which, in turn, gives rise to a torque on the magnetization of the nanomagnet that can switch its magnetization. One obvious drawback of this scheme is that a metallic nanomagnet will shunt the surface current away from the TI and reduce the spin accumulation, which will reduce the torque and increase the current needed to switch. Therefore, this scheme works best with insulating nanomagnets [68]. If the shunting problem can be overcome, then the effective spin Hall angle, defined as the ratio of the spin diffusion current density to the charge current density injected into the TI, can exceed unity.

## 2.3     Domain wall motion with STT and strain

A nanomagnet's magnetization direction can be switched by inducing domain wall motion [69]. Consider, a current that flows through the left most domain of a multi-domain nanomagnet in Fig. 9 (domain's magnetization pointing right). While flowing through this domain, the current gets spin-polarized because the spins of majority electrons in that domain are polarized in the direction of its magnetization. This spin-polarized current then delivers a spin transfer torque on first few atomic layers of the next domain and orients its magnetization in the same direction as the preceding domain, thereby moving the domain wall further to the right. Thus, the domain wall progressively moves to the right. There is at least one report of switching a multi-domain nanomagnet's magnetization in 2 ns by moving domain walls using a current of 0.1 mA [70] using the above method.

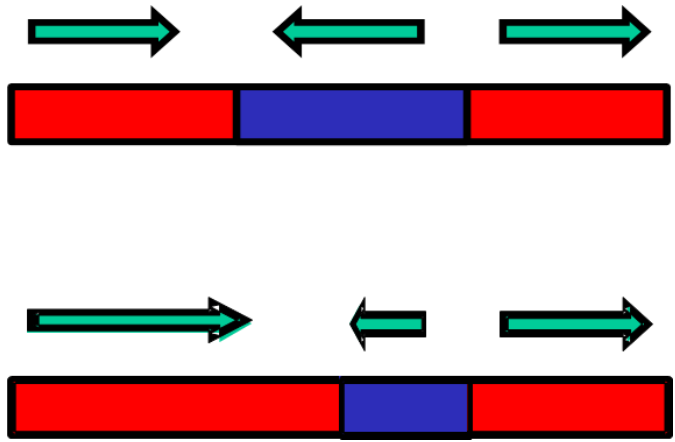

**Fig 9:** Domain wall motion induced with a current flowing from right to left (electrons moving from left to right). The red domains are magnetized in one direction and the blue domain in the opposite direction. The electrons entering the blue domain from the left red domain are spin-polarized in the direction of magnetization of the red domain. They transfer their spin angular momentum to the spins in the blue domain and thus exerts a spin-transfer-torque on the blue domain spins. As the resident blue domain spins begin to turn in the direction of the red domain spins because of the torque, the blue domain shrinks and the red domain expands, pushing the domain wall between them to the right.

Other novel nanowire/nanostrip based memory [71] and logic schemes [72] using magnetic domains and their manipulation have also been experimentally demonstrated. In the latter scheme, a global rotating magnetic field was used to clock the domain wall motion and resulted in NOT, AND, FANOUT and crossover functionalities. Finally, recent experiments have demonstrated that mechanical strain can be used to control the energy barrier for domain wall motion [73] and thus affect its propagation. In these experiments, a multi-domain magnet is fabricated on a piezoelectric substrate. By applying a voltage across the substrate, a strain is generated in the piezoelectric. That strain is partially transferred to the magnet resting on top of the piezoelectric and it modifies the energy barrier for the domain wall motion. This can impede domain wall motion. Such a construct can be used to implement a domain wall gate or stabilize domain walls in memory applications [73].

Recently, a surface acoustic wave, generating periodic strain, has been utilized to induce domain wall motion in a Co/Pt thin film [74]. A standing surface acoustic wave increases the velocity of domain wall motion in these thins films by an order of magnitude compared to magnetic field alone [74]. Furthermore, a recent proposal suggests strain gradient produced by electric fields can move domain walls [75] and thus switch nanomagnets.

## 2.4 Switching with Spin Orbit Torques

The switching schemes discussed in sections 2.2 and 2.3 need a charge current. Therefore, we call these mechanisms "*current controlled-switching*".

There is another important current-controlled nanomagnet switching mechanism that is similar to the GSHE or TI switching mechanisms. It is based on spin-orbit torque generated by the flow of a current. Switching of nanomagnets or magnetic layers with interface Rashba spin orbit torques (associated with the Rashba spin-orbit coupling [56]) have been reported [76, 77]. Rashba spin-orbit interaction arises in a structure with structural inversion asymmetry that produces a non-zero slope of the conduction band profile of a solid [56]. This leads to an effective electric field $\vec{E}$ which gives rise to spin-orbit interaction. Spin-orbit interaction acts like an effective magnetic field on an electron's spin, which can be expressed as

$$\vec{H}_{\text{Rashba}} = \alpha_R \left( \frac{\vec{E}}{|\vec{E}|} \times \langle \vec{k} \rangle \right) \qquad (2.16)$$

where $\alpha_R$ is the Rashba coupling constant (usually proportional to $\vec{E}$) and $\vec{k}$ is an electron's wavevector with the angular bracket denoting ensemble average over all electrons. At equilibrium when no current flows through the solid, $\langle \vec{k} \rangle = 0$ since $+\vec{k}$ and $-\vec{k}$ states are equally populated and hence $\vec{H}_{Rashba} = 0$. When a current flows, $\langle \vec{k} \rangle \neq 0$ and $\vec{H}_{Rashba} \neq 0$. The Rashba magnetic field causes an effective spin accumulation given by the expression in Equation (2.14). This spin accumulation causes spin diffusion into the magnet that generates a spin orbit torque.

The Rashba spin-orbit torque can be used to switch nanomagnets with perpendicular magnetic anisotropy (PMA). However, this usually requires an in-plane magnetic field. The Rashba interaction requires structural inversional asymmetry and this is produced by sandwiching the magnetic layer between two layers of *different* material composition [77]. We call this vertical structural asymmetry. Recently, lateral structural asymmetry was employed to generate a spin–orbit torque that enabled the switching of perpendicular magnetization *without* using an in-plane bias magnetic field [78]. It should be possible for spin-orbit torque to move domain walls as well and indeed this has now been shown to be able to switch magnets [79, 80].

## 2.5    Energy dissipation in current controlled switching

All the schemes that we have described so far require a charge current to flow in order to switch the magnetization of a nanomagnet. This invariably involves an $I^2R$ loss which is determined by the amount of current $I$ needed to switch the nanomagnet in a given time and the resistance $R$ in the path of the current. For conventional STT switching, an approximate analytical expression for the switching time $t_s$ is [81]:

$$t_s = \frac{\tau_0}{I/I_{cr} - 1} \ln\left(\frac{\theta_f}{2\theta_i}\right),$$
$$\tau_0 = \frac{M_s \Omega}{\eta \mu_B} \frac{e}{I_{cr}}$$
(2.17)

where $M_s$ is the saturation magnetization per unit volume, $\Omega$ is the nanomagnet volume, $I_{cr}$ is the critical current for switching, $\eta$ is the spin injection efficiency into the nanomagnet, $\mu_B$ is the Bohr magneton and $\theta_i$ is the polar angle of the magnetization vector in its initial location around one stable orientation along the easy axis. The polar angle for the other stable orientation (final location) is $\theta_f$.

The critical current is approximately expressed as [41]

$$I_{cr} = \frac{2e}{\hbar}\frac{\alpha}{\eta}\left[E_b + 2\mu_0 M_s^2 \Omega\right] \qquad (2.18)$$

where $E_b$ is the energy barrier separating the two stable magnetization states of the magnet (discussed in Section 1.2), $\mu_0$ is the permeability of free space and $\alpha$ is a parameter called the Gilbert damping constant [material constant] that represents the dissipation associated with damping of magnetization rotation.

Using the last two equations, one can derive an expression for the energy dissipation associated with switching:

$$E_d = P_d t_s = I^2 R t_s = I_{cr}^2 \left[\frac{\tau_0}{t_s}\ln\left(\frac{\theta_f}{2\theta_i}\right)+1\right]^2 R t_s = I_{cr}^2 \left[\frac{M_s \Omega}{\eta \mu_B}\frac{e}{I_c t_s}\ln\left(\frac{\theta_f}{2\theta_i}\right)+1\right]^2 R t_s$$

$$= \left[\frac{2e}{\hbar}\frac{\alpha}{\eta}\left(E_b + 2\mu_0 M_s^2 \Omega\right)\right]^2 \left[\frac{M_s \Omega}{\eta \mu_B}\frac{e}{\left(\frac{2e}{\hbar}\frac{\alpha}{\eta}\left[E_b + 2\mu_0 M_s^2 \Omega\right]\right)t_s}\ln\left(\frac{\theta_f}{2\theta_i}\right)+1\right]^2 R t_s \qquad (2.19)$$

where $P_d$ is the power dissipation. For switching with spin transfer torque (STT) generated via the Giant Spin Hall Effect (GSHE), the energy dissipation will decrease by a factor $\beta^2$ where $\beta$ is the gain factor given in Equation (2.7). Furthermore, the resistance ($R$) is much smaller in the GSHE geometry than in the case of applying STT with spin polarized current in a magnetic tunnel junction where the spacer layer acting as the tunnel barrier contributes a large resistance.

Clearly, the energy dissipation depends on the switching time $t_s$. Manipatruni et al. [81] have estimated that for normal STT switching in 1 nsec, the energy dissipation is about ~100 fJ ($2.4\times 10^7$ kT at room temperature), whereas for giant Spin Hall Effect assisted switching, it is ~100 aJ ($2.4\times 10^4$ kT at room temperature) in typical cases. Curiously, Equation (2.19) shows that there is a *minimum* of the energy dissipation and it occurs when the switching time is

$$t_s^{min} = \frac{M_s \Omega}{\eta \mu_B}\frac{e}{\left(\frac{2e}{\hbar}\frac{\alpha}{\eta}\left[E_b + 2\mu_0 M_s^2 \Omega\right]\right)}\ln\left(\frac{\theta_f}{2\theta_i}\right). \qquad (2.20)$$

This is the optimum switching time.

## 2.6 Voltage controlled magnetic anisotropy due to voltage induced modulation of exchange interaction at a nanomagnet and topological insulator interface

A nanomagnet can be also switched with a voltage instead of a current [82, 83]. There are many ways to switch the magnetization of a nanomagnet with a voltage and one of them is by controlling magnetic anisotropy.

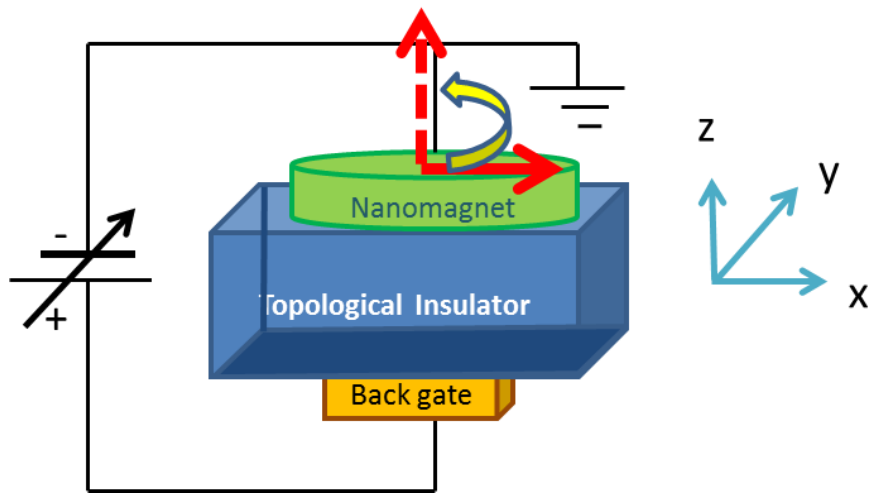

**Fig. 10**: One form of voltage controlled magnetic anisotropy. Applying a back-gate voltage can lift the magnetization from in-plane to out-of-plane, resulting in a $90^0$ rotation of the magnetization vector. This is caused by voltage-controlled modification of the exchange interaction at the interface between the nanomagnet and the topological insulator.

Consider the nanomagnet in Fig. 10 resting on a topological insulator (TI). No current flows through the nanomagnet (either charge or spin), but the topological insulator is back-gated, so the application of an electrostatic potential between the back gate and the grounded nanomagnet varies the Fermi energy of the electrons at the TI-nanomagnet interface. This modulates the exchange interaction at the interface and hence the free energy of the combined system. The spin-momentum interlocking of the TI surface electrons can change the potential energy landscape of the nanomagnet when the back-gate potential is varied to change the Fermi energy. An IPA nanomagnet's easy axis will be in-plane along the major axis of the ellipse (*x*-axis), i.e., the potential energy minima will correspond to the magnetization orienting in the $\pm x$ direction. However, when the Fermi energy is varied near the Dirac point of the TI, the potential energy minimum can shift to the location corresponding to the magnetization orienting in the $\pm z$ direction. In that case, the magnetization will lift out of the plane (away from the $\pm x$ direction towards the $\pm z$ direction) and align itself perpendicular to the plane (along the $\pm z$ direction), resulting in a $90^0$ rotation of the magnetization vector in a time scale of nanoseconds [68]. This corresponds to changing the magnetic anisotropy from in-plane to perpendicular-to-plane and is shown in Fig. 10.

In order to understand how this happens, consider the fact that the potential energy of the nanomagnet is given by the expression [68]

$$U = 2\pi\Omega M_s^2 \sum_{i=x,y,x} N_i \hat{m}_i^2 + \upsilon(E_F,T)\left(1-\hat{m}_z^2\right)$$

$$\sum_i N_i = 1 \qquad (2.21)$$

where $\upsilon(E_F,T)$ is an exchange interaction term that can be varied by varying the Fermi energy $E_F$ with the back-gate potential, $\hat{m}_i$ is the normalized component of the magnetization vector along the $i$-th coordinate axis, and $N_i$ is the demagnetizing factor in the $i$-th direction that depends on the shape of the nanomagnet. It is given by [68]

$$N_i = \frac{d_x d_y d_z}{2} \int_0^\infty d\xi \frac{1}{\Xi_i \sqrt{\Xi_x \Xi_y \Xi_z}}$$

$$\Xi_i = \xi + d_i^2 \qquad (2.22)$$

where $d_x$, $d_y$ and $d_z$ are the dimensions of the elliptical disk magnet along the three principal axes and $\xi$ is a dummy variable for integration. Equation (2.21) shows that by varying $\upsilon(E_F,T)$, we can move the minimum of the potential energy $U$ (and hence the easy axis of magnetization) from the $x$-axis to the $z$-axis.

## 2.6.1. Voltage controlled magnetization reversal in a ferromagnet layer-multiferroic (with coupled antiferromagnetic and polarization states) heterostructure due to exchange coupling

In a heterostructure consisting of a ferromagnet deposited on a single-phase multiferroic layer whose antiferromagnetic and ferroelectric polarization states are coupled, an electric field applied to the multiferroic layer can change the polarization of the ferroelectric domains in the multiferroic. This can result in a rotation of the magnetization in the antiferromagnetic magnetic planes of the multiferroic resulting in an in-plane rotation of the canted moment in the atomic plane adjacent to the ferromagnetic layer. The ferromagnet's magnetic moment exchange couples to the canted moment in the multiferroic antiferromagnet's atomic plane adjacent to it and hence rotates when the latter rotates. This results in magnetization rotation in the ferrromagnet owing to the voltage applied across the multiferroic and has been demonstrated in a $Co_{0.9}Fe_{0.1}$/$BiFeO_3$ heterostructure [84].

## 2.6.2 Voltage controlled magnetization reversal through electrical field control of the Dzyaloshinskii-Moriya (DM) vector

Recent work has shown deterministic $180^0$ rotation of the canted moment in $BiFeO_3$ under the influence of an electric field at room temperature [85, 86]. Here, the ferromagnetism in $BiFeO_3$ is due to the Dzyaloshinskii-Moriya (DM) interaction [85, 86] and an electric field applied to $BiFeO_3$ can rotate the DM vector by $180^0$, resulting in complete magnetization reversal at an energy cost that is one order of magnitude smaller than that associated with spin transfer torque switching. This feature has been exploited to switch the resistance of a spin valve device [87].

## 2.6.3 Voltage control of magnetic anisotropy at a magnet-tunnel barrier interface due to band filling in the magnet and spin-orbit interaction

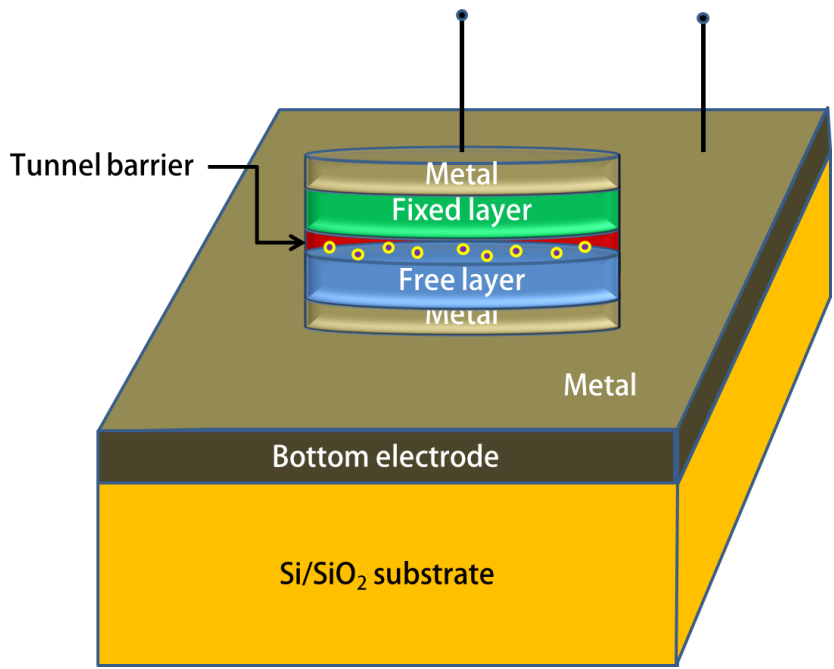

**Fig. 11**: Voltage controlled switching of magnetic anisotropy due to band filling in the soft ferromagnet and change in its magnetic anisotropy as a result of spin-orbit interaction.

Consider the magneto-tunneling junction (MTJ) structure shown in Fig. 11. A potential applied between the two electrodes shown will inject electrons into the spacer layer from either the fixed layer or the free layer, depending on the voltage polarity. Some of these electrons will accumulate in the spacer layer and modify the occupation of the "d-like" bands in the transition metal free layer. Because of spin-orbit interaction and/or spin-dependent screening, this will modify the magnetic anisotropy perpendicular to the interface in the free layer [88-91]. Such an effect is similar to the surface magneto-electric effect where an electric field modifies the magnetocrystalline anisotropy and magnetization at the interface of a ferromagnetic metal and dielectric owing to spin-dependent screening [92, 93] or change in band structure [94]. In other words, the voltage applied between the two electrodes will change the surface anisotropy constant $K_s$ (discussed in Section 2.4) within the free layer. The change is expressed through a linear relationship of the form [91]

$$K_s = K_{s0} + \frac{C_s V}{t_b} \quad , \tag{2.23}$$

where $C_s$ is the so-called VCMA coefficient, $K_{s0}$ is a constant surface anisotropy energy per unit volume, $V$ is the applied voltage and $t_b$ is the spacer layer thickness.

If the applied voltage increases $K_s$ ($C_s V$ product is positive), making $K > 0$ [recall Equation (1.5)], then the easy axis will become perpendicular to the nanomagnet's plane. On the other hand, if it decreases $K_s$ ($C_s V$ product is negative), making $K < 0$, then the easy axis will lie in the nanomagnet's plane along the major axis of the ellipse (in case of a circular nanomagnet, all in-plane directions would be equally probable). Therefore, the easy axis can transition from in-plane to out-of-plane, or vice versa, depending on the initial anisotropy (anisotropy at $V=0$) due to perpendicular magnetic anisotropy and shape anisotropy and change in surface anisotropy produced upon application of the voltage $V$. This results in changing the angle between the magnetization of the soft magnet and the hard magnet by $90^0$, which will change the MTJ resistance and accomplish switching of the resistance between two values.

Voltage control of magnetic anisotropy (VCMA) in a Fe/MgO interface [95] and in a MgO/CoFeB/Ta structure [96] has been demonstrated, as well as in MgO-based magneto-tunneling junctions [97-101]. VCMA has also been studied in monodomain nanomagnets, extending the phenomenon to the nanoscale [102]. VCMA based switching of the magnetization of soft layers in MTJ-s (and hence the switching of the MTJ resistance) can be accomplished in < 1 ns with an energy dissipation of < 40 fJ/bit [103]. Recently, giant VCMA coefficient of 1800 fJ $V^{-1}$ $m^{-1}$ in a Au/FeCo/MgO heterostructure was reported [104] and this could improve these figures further. By increasing the spacer layer thickness, it is possible to reduce the tunneling current in an MTJ when a voltage is applied to induce VCMA and this has reduced the energy dissipation in VCMA switching of MTJs down to ~ 6 fJ/bit. [105].

The VCMA mode of switching is most frequently employed in MTJs whose fixed and free layers have perpendicular magnetic anisotropy ($K > 0$) so that the easy axis of magnetization (of both fixed and free layers) is perpendicular to the plane of the MTJ when $V = 0$. Such MTJ's can have smaller cross-section or footprint than MTJs with in-plane magnetic anisotropy. Therefore, they are preferred for memory applications where high density is of paramount importance. When such perpendicular anisotropy MTJs (p-MTJ) are switched with VCMA, the magnetization of the free layer switches from out-of-plane to in-plane, meaning that the magnetization rotates by $90^0$ and not full $180^0$. This would reduce the resistance on/off ratio of the p-MTJ (sometimes referred to as the "tunneling magnetoresistance ratio" or TMR) and hinder unambiguous reading of the MTJ resistance (and hence the stored bit). More importantly, such a device would be unreliable since when the VCMA voltage is withdrawn, the magnetization will find itself in an unstable state. Thereafter, it will either return to the original state or to the state anti-parallel to the original state. If the dipole coupling between the hard and soft layer is weak due to the use of synthetic antiferromagnets for the hard layer, then the probability of returning to either state is ~50%. On the other hand, if there is significant dipole coupling, then the probabilities will be unequal (the state favored by dipole coupling will be more likely). In any case, the probability of switching correctly is far less than 100%, making this paradigm unacceptably error-prone.

The problem was overcome in ref. [100, 101] by using an in-plane bias magnetic field in circular nanomagnet discs. When the voltage pulse inducing VCMA is turned on to dislodge the magnetization vector of the free layer from the normal-to-plane orientation, it begins to precess around the bias magnetic field. At this point, the magnetization vector experiences two torques: a precessional torque that will tend to take it past a $90^0$ rotation, and a damped torque that will tend to make it settle into an in-plane orientation ($90^0$ rotation). The former torque can be made stronger by increasing the strength of the bias magnetic field. The VCMA voltage pulse duration is adjusted to approximately one-half of the precession period, which means that the voltage is withdrawn when the magnetization vector approaches the opposite normal-to-plane orientation ($180^0$ rotation). Since the opposite normal-to-plane orientation coincides with the easy axis in the absence of the voltage, the magnetization will settle into this orientation at the end of the voltage pulse, completing the $180^0$ rotation. Such an idea was proposed theoretically in ref. [106] earlier.

At first glance, this mode of switching may not appear very reliable at room temperature. In the presence of thermal noise and other perturbations, the precessional period *varies* from cycle to cycle, which means that there is a significant spread in the precessional period. Thermal noise can actually return the magnetization vector to the initial orientation after the end of the voltage pulse ($0^0$ rotation), resulting in switching failure. Fortunately, the switching is not all that unreliable. Since the opposite normal-to-plane direction is a "stable" state, as long as the magnetization comes *close* to it at the end of the voltage pulse, it will settle into this state in the end with high probability. This makes the switching fairly reliable. The reliability depends on the duration of the pulse and the in-plane magnetic field. These two parameters are independently adjusted to obtain very high switching probability [107]. Nonetheless, the disadvantage of this approach is the requirement for the external in-plane magnetic field, which is undesirable in a chip.

Recently, there has been a proposal to replace the in-plane magnetic field with an effective magnetic field due to mechanical stress [108]. In a magnetostrictive magnet, stress can act like an effective magnetic field and hence mimic the in-plane magnetic field. The stress can be generated electrically by making the free layer out of a two-phase (magnetostrictive/piezoelectric) multiferroic. A voltage applied across the piezoelectric generates strain in it. This strain is transferred to the magnetostrictive layer and acts as an in-plane magnetic field. The magnetization vector precesses about this effective magnetic field in the same way as if this was a real magnetic field. The advantage of this approach is that it is an *all-electric* implementation that eliminates the need for an on-chip bias magnetic field. The magnitude of the stress and the voltage pulse width are independently adjusted to achieve a high switching error probability.

There are other potential ways in which in-plane magnetic fields can be avoided and the switching made resilient to thermal noise and defects. A recent proposal shows that adding a heavy metal interface (to introduce Dzialoshinskii-Moriya (DM) interaction) can induce a skyrmion state when the PMA is reduced on application of VCMA [109, 110]. This intermediary skyrmion state can provide a pathway for energy efficient and robust reversal of the magnetization of a perpendicular-magnetic tunnel junction (p-MTJ) in the presence of thermal noise and defects [111]. One could also use a combination of VCMA and spin transfer torque (STT) to achieve an energy efficient reversal with reduced spin current requirements [112], provided the VCMA is an even function of electric field [113].

### 2.6.4 Voltage controlled domain wall dynamics

As discussed in Section 2.3, domain wall dynamics can switch the magnetization of a nanomagnet. Voltage controlled magnetic anisotropy modulation of domain wall velocity [114] and nucleation [115] has been shown and recently there has been a prediction of moving domain walls purely by an electric field. An electric field controls magnetic anisotropy through spin-orbit coupling. The equilibrium magnetic texture can be tuned between Néel and Bloch domain walls and near the Néel to Bloch transition. A pulsed electric field can cause precessional domain wall motion, which can be utilized to reverse the chirality of a Néel wall or depin it [116]. The electric field induced domain wall motion is usually less dissipative than the current induced domain wall motion discussed in Section 2.3 and therefore preferred for technological applications.

# 3. Hybrid spintronics-straintronics: Rotating the magnetization of nanomagnets with voltage-generated strain in two-phase multiferroics consisting of piezoelectric/magnetostrictive heterostructures

This section is devoted to the mainstay of this article – straintronic switching of nanomagnets. It is potentially one of the most energy efficient approaches to switching any binary switch with energy dissipation ~ 10aJ/bit – magnetic or non-magnetic – and employs magnetoelastic effects.

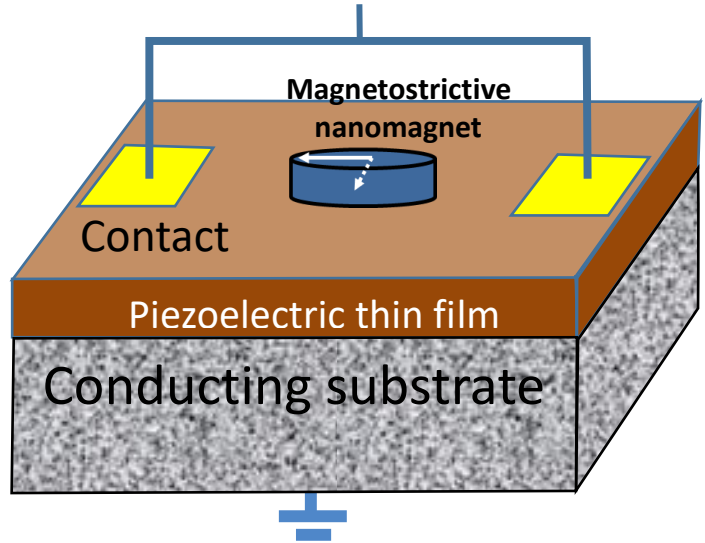

**Fig. 12**: A strain-coupled system consisting of a magnetostrictive shape-anisotropic nanomagnet delineated on a piezoelectric thin film, forming a two-phase multiferroic. The electrode scheme was proposed in Ref [27-29]. The thickness of the nanomagnet is much smaller than that of the piezoelectric thin film.

Consider the structure shown in Fig. 12. It consists of a *magnetostrictive* nanomagnet in the shape of an elliptical disk fabricated on a *piezoelectric* thin film deposited on a conducting substrate. The magnetostrictive/piezoelectric constituents form a strain-coupled *two-phase multiferroic*. Because of the elliptical shape, the nanomagnet's magnetization has two stable orientations (left and right-pointing) along the major axis of the ellipse (or the "easy axis"). Here, we are assuming that the in-plane anisotropy dominates over the surface anisotropy, which is why the easy axis is in-plane.

Two contact pads are delineated on the surface of the piezoelectric film and the line joining their centers is collinear with the major axis of the elliptical nanomagnet. The lateral dimensions of the contact pads, the separation between the edges of the contact pads and the nearest edge of the nanomagnet, and the piezoelectric film thickness, are all of the same order [27-29]. The two contact pads are electrically shorted and an electrostatic potential is applied between the pads and the grounded conducting substrate to produce a vertical electric field in the piezoelectric film.

The electrostatic potential generates biaxial strain in the pieozoelectric layer (compression in the direction of the major axis of the elliptical nanomagnet and tension along the minor axis, or vice versa, depending on the polarity of the electrostatic potential relative to the direction in which the piezoelectric film has been poled), overcoming some of the substrate clamping [27-29]. This strain is partially transferred to the magnetostrictive layer – the amount of transfer depends on how thin the magnetostrictive layer is compared to the piezoelectric layer, as well as the aspect ratio of the magnetostrictive nanomagnet. The strain thus transferred can rotate the magnetization of the magnetostrictive nanomagnet away from its stable orientation along the major axis toward the minor axis. If the magnetostriction coefficient of the nanomagnet is *positive* (examples are FeGa and Terfenol-D), then compressive stress along the major axis and tensile stress along the minor axis of the ellipse will cause the rotation, while if the magnetostriction coefficient is *negative* (examples are Co and Ni), then stresses of the opposite sign along the respective axes will cause the rotation to occur. The signs of the stresses can be reversed by reversing the polarity of the applied voltage. The maximum rotation is $90^0$ (i.e. the magnetization vector can be made to align along the minor axis of the ellipse or the so-called "hard axis"), although there are ways to make it exceed $90^0$ as will be discussed later. The effect that causes this rotation is the *Villari effect* and is best understood by considering the change in the potential energy profile of the nanomagnet under stress as shown in Fig. 13. In the presence of stress of the appropriate sign along the major and minor axes of the ellipse, the potential energy minimum moves from $(\phi=0^0, 180^0)$ to $(90^0, 270^0)$, i.e. the minor axis becomes the easy axis and the major axis the hard axis. That is why the magnetization will rotate by $90^0$ from the major to the minor axis of the ellipse if the stress is maintained for a sufficiently long time.

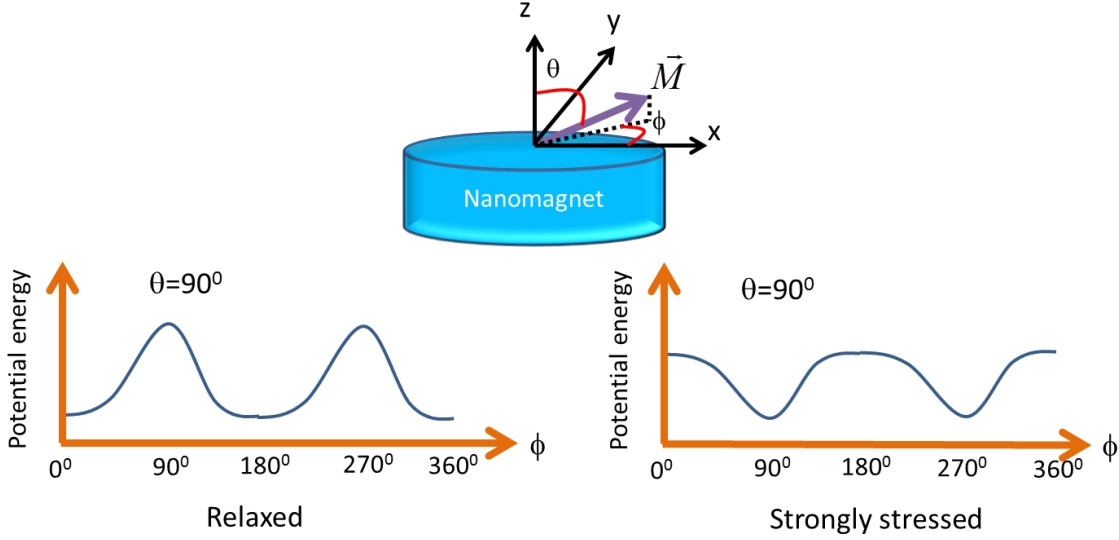

**Fig. 13**: A magnetostrictive nanomagnet shaped like an elliptical disk whose magnetization vector is $\vec{M}$ that has a polar angle θ and azimuthal angle ϕ. The potential energy profile in the plane of the magnet (θ = $90^0$) is shown in the relaxed and strongly stressed conditions. In the relaxed state, the potential profile is bistable with two degenerate minima at ϕ = $0^0$ (or $360^0$) and $180^0$. In the strongly stressed condition, the two energy minima have moved to ϕ = $90^0$ and $270^0$. Therefore, in the relaxed state, the stable magnetization orientations are along the major axis of the ellipse, and in the strongly stressed condition, the stable orientations are along the minor axis.

Note that here the voltage applied to the two-phase multiferroic rotates the magnetization vector. Therefore, this effect can be utilized to change the resistance of a magneto-tunneling junction (MTJ) if the soft layer is made of the two-phase multiferroic. There are experimental reports of MTJs being switched in this fashion [35, 117]. We will show later that the voltage that needs to be applied across the piezoelectric thin film can be miniscule – few to few tens of mV – if the piezoelectric film is ~100 nm thick. Therefore, the energy dissipation in this mode of switching could be very small and typically much smaller than in most current-mode switching or even VCMA. That motivates the interest in this switching modality. The term "hybrid spintronics/straintronics" was coined to describe this methodology and was inspired by the fact that strain reorients the spins in the nanomagnet and makes the magnetization vector rotate [118].

We also point out that making high quality thin film piezoelectrics is a materials challenge. Our experience has been that ~100 nm thin films tend to be grainy and annealing at reasonable temperatures does not improve the situation much. When a metallic nanomagnet is deposited on such a film, it tends to diffuse through the grain boundaries and electrically short the nanomagnet to the underlying conducting substrate. Perhaps a diffusion barrier between the nanomagnet and the film can help, but it will also impede strain transfer from the piezoelectric to the nanomagnet. This is a processing challenge and hopefully it can be overcome in near future.

**3.1 Controlling magnetization in nanoscale magnetostrictive materials with strain**

Several groups have experimentally studied the control of magnetization in magnetostrictive films using voltage generated strain in a piezoelectric film [119], demonstrating reversible control of nanomagnetic domains [120], repeatable reversal of perpendicular magnetization in the absence of a magnetic field in regions of a Ni film [121], and strain assisted reversal of perpendicular magnetization in Co/Ni multilayers [122]. Others demonstrate the use of strain control of magnetization orientation in LSMO films [123, 124], iron films [125], TbCo$_2$/FeCo multilayers [126] and strain control of magnetic properties of FeGa/NiFe multilayer films [127] and FeGa films [128].

Strain has been shown to reorient magnetization in magnetostrictive Ni rings [129, 130] and Ni squares of 2 microns side [131] and soft layer of Magnetic Tunnel Junctions (MTJs) of lateral dimensions 20 microns × 40 microns [117]. In another work, the magneto-electric effect is used to read the magnetization orientation in a composite multiferroic heterostructure [N×(TbCo$_2$/FeCo)]/[Pb(Mg$_{1/3}$Nb$_{2/3}$)O$_3$]$_{1-x}$ [PbTiO$_3$]$_x$ [132].

Some groups have demonstrated control of magnetization in nanomagnets deposited on piezoelectric substrates. For example, an electric field induced stress mediated reversible control of magnetization orientation (see magnetic force microscopy images in Fig 14) in nanomagnets of nominal lateral dimensions 380 nm × 150 nm deposited on a 1.28 micron PZT thin film was demonstrated with the application of 1.5 V to the PZT film [133]. Futhermore, building on individual control of magnetoelectric heterostructures with localized strain to reorient the magnetization in a Ni ring of 1000 nm outer diameter, 700nm inner diameter, and 15 nm thickness, deterministic multistep reorientation of magnetization in a 400 nm Ni dot of 15 nm thickness has been reported [134]. Fig 15 shows that the strain profile (generated when voltage was applied sequentially across multiple electrode pairs) rotated the magnetization of the Ni dot by 180º through multiple steps (each rotating the magnetization through a small angle).

Uniform magnetization rotation through 90º has also been demonstrated through imaging with X-ray photoemission electron microscopy (X-PEEM) and X-ray magnetic circular dichroism (XMCD) in elliptical nanomagnets of dimensions of nominal lateral dimensions ~100 nm ×150 nm [135].

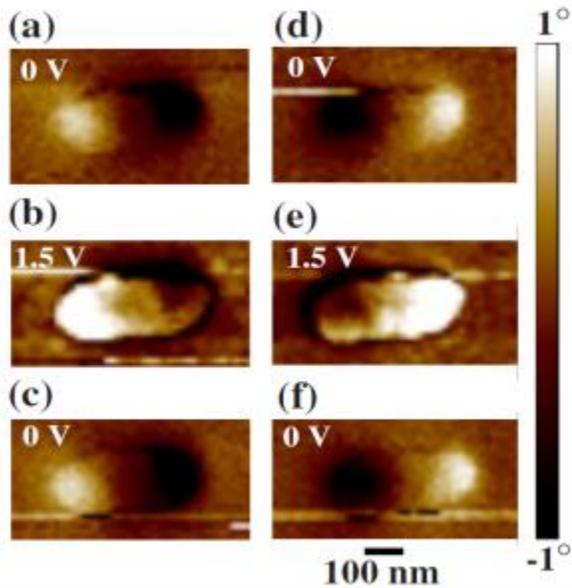

**Fig 14:** Magnetization reorientation due to voltage induced strain in a nanomagnet of nominal lateral dimensions 380 nm × 150 nm deposited on a 1.28 micron PZT thin film. Reprinted from Chung T K, Keller S and Carman G P 2009 Electric-field-induced reversible magnetic single-domain evolution in a magnetoelectric thin film *Appl. Phys. Lett.* **94**, 132501 with permission of AIP Publishing.

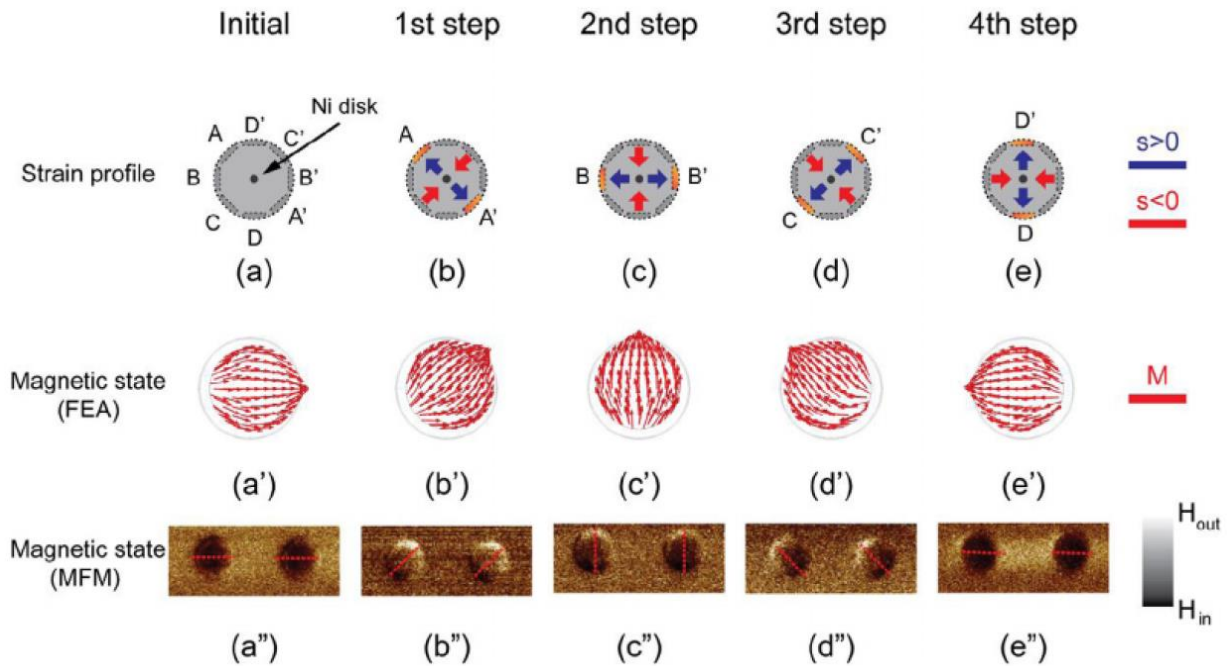

**Fig 15:** Magnetization reorientation due to voltage induced strain in nanomagnets of nominal lateral dimensions 380 nm × 150 nm deposited on a 1.28 micron PZT thin film. Reprinted from Sohn H, Liang C yen, Nowakowski M E, Hwang Y, Han S, Bokor J, Carman G P and Candler R N. Deterministic multi-step rotation of magnetic single-domain state in Nickel nanodisks using multiferroic magnetoelastic coupling *J. Magn. Magn. Mater.* **439** 196 (2017) with permission from Elsevier.

## 3.2 Complete $180^0$ rotation of magnetization with strain

By applying strain along only one axis, the magnetization vector can only be rotated by $90^0$ and not full $180^0$ because stress moves the energy minimum in the potential energy profile of Fig. 13 from $\phi = 0^0$, $180^0$ to $\phi = 90^0, 270^0$. That causes a problem. When stress is relaxed, the energy minima move back to $\phi = 0^0$ and $180^0$. Hence, the magnetization will settle into *either* the right pointing orientation along the major axis of the ellipse (+*x*) or the left pointing orientation (-*x*) with equal probability. Let us say that the +*x* ($\phi = 0^0$) orientation encodes the bit 1 and the –*x* ($\phi = 180^0$) orientation encodes the bit 0. Assume that the initial stored bit was 1 and we wish to write the bit 0. If we apply strain and then withdraw it, we will successfully write the bit 0 with only ~50% probability and not ~100% probability. This level of error cannot be tolerated. Of course, we can decide to encode bit 1 in the $\phi = 0^0$ orientation and the bit 0 in the $\phi = \pm\ 90^0$ orientation. Then we can write the bit 0 successfully and store it only as long as we keep the stress on. If we withdraw stress, we will revert back to either bit 1 ($\phi = 0^0$) or an undefined bit ($\phi = 180^0$) [again, with equal probability]. This makes the memory element "volatile" because we have to keep the stress on to store and retain bit 0, and we will lose the bit if we withdraw stress. Note that in this case, we can write the bit 0 with almost unit probability if we keep the stress on, but we will not be able to write the bit 1 with unit probability when we withdraw the stress. That latter probability will be ~0.5.

One twist to this is to have a bias magnetic field in the +*x*-direction ($\phi = 0^0$). Then, if we apply sufficient stress, we will overcome the bias field and rotate the magnetization to the $\phi = \pm\ 90^0$ orientation (write bit 0 with almost unit probability) and when we withdraw stress, the magnetization will return to the $\phi = 0^0$ orientation because of the bias field (write bit 1 with almost unit probability). The bias magnetic field will allow us to write both bits 0 and 1 with very high probabilities, but the memory element is still volatile since we have to keep the stress on to write and retain the bit 0.

This problem afflicts VCMA switching as well, but there it is resolved by applying an in-plane magnetic field. The field induces precession of the out-of-plane magnetization vector when it is dislodged from the out-of-plane direction by a voltage and the voltage pulse is adjusted to one-half of the precession period to complete $180^0$ switching.

In principle, a similar approach can be taken in the case of straintronic switching. We can apply a magnetic field out-of-plane (in the *z*-direction in Fig. 13) that will induce precession of the magnetization vector around it when the vector is dislodged from the easy axis by a voltage pulse that generates strain.

We can then adjust the voltage pulse width to one-half of the precession period to flip the magnetization by $180^0$. But this is hard to do in the in-plane geometry. There are other ways of doing this, which do not require a magnetic field. We discuss them below.

i. *Dynamic approach*: One approach is to make the voltage pulse width equal to the time it takes for the magnetization vector's projection on the nanomagnet's plane to just complete $90^0$ rotation under voltage-generated stress. This will ensure that the stress is removed as soon as the projection of the magnetization vector on the plane of the nanomagnet coincides with the minor axis of the elliptical nanomagnet. In that case, the magnetization vector will continue to rotate past $90^0$ and complete $180^0$ rotation [30].

> Why this happens can be explained succinctly as follows. When the magnetization vector rotates, it also lifts out of the nanomagnet's plane. The out-of-plane component results in a torque that will make the magnetization vector rotate past $90^0$ if the stress is removed as soon as the $90^0$ rotation is completed. Removal of stress makes the minor axis direction the maximum energy state as opposed to the minimum energy state (see Fig. 13). Therefore, if the stress is removed at the precise juncture when the magnetization vector's projection aligns along the minor axis, the magnetization vector will not settle along the minor axis (since it has become the energy maximum and hence unstable) but will continue to rotate further and settle into the opposite direction along the major axis because of the torque.

> There is, of course, a slight probability that in the presence of thermal noise, the torque can reverse itself and make the magnetization vector rotate in the opposite direction and complete a $0^0$ rotation instead of $180^0$ rotation. Simulations have shown that this probability is very small, typically $< 10^{-4}$ at room temperature [30].

> The reason why this approach is not preferred is because it requires precise knowledge of how long it takes for the projection of the magnetization vector on the nanomagnet's plane to rotate through $90^0$. That time is uncertain in the presence of thermal noise. Therefore, it is impossible to make the voltage pulse width always exactly equal to this time. There have been proposals of using some kind of feedback network that monitors the rotation of the magnetization vector continuously, feeds that information back to the stress generator, which withdraws the stress at the right moment [136]. Needless to say, this is not very practical and moreover the feedback circuit would increase energy dissipation significantly.

ii. *Static approaches*: There are a number of approaches to achieve $180^0$ rotation that do not require precise timing of the stress pulse or the presence of any magnetic field. One of the earliest approaches was due to Novosad et al. [137] where two pairs of electrodes were used to apply a local electric field on a two-phase multiferroic nanomagnet. The two pairs are simultaneously activated and allows control over the direction and amplitude of the electric field by varying the polarity and amplitude of the voltages. The in-plane energy minimum, corresponding to the easy axis, follows the local electric field. By rotating the electric field, one can effectively rotate the stress and implement complete magnetization reversal ($180^0$ switching).

A simpler scheme for $180^0$ rotation with stress is illustrated in Fig. 16 [138]. It requires applying uniaxial stress along two different directions sequentially, and that rotates the magnetization through $180^0$ in *two steps*. Refer to the inset of Fig. 16 and assume that the magnetization is initially along the $+z$ direction and we wish to flip it to the $-z$ direction. In the first step, the applied uniaxial stress (which is applied at an acute angle with the $+z$ axis) transforms the potential energy profile of the nanomagnet into a monostable well with a single energy minimum located somewhere between $\theta = 0^0$ and $90^0$ where $\theta$ is the angle that the magnetization subtends with the $+z$ axis. Next, stress is applied along a second direction that subtends an obtuse angle with the $+z$ direction while the stress along the initial direction is relaxed. This moves the energy minimum to a new location between $\theta = 90^0$ and $180^0$ that makes the magnetization subtend an obtuse angle with the $+z$ direction (and hence an acute angle with the $-z$ direction). Finally, when all stresses are removed, the magnetization settles into the nearest energy minimum which corresponds to orientation along the $-z$ direction. This results in complete magnetization reversal. The mechanism is illustrated in Fig. 16.

The advantage of this approach is that no precise timing of anything is required and there is no need for any magnetic field. This modality of complete magnetization reversal ($180^0$ switching) has been demonstrated experimentally [139]. Some of the experimental results showing strain induced complete magnetization reversal of Co nanomagnets deposited on a piezoelectric PMN-PT substrate due to two step rotation (induced by two pairs of electrodes that are sequentially activated to generate two different strain profiles) are shown in Figs. 17 and 18. Unfortunately, a key disadvantage of this approach is that if we use this construct for a non-volatile memory cell, then we will inevitably end up with a 4-terminal memory cell (separate read and write paths), which is not preferred. The electrode pairs take up additional space on a chip, thus severely decreasing the memory density. This is a serious shortcoming in memory applications.

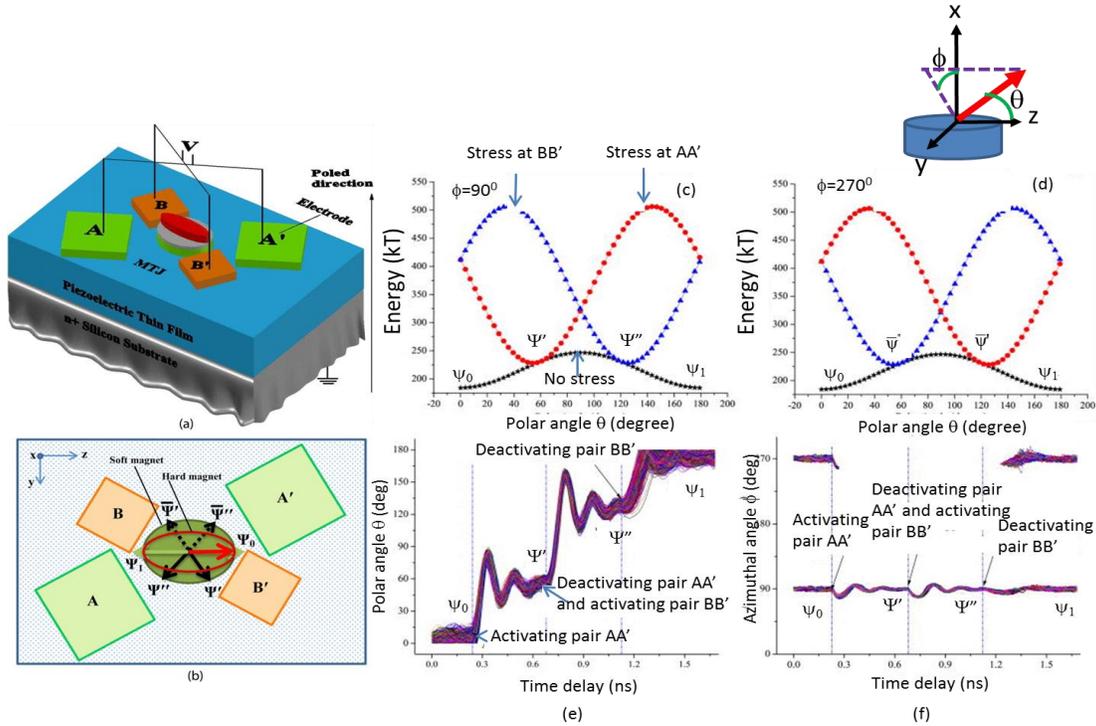

**Fig. 16**: (a) An MTJ whose soft layer is magnetostrictive and in contact with a piezoelectric thin film. Two pairs of electrodes AA' and BB' are delineated on the surface of the piezoelectric and apply effectively uniaxial stresses along the lines joining their centers when they are activated. (b) The initial direction of magnetization is along $\Psi_0$. When pair AA' is activated, the magnetization rotates through an acute angle to $\Psi'$. Then pair BB' is activated and AA' deactivated, whereupon the magnetization rotates to $\Psi''$. Finally when BB' is deactivated, the magnetization rotates to $\Psi_1$, completing $180^0$ rotation. (c) and (d) The in-plane energy profiles of the nanomagnet as a function of the polar angle of the magnetization vector under various stressing scenarios. (e) polar angle and (f) azimuthal angle of the magnetization vector as a function of time. Reprinted from Biswas A. K., Bandyopadhyay S. and Atulasimha J. 2014 Complete magnetization reversal in a magnetostrictive nanomagnet with voltage-generated stress: A reliable energy-efficient non-volatile magneto-elastic memory *Appl. Phys. Lett*. **105** 072408 with permission of AIP Publishing.

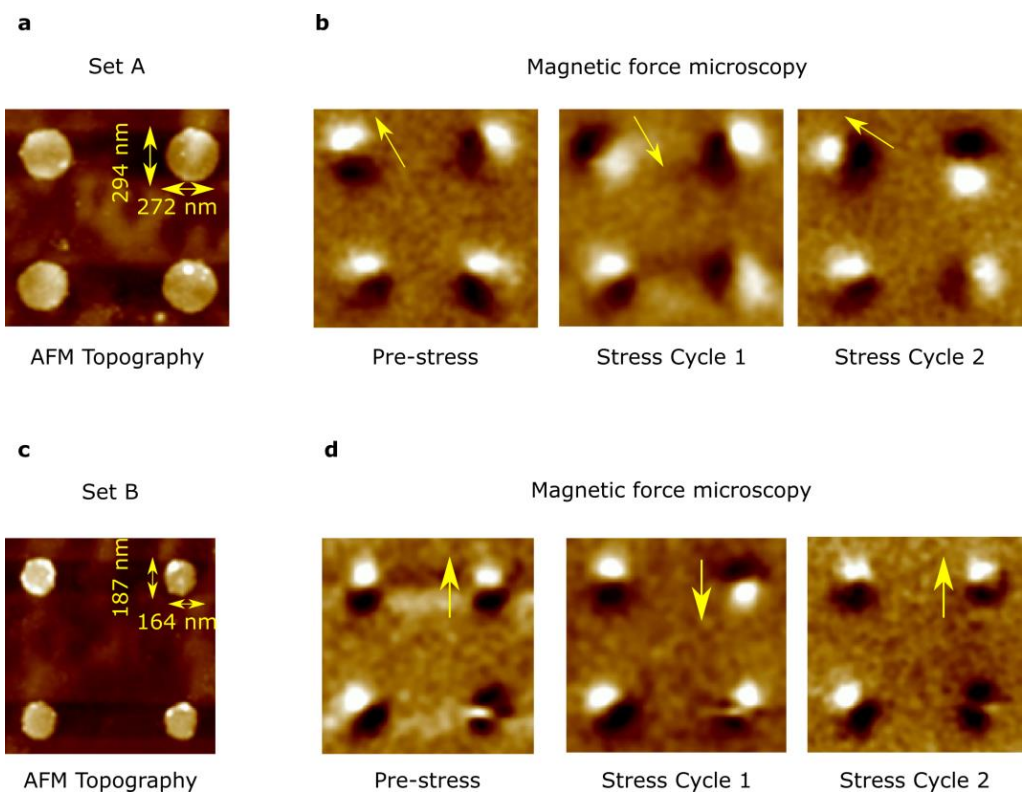

**Fig. 17**: AFM and MFM images of two sets of magnetostrictive Co nanomagnets (Set A and Set B) delineated on a piezoelectric PMN−PT substrate showing how their magnetizations react to two consecutive cycles of stress. Set A: four nanomagnets with major axis 294 nm and minor axis 272 nm. Set B: four nanomagnets with major axis 187 nm and minor axis 164 nm. (a,c) AFM image showing the topography of the four isolated nanomagnets. (b, d) The left panels show the MFM images of the pre-stress initial states; the center panels show the MFM image after one sequential stress cycle indicating that the nanomagnets (marked by yellow arrows in b and d) experienced complete 180° rotation; the right panels show that the same nanomagnets marked with yellow arrows have undergone another ~180° rotation and hence returned to their initial orientation after the second sequential stress cycle. After each sequential stress cycle, the nanomagnets undergo complete magnetization reversal ($180^0$ rotation of the magnetization vector). Reprinted with permission from Biswas A. K., Ahmad H., Atulasimha J. and Bandyopadhyay S. 2017 Experimental demonstration of complete $180^0$ reversal of magnetization in isolated Co nanomagnets on a PMN-PT substrate with voltage generated strain *Nano Letters,* **17,** 3478. Copyright 2017 American Chemical Society.

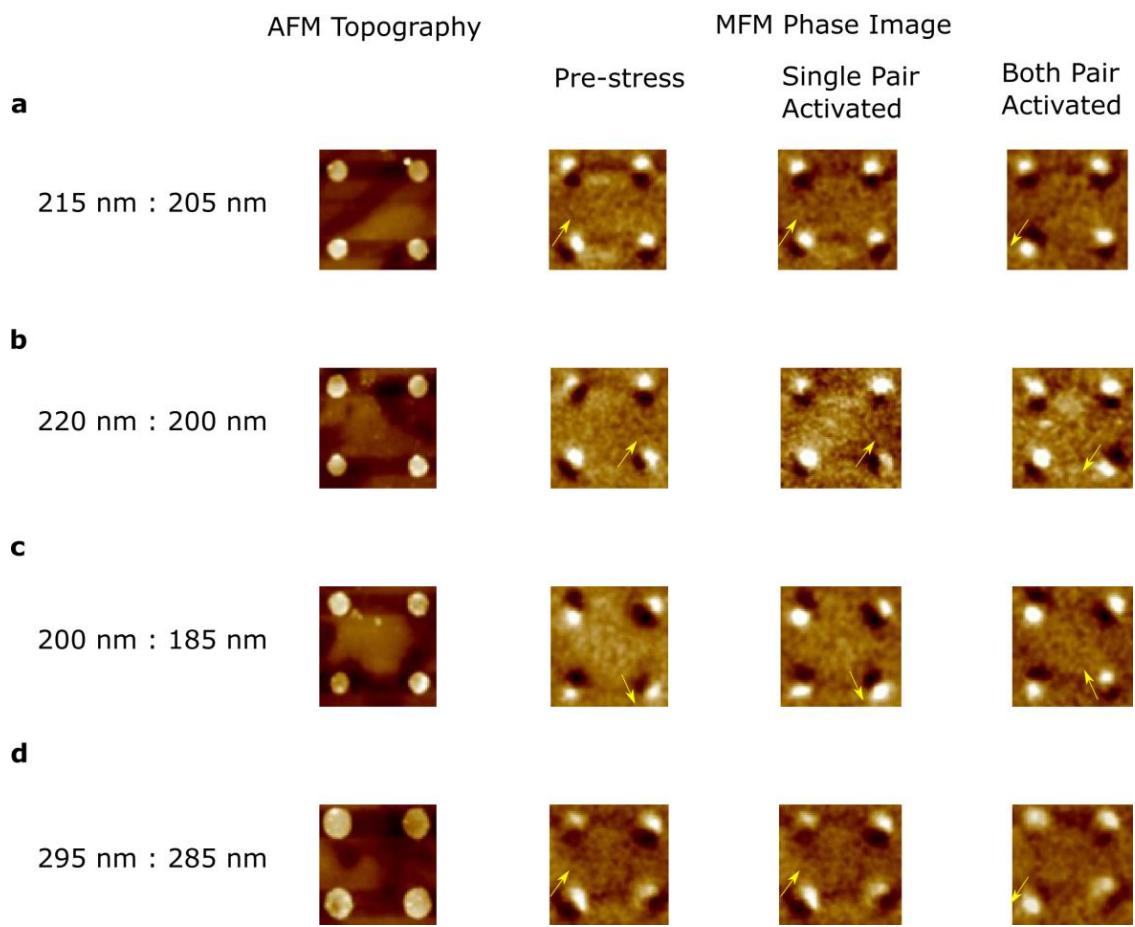

**Fig. 18**: AFM and MFM micrographs of four sets of magnetostrictive Co nanomagnets of different sizes and ellipticity on a PMN-PT substrate showing how their magnetizations evolve when one and both pairs of electrodes are activated to generate stresses along one and two different directions, respectively. The nominal dimensions (major and minor axes) are shown on the left in each horizontal panel. The calculated in-plane shape anisotropy energy barriers in these four sets are, respectively, 2.84 eV, 5.728 eV, 4.202 and 3.099 eV. The first column shows the topography of the four sets of nanomagnets, the second shows the initial magnetization states after magnetizing with a global magnetic field directed vertically up in this figure, the third shows the magnetization states after one pair of metal pads on the PMN-PT substrate is activated (to generate stress in one direction) and then deactivated, while the fourth shows the magnetization states after both pairs are activated successively (to generate stresses in two different directions sequentially) and deactivated successively. Reprinted with permission from Biswas A. K., Ahmad H., Atulasimha J. and Bandyopadhyay S. 2017 Experimental demonstration of complete $180^0$ reversal of magnetization in isolated Co nanomagnets on a PMN-PT substrate with voltage generated strain *Nano Letters,* **17,** 3478. Copyright 2017 American Chemical Society.

An additional issue, which is obvious in Figs. 17 and 18, is that only a small fraction of the nanomagnets underwent magnetization reversal in the experiment. Only a fraction of the nanomagnets flipped magnetization and the rest were non-responsive. This is not a limitation of the switching scheme but is due to the low magneto mechanical coupling in the magnetostrictive Co used in fabricating the nanomagnets. The low effective field due to stress anisotropy (given the low magnetostriction of Co) may not be able to overcome the demagnetizing field due to shape anisotropy, especially when unintentional variations in the nanomagnet shape that are introduced during lithography increase the shape anisotropy. This issue is discussed in more detail later.

## 3.3. Straintronic switching of dipole coupled nanomagnet for Boolean NOT gate operation

The simplest logic gate is the Boolean inverter (or NOT gate). It is a 1-input and 1-output logic gate where the output is the logic complement of the input. Strain switched nanomagnets can easily implement an inverter as shown in Fig. 19. It consists of two elliptical nanomagnets, one having larger eccentricity than the other. Since both nanomagnets are elliptical, their magnetization orientations are bistable, i.e. each can point in one of two directions along the respective major axis. The magnetization of the more eccentric one encodes the input bit and that of the less eccentric one encodes the output bit.

As long as the two nanomagnets are placed close to each other to have significant inter-magnet dipole interaction, their magnetizations will tend to be mutually antiparallel if the line joining the centers of the two ellipses are collinear with the minor axis, as shown in Fig. 19.

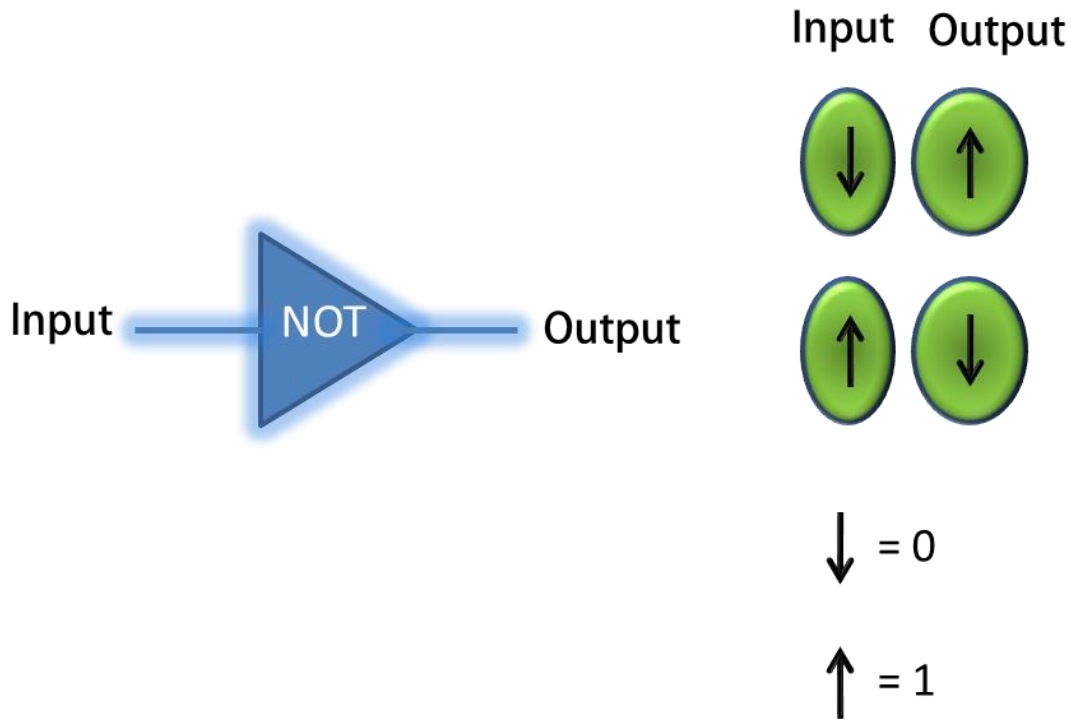

**Fig. 19**: (Left) An inverter; (Right) An inverter implemented with two dipole coupled logic gates. The input bit is encoded in the magnetization orientation of the more eccentric nanomagnet while the output bit is encoded in the magnetization orientation of the less eccentric nanomagnet. Because of dipole coupling between them, the magnetizations of the two nanomagnets will be anti-parallel in the ground state of the system which means that the output bit is the logic complement of the input bit. If the input bit is 0, the output bit is 1, and vice versa.

Let us now consider the 2-nanomagnet system in Fig. 20 and assume that the magnetizations of both have been oriented in the same direction by an external magnetic field. In this case, the input and output bits are the same and the NOT operation is not realized. We may expect that dipole coupling will flip the output bit (since it is encoded in the less eccentric nanomagnet which has the lower shape anisotropy energy barrier; magnetization has to transcend this barrier to flip). However, the nanomagnet separation is usually such (due to lithographic tolerances) that the dipole coupling is not strong enough to overcome the shape anisotropy barrier in the nanomagnet hosting the output bit and make its magnetization flip and assume an orientation anti-parallel to that of its neighbor's.

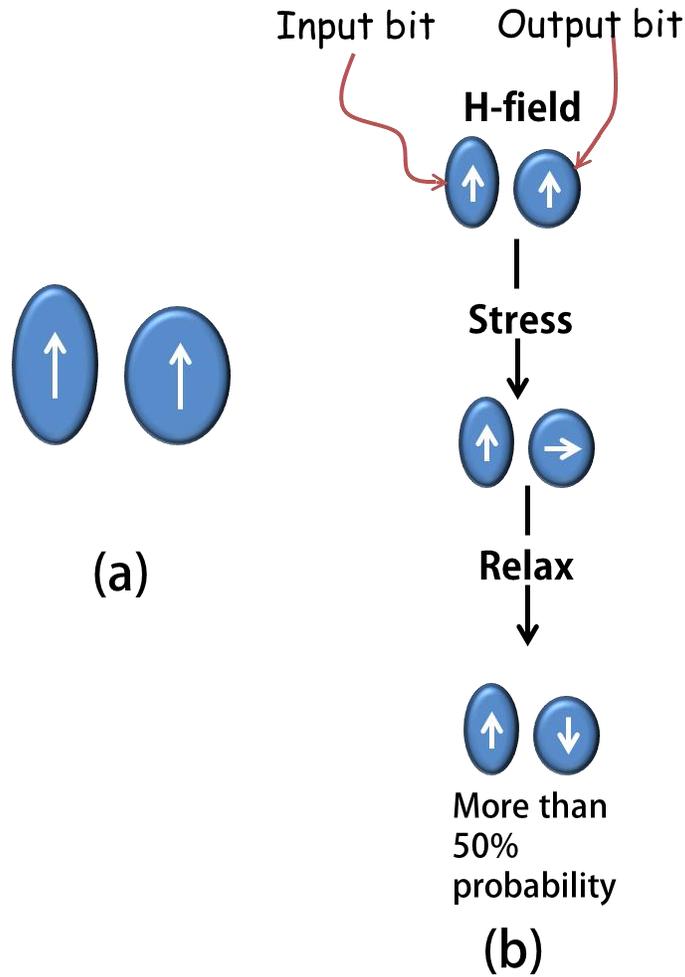

**Fig. 20**: (a) The magnetizations of two dipole coupled nanomagnets implementing an inverter are oriented in the same direction with an external agent; (b) triggering the NOT operation with stress.

To trigger the NOT operation, we can subject both nanomagnets to stress. The stress magnitude is chosen such that it can invert the energy barrier (as in Fig. 13b) within the output nanomagnet but not within the input nanomagnet which is much too anisotropic. In that case, stress will make the magnetizations of the two nanomagnets almost mutually perpendicular as shown in Fig. 20b. When stress is relaxed, the magnetization of the output nanomagnet will go back to one of the two stable states – either pointing vertically up or vertically down. Because of dipole coupling, it will now prefer to orient vertically down (with much larger than 50% probability), thereby implementing the NOT operation. We note that in an actual circuit with nanomagnets, one cannot have nanomagnets of different shapes/eccentricity, as we have considered in the proof of concept experiment below. In a practical application, we have to contact electrodes placed around the output nanomagnet and locally stress only this nanomagnet with a voltrage without stressing the input nanomagnet. We will carry out this local selective stressing of the output nanomagnet alone to implement the NOT operation, instead of relying on designs with differently shaped nanomagnets.

This stress-induced triggering action was experimentally demonstrated with sets of two Co nanomagnets delineated on a piezoelectric PMN-PT substrate. Each set constituted an inverter. The magnetizations of all nanomagnets were initially oriented in the same direction with an external magnetic field and then global stress was generated in the PMN-PT substrate with a voltage. The output bit flipped to implement the NOT operation. In Fig. 21, we show SEM and MFM micrographs of multiple sets. In the MFM images, the bright and dark regions can be viewed as opposite poles (in reality the phase contrast in a MFM image corresponds to the out of plane direction of the stray magnetic field from which the in-plane magnetization directions can be inferred). An external magnetic field magnetized all nanomagnets such that the north poles were pointing up and in each pair the output bit was identical to the input bit. The magnetic field was removed and stress was applied and withdrawn. The output bit in one pair (1 out of 9) flipped to implement the NOT function. Why only 1 out of 9 will be discussed later.

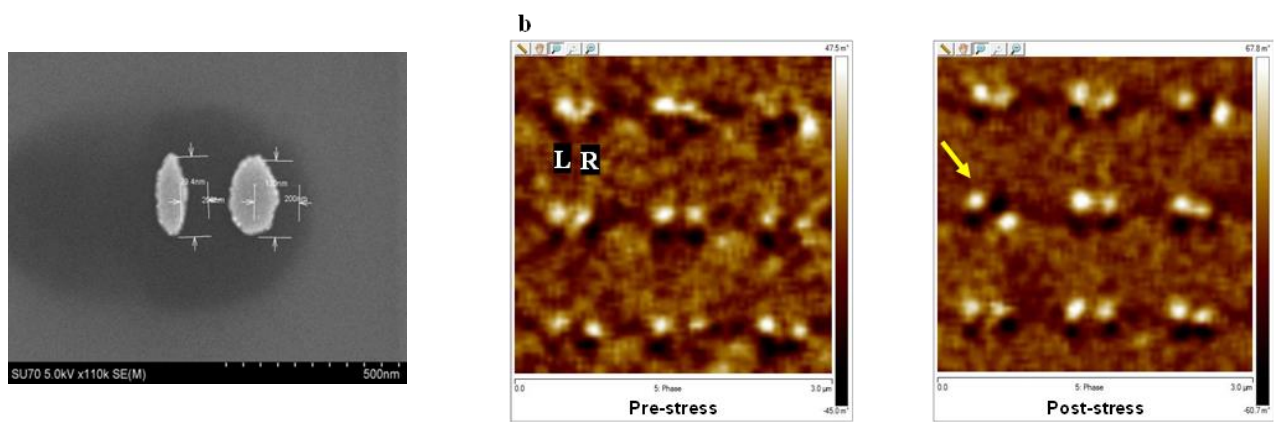

**Fig. 21**: Clocking of dipole-coupled single-domain magnetostrictive Co nanomagnets on a piezoelectric PMN-PT substrate implementing a Boolean NOT logic gate. (a) SEM image of a pair of elliptical Co nanomagnets where the one encoding the input bit is on the left (more eccentric) and the one encoding the output bit is on the right (less eccentric) ; (b) A dipole-coupled nanomagnet pair (L, R) is initialized in the "down" direction (north pole up and south pole down) by a magnetic field in the left panel. The nanomagnet dimensions are (L~250✕150✕12 nm, R~200✕175✕12 nm) having a center-to-center separation of ~300 nm. Upon stressing with a global tensile stress of 80 MPa generated with a voltage applied across the poled PMN-PT substrate, and subsequent stress removal, the magnetization of R in one set (out of nine) flips to implement the NOT operation. This set is identified with an arrow in the right panel. The reason that only one out of nine pairs shows successful operation is that Co is not sufficiently magnetostrictive and hence not enough effective magnetic field due to stress was generated in every pair to cause switching of the output bit. (b) is reprinted with permission from D'Souza N., Fashami M. S., Bandyopadhyay S. and Atulasimha J. 2016 Experimental clocking of nanomagnets with strain for ultralow power Boolean logic *Nano Lett*. **16** 1069. Copyright 2017 American Chemical Society.

## 3.4 Straintronic switching of dipole coupled nanomagnet for Bennett clocking

In Boolean circuits, a bit will have to be transported from an output stage to the next input stage to carry out the circuit operation. This is easy to do in electronic circuits where bits are encoded in voltages and hence a bit can be transported by simply connecting a wire between the two stages. In nanomagnetic circuits, where magnetization states (not voltages) encode bits, this obviously will not work. A logic wire for transporting bits unidirectionally is implemented with a string of dipole coupled nanomagnets containing an odd number of nanomagnets. Their magnetizations assume artificial anti-ferromagnetic ordering when the array is in the ground state, meaning nearest neighbors have antiparallel orientations, as shown in Fig. 22a. This is a consequence of dipole coupling between nearest neighbors. The input bit, encoded in the first nanomagnet on the left, is reproduced in every odd numbered nanomagnet and hence the bit is transported from one location to another.

A problem arises when the input bit is flipped. One would expect that all succeeding bits will flip in a domino-like fashion to implement the logic wire, but this cannot happen. After the first nanomagnet flips, the second goes into a "tie-state" because the influence it feels from dipole coupling with its left neighbor is equal and opposite to the influence it feels from dipole coupling with its right neighbor. This is shown in Fig. 22b. Therefore, the second nanomagnet goes into an indeterminate magnetization state and the input bit does not propagate down the chain.

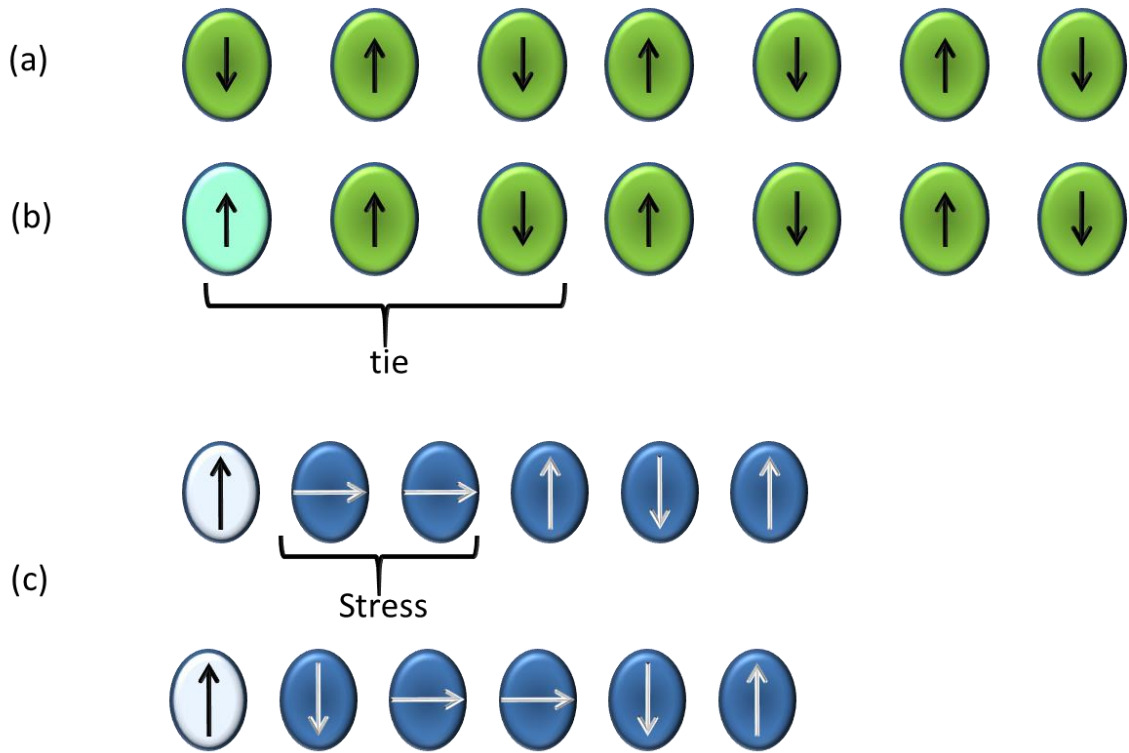

**Fig. 22:** (a) An array of dipole coupled elliptical nanomagnets where the line joining the centers is collinear with the minor axes. The ordering of the magnetizations is anti-ferromagnetic when the array is in the ground state, i.e. nearest neighbors have antiparallel magnetizations. (b). When the magnetization of the input nanomagnet (first one on left) is flipped, the second one goes into a tie state since dipole couplings with both left and right neighbors are equally strong. (c) The nanomagnets are stressed sequentially pairwise to rotate their magnetizations by 90°. This breaks the tie and allows the input bit to propagate down the chain so that ultimately the input bit is reproduced in every odd numbered nanomagnet.

The tie can be broken by pairwise stressing the nanomagnets, starting with the second nanomagnet, which will turn the magnetizations of the stressed pair by $90^0$ to align their magnetizations along their minor axis. The stress is then shifted to the right by one cell, as shown in the second row of Fig. 22c. After stress is shifted, the first member of the original stressed pair is relaxed and feels *unequal* dipole coupling from its left and right neighbors since one has its magnetization pointing along the major axis and the other has its magnetization pointing along the minor axis. The dipole influence from the left is stronger and hence the second nanomagnet will obey its left neighbor and align to an orientation antiparallel to that of the first. By repeating this process, i.e. by shifting the stress one cell at a time, the magnetizations of all magnets can be flipped sequentially such that the input bit is again reproduced in every odd numbered nanomagnet [31]. This is one variant of Bennett clocking that requires *local* stress generation to stress one pair at a time. Note that it allows *pipelining* of data since the magnetization of the first nanomagnet can be flipped again as soon as the second nanomagnet's magnetization has completed flipping. One does not have to wait for the input bit to propagate all the way down the chain before it can be changed.

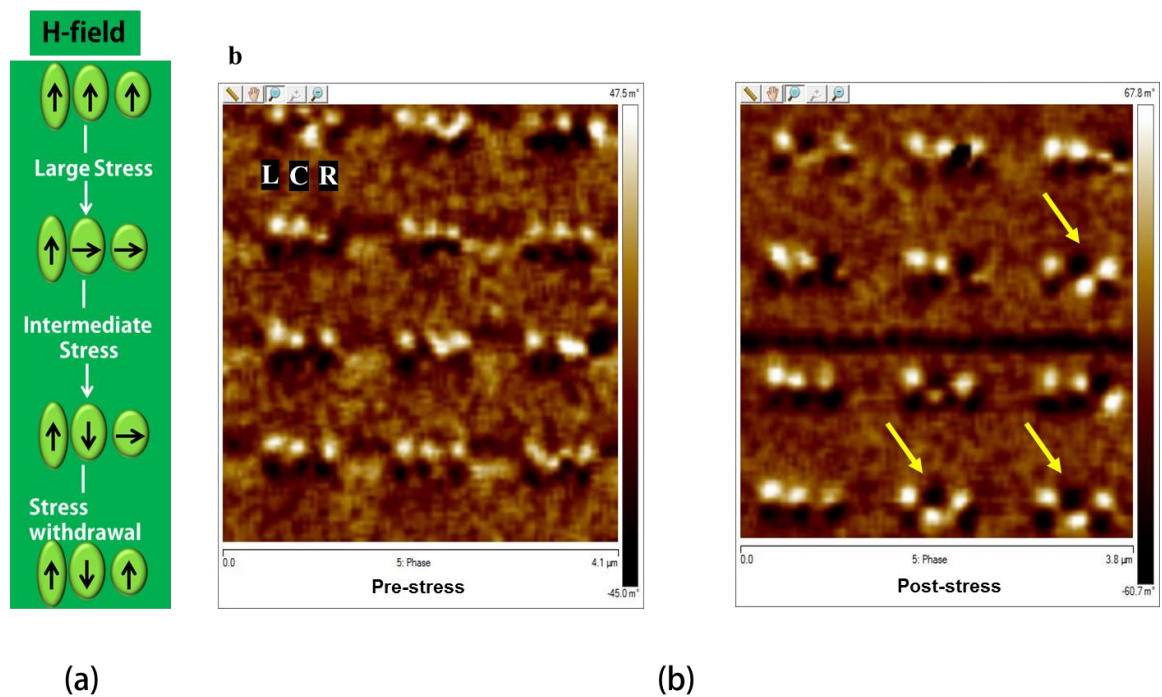

(a)                                  (b)

Fig. 23: (a) Bennett clocking of three dipole-coupled nanomagnets; (b) [left panel] MFM images of elliptical magnetostrictive Co nanomagnets, with decreasing eccentricity from left to right, delineated on a piezoelectric PMN-PT substrate. The images show the magnetizations of 12 sets of trios which have been all magnetized in the same direction (north pole pointing up) by an external magnetic field; [right panel] after stress application with a voltage dropped across the PMN-PT substrate and subsequent removal of the stress, the magnetizations of three sets (marked by the arrows) have assumed anti-ferromagnetic ordering, i.e. the bit encoded in the first nanomagnet has been inverted in the second and reproduced in the third in these sets. This is a variation of Bennett clocking. The reason why only 25% of the sets exhibits this behavior is that Co is not sufficiently magnetostrictive to generate a large enough effective magnetic field due to the stress applied. (b) is reprinted with permission from D'Souza N., Fashami M. S., Bandyopadhyay S. and Atulasimha J. 2016 Experimental clocking of nanomagnets with strain for ultralow power Boolean logic *Nano Lett*. **16** 1069. Copyright 2017 American Chemical Society.

Another variant of Bennett clocking is shown in Fig. 23a and it can work with *global* stress which is much easier to generate than *local* stress. Three dipole-coupled magnetosrtrictive nanomagnets, with decreasing eccentricity from left to right, are deposited on a piezoelectric substrate. An external magnetic field aligns their magnetizations in the same direction to produce artificial ferromagnetic ordering of the magnetizations. This is a metastable state since the ground state should be anti-ferromagnetic. The ground state is not reached automatically since the shape anisotropy energy barrier in the second nanomagnet will prevent its magnetization from flipping spontaneously. To drive the system out of the metastable state and into the ground state, the nanomagnets will have to be stressed to erode or invert the shape anisotropy energy barrier(s) in the nanomagnet(s).

We can first apply a large stress such that the stress anisotropy energy will exceed the shape anisotropy energies of all but the first nanomagnet which is most shape anisotropic. In that case, the magnetization of the far left nanmagnet will not rotate by much, if at all, but the magnetizations of the other two (less shape anisotropic) nanomagnets will rotate by $90^0$. When the stress magnitude is reduced, the stress anisotropy can no longer beat the shape anisotropy barrier in the second nanomagnet, but can still beat it in the third. Hence, the second nanomagnet will revert to a stable orientation along the major axis which is antiparallel to the magnetization of the first because of dipole coupling. Finally, when stress is completely removed, the last nanomagnet also reverts to a stable orientation which is antiparallel to the magnetization of the second nanomagnet. This results in anti-ferromagnetic ordering and effective Bennett clocking. In Fig. 23b, we show experimental demonstration of this principle.

### 3.5    Poor switching statistics in straintronic switching

One consistent problem with straintronic switching experiments is the poor yield, i.e. only a small fraction of the fabricated sets shows the correct operation while the majority appears to be unaffected by stress. This is possibly due to the use of Co nanomagnets which have low magnetostriction. We can estimate an effective magnetic field due to stress by equating the stress anisotropy energy to the magnetostatic energy due to the effective magnetic field:

$$\mu_0 M_s H_{eff} = (3/2) \lambda_s \sigma$$
$$\Rightarrow H_{eff} = \frac{(3/2) \lambda_s \sigma}{\mu_0 M_s} \quad . \tag{3.1}$$

The saturation magnetization of Co is $14.22 \times 10^5$ A/m, the saturation magnetostriction of Co is ~50 ppm and the stress that could be generated in the experiments is ~80 MPa. Therefore, $H_{eff}$ ~ 30 Oe, which is much too small to beat the effective magnetic field due to shape anisotropy energy in every nanomagnet. The latter ($H_k$) can be found by equating the magnetostatic energy associated with it to the shape anisotropy energy barrier:

$$\mu_0 M_s H_k \Omega = E_b \qquad (3.2)$$

where $\Omega$ is the volume of the nanomagnet and $E_b$ is the shape anisotropy energy barrier. While we design the nanomagnets such that stress anisotropy would be able to overcome the shape anisotropy barrier ($E_b$) when the nanomagnet is stressed, we recognize that a small variation in the nanomagnet dimensions due to lithographic imperfections could vastly increase $E_b$ (and consequently $H_k$), thereby rendering the applied stress insufficient to switch the nanomagnet. That makes it difficult to enforce the condition $H_{eff} > H_k$ in most of the nanomagnets. This is likely to be the reason for the poor switching statistics. In some nanomagnets, there may be pinning sites and that too contributes to the poor yield.

In order to improve the switching statistics, FeGa was used to replace Co since it has a higher magnetostriction of 300 ppm. The switching of FeGa inverter pairs is shown in Fig. 24, which is reproduced from ref. [140].

FeGa however has its own problems. Since it is a binary alloy and has numerous phases, it tends to have more pinning sites for the magnetization and ultimately may not be significantly superior to Co. Thus, there are serious materials issues with straintronic switches and these have to be overcome before the technology can mature.

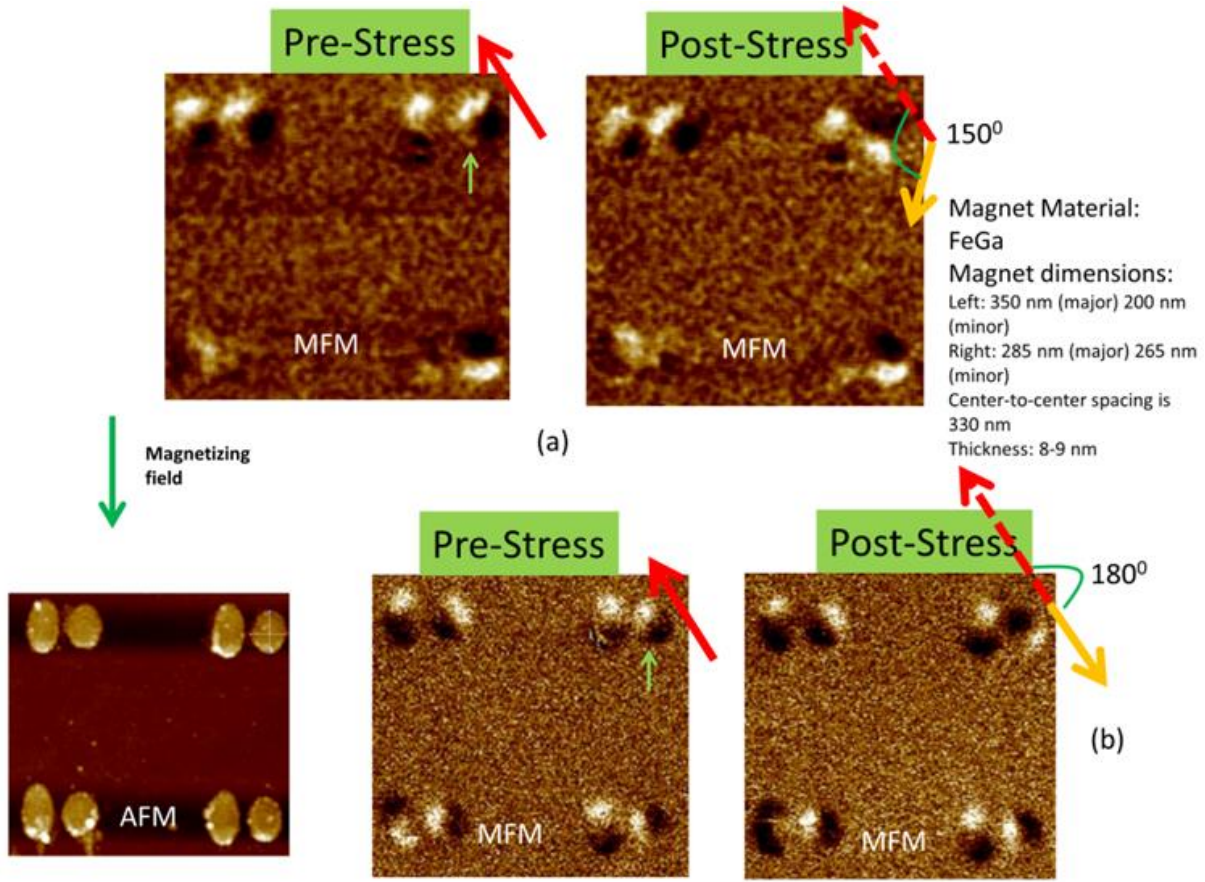

**Fig. 24**: Operations of FeGa inverter pairs. Pre- and post-stress MFM images of two different samples are shown. Reproduced from ref. [141] with permission from the Institute of Physics.

### 3.6   Energy dissipation in scaled straintronic switches – estimate from experiments

In the experiments with Co in ref. [34], the electric field that had to be generated in the PMN-PT substrate to produce 80 MPa of stress for switching was 0.6 MV/m. In a 100 nm thin PMN-PT film, this would translate to a voltage of $0.6 \times 10^6 \times 100 \times 10^{-9} = 60$ mV. The primary component in the energy dissipation during switching is the $CV^2$ dissipation, where $C$ is the capacitance of the electrode pads used to apply the voltage across the piezoelectric substrate and $V$ is the voltage needed to generate 80 MPa of stress. Since $C \sim 1$ fF and $V \sim 60$ mV, the energy dissipation is 3-4 aJ, which is exceptionally low. Accounting for non-linearity and other losses, ~10 aJ would be a conservative estimate for energy dissipated per bit, which is consistent with our earlier theoretical estimates.

## 3.7 Switching multiferroic nanomagnetic switches with bulk and surface acoustic waves

So far, we have discussed straintronic switching of a magnetostrictive nanomagnet with a *static* strain generated with a static voltage applied across an underlying piezoelectric substrate or film in elastic contact with the nanomagnet. In this section, we discuss dynamic (time-varying) stress generated with a bulk or surface acoustic wave. As the wave propagates, it generates periodic compressive and tensile stress at any given location.

The use of an acoustic wave to generate time varying stress has distinct advantages. For example, consider the Bennet clocking scenario in Fig. 22c. In order to propagate the input bit down the line, we have to use local gate pads surrounding each nanomagnet and activate them (pairwise) with a voltage pulse sequentially, using a multiphase clock. This calls for complex lithography (placing gate pads in precise alignment with the nanomagnets) and then contacting them for connection to the outside world. We can instead use an acoustic wave to propagate stress down the line, which will also sequentially stress the nanomagnets and serve our purpose. This will eliminate the need for the contact pads (which are a daunting lithographic challenge), but it comes with its own challenges. First, the spacing between the nanomagnets has to be one quarter of the wavelength in order to generate the right sequence of stress. The spacing may be ~300 nm (any larger spacing may make the dipole coupling too weak) and hence the wavelength $\lambda$ of the acoustic wave has to be no more than ~1.2 μm. Acoustic waves with frequency $f$ exceeding ~1 GHz are lossy in standard piezoelectric substrates and hence let us assume that we are constrained to a frequency of 1 GHz. A higher frequency may not work anyway since the switching delay of nanomagnets is not shorter than ~1 ns. A frequency of 1 GHz would require the acoustic wave velocity to be $v = \lambda f = 1.2 \times 10^{-6} \times 10^9 = 1200$ m/sec, which is very low. Typical acoustic wave velocities are 3-5000 m/sec in most piezoelectric substrates and hence we are off by a factor of ~4. There are procedures to produce slow acoustic waves, with velocities of the order of 100 m/sec [141], but they are complicated. In reality, we may want to work at frequencies considerably lower than 1 GHz to give the nanomagnets ample time to switch, and this would require an even slower acoustic wave. Thus, there are challenges associated with acoustic wave clocking as well. Finally, this kind of clocking can work well only with very simple nanomagnet geometries. More complicated geometries will require propagating acoustic waves in different directions and these may interfere with each other to complicate matters further. We mention acoustic wave clocking (for such functions as Bennett clocking) merely as an enticing prospect, while recognizing that its actual implementation is going to be certainly difficult.

While Bennett clocking with acoustic waves faces some hurdles, there are reports of switching the magnetizations of isolated and dipole-coupled nanomagnets with acoustic waves [36, 37]. Recent work in this area has been motivated by the realization that "mixed mode" switching of nanomagnets, where both spin transfer torque and stress produced with an acoustic wave are used to switch the magnetization of an elliptical nanomagnet with in-plane anisotropy, can reduce the switching energy dissipation compared to switching with spin transfer torque alone [142, 143]. Periodic switching of magnetization between the hard and easy axis of 40 μm × 10 μm × 10 nm Co bars sputtered on $LiNbO_3$ has been shown [144]. Other authors have studied acoustically induced switching in thin films [145] including focusing surface acoustic waves (SAW) to switch a specific spot in an iron-gallium film [146] as shown in Fig. 25. Several proposals suggest a complete 180º rotation with an appropriately timed acoustic pulse [147]. Stroboscopic X-ray techniques have been used to study strain waves and magnetization at the nanoscale [148].

Excitation of spin wave modes in GaMnAs layers by a picosecond strain pulse [149] as well as magnetization dynamics in GaMnAs [150] and GaMn(As,P) [151] have been demonstrated. In in-plane magnetized systems, surface acoustic waves have been utilized to drive ferromagnetic resonance in thin Ni films [152, 153]. Resonant effects have also been studied by spatial mapping of focused SAWs [154]. There are theoretical studies of the possibility of complete magnetization reversal in a nanomagnet subjected to acoustic wave pulses [147, 155]. Interestingly, for high frequency excitation of extremely small nanomagnets, the Einstein De Haas effect seems to dominate as has been proposed [156] and experimentally demonstrated [157].

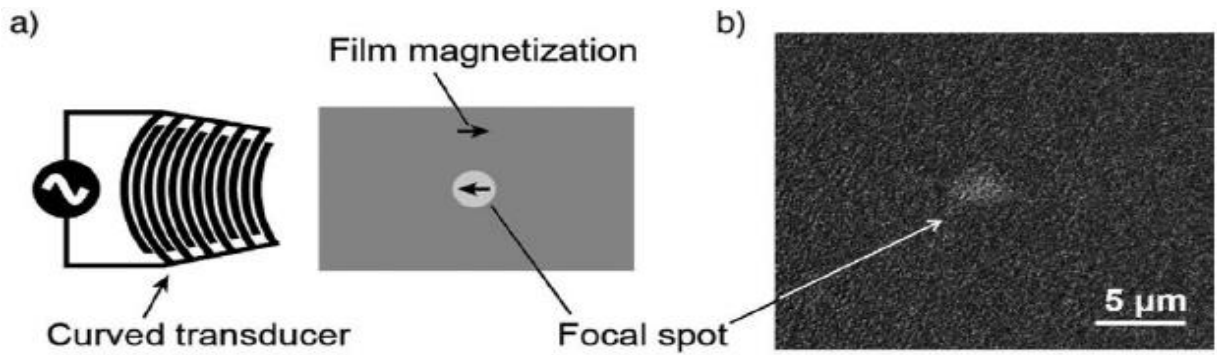

**Fig 25.** (a) A scheme for using a curved transducer to switch a magnetization of a focused spot (b) Kerr image showing the reversed magnetization. Reprinted from Li W, Buford B, Jander A and Dhagat P 2014 Writing magnetic patterns with surface acoustic waves *J. Appl. Phys.* **115** 17E307 with permission of AIP Publishing.

We have studied switching of the magnetization of nanomagnets under the influence of surface acoustic waves [37]. The magnetic states of elliptical cobalt nanomagnets (with nominal dimensions of ∼340 nm✕270 nm✕12 nm) delineated on bulk 128° Y-cut lithium niobate were changed with acoustic waves (AWs) launched in the lithium niobate substrate. Isolated nanomagnets that are initially magnetized with a magnetic field to a single-domain state, with the magnetization aligned along the major axis of the ellipse, are driven into a vortex state by acoustic waves that modulate the stress anisotropy of these nanomagnets. The nanomagnets remain in the vortex state until their magnetizations are realigned by a strong magnetic field to the initial single-domain state, making the vortex state nonvolatile. A diagram showing the experimental structure is given in Fig. 26.

We have also studied a NOT gate whose operation is triggered by an acoustic wave. Dipole coupled pairs of elliptical Co nanomagnets were delineated on a $LiNbO_3$ substrate with one member of the pair more eccentric than the other [36]. The first encodes the input bit in its magnetization state and the second the output bit. As usual, both members are magnetized in the same direction with an external magnetic field to represent the (1, 1) states of the input and output bits. After exciting both nanomagnets with an acoustic wave, the less eccentric nanomagnet flips its magnetization to assume a state where its magnetization is antiparallel to that of the more eccentric nanomagnet. This represents the logic state (1, 0) which is the correct state for an inverter. The results are shown in Fig. 27.

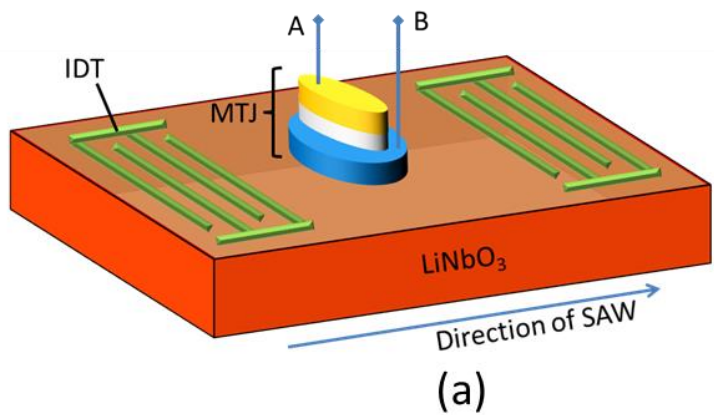

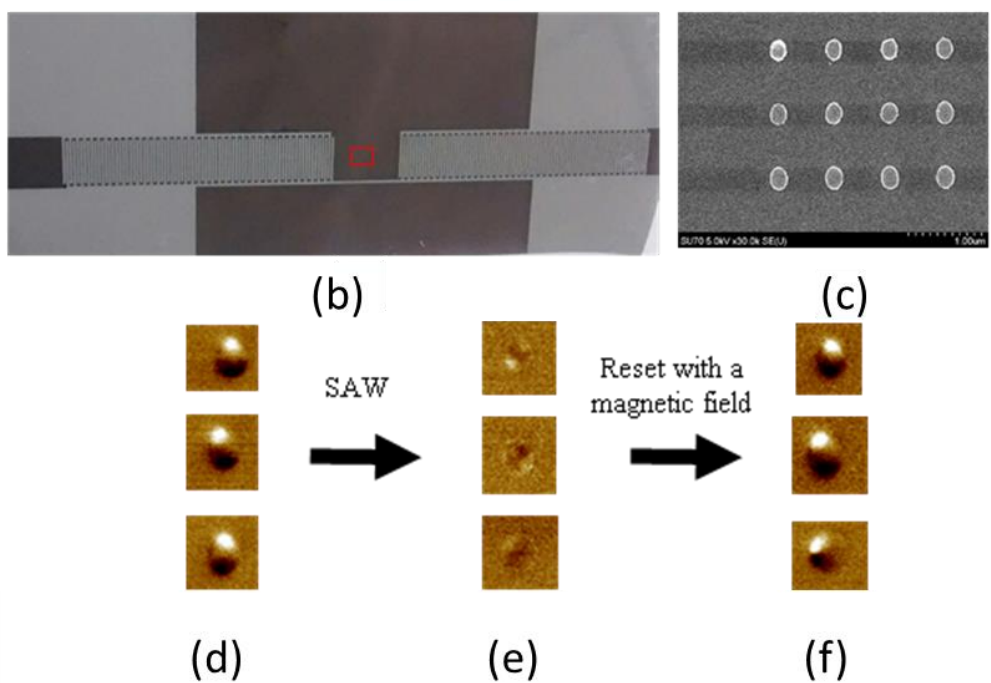

**Fig. 26**: (a) An illustration of an MTJ whose soft layer is magnetostrictive fabricated between interdigitated transducers (IDT) that launch acoustic waves in the $LiNbO_3$ substrate. In the experimental devices in (b) and (c), no MTJ was fabricated and only magnetostrictive nanomagnets were deposited on the piezoelectric substrate. (b) a micrograph of a delay line consisting of the IDTs (magnetostrictive Co nanomagnets are fabricated in the area enclosed by the small red square; (c) scanning electron micrograph of the nanomagnets; (d) magnetic force micrograph of three different elliptical nanomagnets magnetized into single domain states by a magnetic field applied along the major axes, (e) the nanomagnets are driven into a vortex state as shown by the magnetic force micrograph of the state after experiencing the acoustic wave; (f) the initial single domain state is recovered by applying an external magnetic field. (b)-(f) are reprinted with permission from Sampath V., D'Souza N., Bhattacharya D., Atkinson G. M., Bandyopadhyay S. and Atulasimha J. 2016 Acoustic-wave-induced magnetization switching of magnetostrictive nanomagnets from single-domain to non-volatile vortex states *Nano Lett.* **16** 5681 Copyright 2016 American Chemical Society.

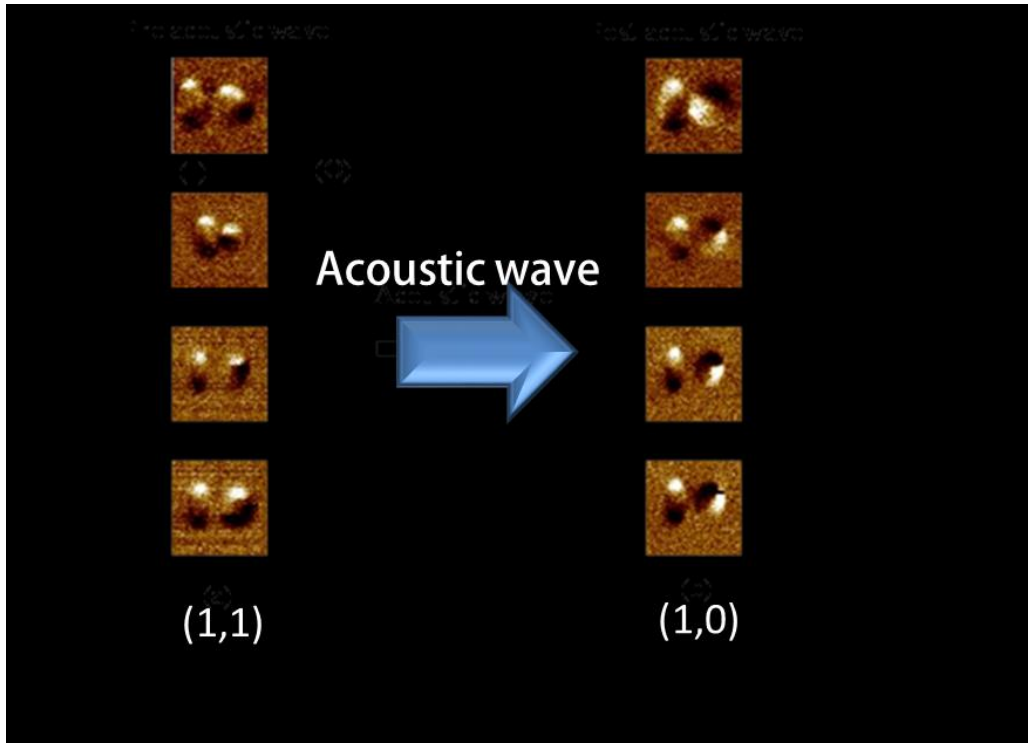

**Fig. 27**: Magnetic force micrographs of four different pairs of elliptical Co nanomagnets fabricated on a LiNbO$_3$ substrate. The left member is more eccentric than the right member in each pair. The magnetization state of the left member represents the input bit and that of the right member represents the output bit. (a) Both members are magnetized to the (1, 1) state with an external magnetic field that aligns the magnetizations in the "down" direction. (b) After passage of the acoustic wave, the magnetization of the output nanomagnet flips to produce the (1, 0) configuration that corresponds to the operation of a NOT gate. Reprinted from Sampath V., D'Souza N., Atkinson G. M., Bandyopadhyay S. and Atulasimha J. 2016 Experimental demonstration of acoustic wave induced magnetization switching in dipole coupled magnetostrictive nanomagnets for ultralow power computing *Appl. Phys. Lett.* **109** 102403 with permission of AIP Publishing.

### 3.8    Summary of "straintronic" switches

This section summarized some of the recent work in the field of "straintronics", with emphasis on device applications. There is now more than sufficient evidence in the literature that the magnetization in nanoscale magnetostrictive structures can indeed be controlled by strain. Dipolar interaction between closely spaced neighboring nanomagnets can implement nanomagnetic logic devices clocked by voltage generated strain. However, such gates are error-prone and the dynamic error rates (the error in switching a nanomagnet's magnetization to the desired stable state) in such gates are too high for logic applications. Recent experimental work performed in conjunction with modeling [158,159] as well as other experimental [160] and modeling effort [161] explain the role of thermal noise and defects (that pin the magnetization) causing large switching errors.

There are two potential ways of reducing the switching error rate:
1. Developing nanostructures of materials such as Terfenol-D that have high magnetomechanical coupling, so it is possible to generate a high $H_{eff}$ (see Equation 3.1 and associated explanation) and possibly overcome the pinning due to defects. Recently, highly magnetostrictive Terfenol-D films grown at CMOS compatible temperature has been demonstrated [162] but patterning down to nanoscale or detailed study of switching analysis still needs to be performed.
2. Designing device architectures that do not rely (or rely very little) on dipole coupling to elicit logic functionality. Some such devices proposals are discussed next in Section 4.

**Section 4: Proposals for Boolean Logic and Memory Devices based on "Strainronic MTJs"**

While Section 3 discussed various experimental work based on strain mediated switching of the magnetization of nanomagents, this section highlights two device proposals that employ strain to switch the soft layer of a magnetic tunnel junction (MTJ): one for logic and the other for memory.

**4.1    Straintronic Boolean Universal Logic Gate**

Most digital computation and signal processing today is carried out with Boolean logic gates. Almost all logic gates in existence at this time are realized with transistors and almost none with nanomagnets. Nevertheless, there has been some interest in implementing Boolean logic gates with nanomagnets because they may have the potential to be more energy efficient, but much more importantly, they are "non-volatile". These gates can perform a Boolean operation and then store the result of the operation locally, in-situ, and not have to store them in a remote memory. The gate retains the information even after the circuit is powered off, and this allows certain types of circuits to be implemented, which may exhibit superior performance compared to circuits implemented with volatile logic gates built with transistors.

The ability to store and process information with the same device could afford immense flexibility in designing computing architectures. Non-volatile-logic-based architectures can reduce overall energy dissipation by eliminating refresh clock cycles, improve system reliability and produce 'instant-on' computers with virtually no boot delay. A number of non-volatile universal logic gates implemented with nanomagnets have been proposed to date [163-168] but not all of them satisfy all the requirements for a logic gate [168, 169] and therefore may not be usable in all circumstances. Furthermore, those that rely on dipole interaction between nanomagnets [163] are extremely error-prone [170-172].

Most nonvolatile nanomagnetic logic schemes that have been proposed and analyzed so far exhibit poor energy-delay product. This happens because the methods adopted to switch the nanomagnets in these schemes are sub-optimal. The scheme in Ref. [163], for example, uses current generated magnetic fields to switch magnets and hence would dissipate enormous amount of energy, orders of magnitude more than transistor-based logic [24]. A recent experiment that used on-chip current-generated magnetic fields to switch magnets dissipated approximately $10^{12}$ kT of energy (4 nJ) per switching event, despite switching very slowly in ~1 $\mu$s (energy-delay product = $4 \times 10^{-15}$ J-s) [173]. Extremely energy-inefficient switching is the primary reason why these logic schemes end up wanting. In fact, most non-volatile nanomagnetic schemes that have been critically examined so far appear to be inferior to transistor-based logic in energy-delay product [174-176], despite the promise of nanomagnets. This, and the high error rates, have prevented the widespread application of nanomagnetic logic, despite the attractive property of non-volatility.

In the past, we had proposed a straintronic non-volatile NAND gate implemented with a "skewed" magneto-tunneling junction (MTJ) [176]. A skewed MTJ (s-MTJ) is one where the easy axes of the hard and soft elliptical layers are non-collinear. This does not require making the major axes of the two layers non-collinear (which is a difficult fabrication feat). We can make the easy axis of the hard layer non-collinear with its own major axis by applying a strong magnetic field non-collinear with the major axis during annealing of the hard layer which is implemented with a synthetic anti-ferromagnet. This pins its magnetization in the direction of the applied magnetic field. As a result, the easy axis of the hard layer and that of the soft layer become non-collinear since the latter is still along the major axis of the soft layer. The soft layer is a magnetostrictive material with a piezoelectric layer integrated vertically underneath. The gate is shown in Fig. 28.

The two input bits are encoded in voltages that generate strain in the soft layer. The gate is reset to a state where the output voltage of the MTJ is high and encodes the output bit 1. When the inputs are (0, 0), (0, 1) or (1, 0), the voltage appearing across the piezoelectric layer does not generate enough strain to beat the shape anisotropy energy of the soft layer and rotate its magnetization, so the output remains high and represents the bit 1. Only when the inputs are in the state (1, 1), the voltage appearing across the piezoelectric layer generates enough strain to rotate the magnetization of the soft layer and make the resistance of the MTJ switch. This switches the output voltage to a state that represents the bit 0. This then implements the truth table of the NAND gate.

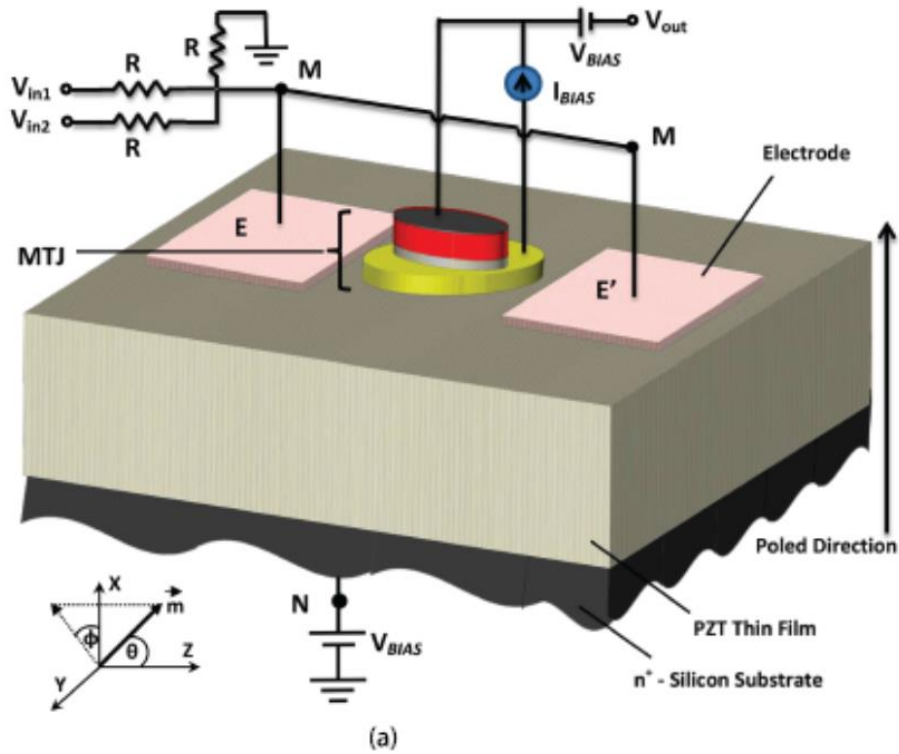

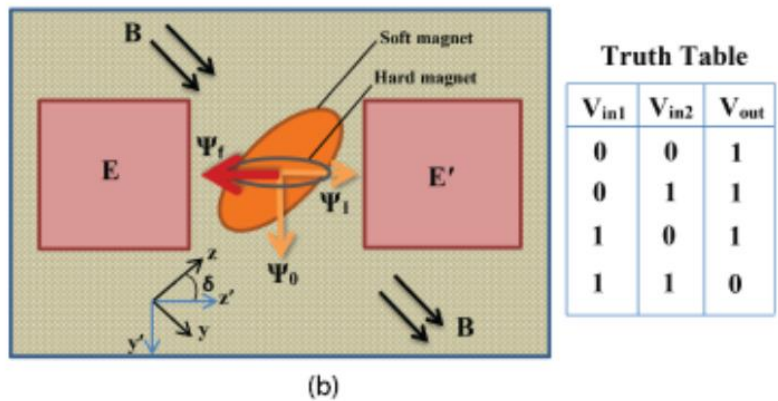

**Fig. 28**: (a) Structure of the straintronic NAND gate; (b) top view and truth table. Reproduced from ref. [179] with permission from Nature Publishing Group (license link http://creativecommons.org/licenses/by/4.0/).

The calculated energy-delay product of this proposed gate is $1.6 \times 10^{-26}$ J-sec [176], which potentially makes it comparable to a transistor-based NAND gate in energy-delay product. This gate also satisfies the essential requirements of logic, namely concatenability, non-linearity, isolation between input and output, gain, logic universality and scalability, but it fails in error resilience. The calculated dynamic bit error probability was $10^{-8}$, which is much too high for logic. Logic has much more stringent requirements for error-resilience than memory. In memory, if a single bit is corrupted, it does not affect other bits, but in logic, if a single bit is corrupted and is fed as input to a succeeding gate, then the output of that gate is corrupted, and so on. Thus, error in logic is "contagious" and propagates. The bit error probability in a switching operation should be no more than, say, $10^{-15}$. If the gate is switching once in every 1 ns, then the mean time between errors will be $10^6$ seconds, or 11.6 days. Even this may be a little too much. Unfortunately, magnetic logic gates are vulnerable to errors since magnetization dynamics is easily disrupted by thermal noise which acts like a random magnetic field. There are certain digital applications, such as stochastic computing, that may be able to tolerate high error probabilities, but conventional Boolean logic cannot. That is why we believe that despite the non-volatility, nanomagnetic devices are not very attractive for Boolean logic. There are many other application areas, some of which we will discuss later in this article, where nanomagnetic information processing devices (not necessarily logic gates) can be very attractive.

## 4.2 Straintronic Memory

A straintronic memory cell is straightforward to implement and is shown in Fig. 29. It consists of a MTJ whose soft layer is magnetostrictive and in elastic contact with a thin piezoelectric film through an ultrathin metal layer. The hard layer is a synthetic anti-ferromagnet whose magnetization is pinned. The fabrication of this device is very similar to that of the device reported in ref. [35] and does not pose any additional challenge.

The four corner electrodes are shorted pairwise and are used to apply stress in two different directions to switch the magnetization of the soft layer by $180^0$. The stresses are generated by applying a voltage between a shorted pair and ground, which drops a voltage across the piezoelectric layer and generates biaxial strain in the soft layer (compressive along the line joining the pair and tensile in the perpendicular direction, or vice versa, depending on the voltage polarity and the direction in which the piezoelectric thin film was poled). These stresses flip the magnetization of the soft layer. We note that the key ideas on which this the proposed device is predicated are: (1) Experimental demonstration of rotation of magnetization in the magnetostrictive layer by $180º$ with strain applied successively in different directions; (2) Theoretical simulations showing the potential of this device to switch with error rates ~$10^{-6}$ or lower in the presence of thermal noise (provided excessive pinning of magnetization by defects is avoided and nanomagnets with materials that exhibit large magnetostriction can be developed). These were discussed in Section 3. Here, we develop this device concept further by proposing the incorporation of a MTJ layer for read out.

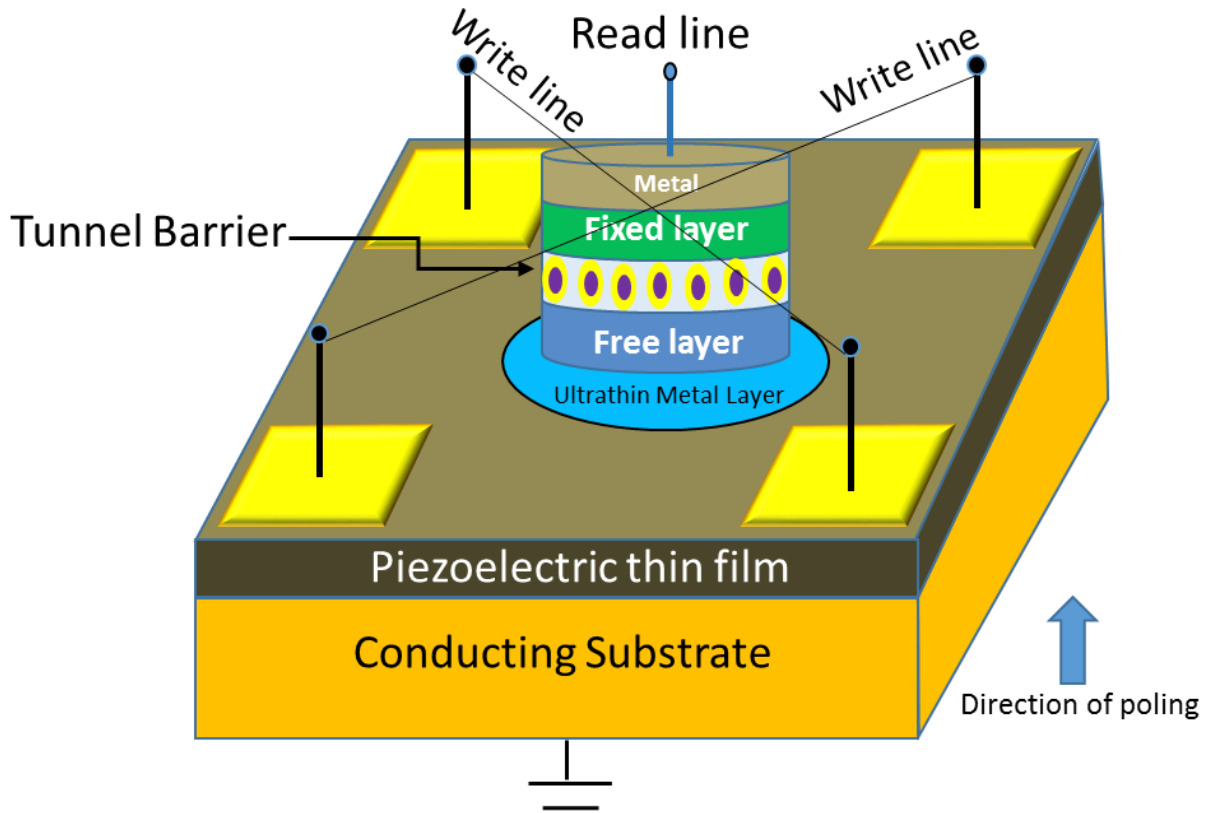

**Fig. 29**: A straintronic toggle memory cell.

In order to "write" a bit, the MTJ resistance is measured between the two floating lines to "read" the stored bit. If the stored bit is the desired bit, then no action is taken. If not, then the write lines are activated to generate stress sequentially in two different directions to flip the magnetization of the soft layer and write the desired bit.

The ultrathin metal layer serves two purposes: it adheres the MTJ to the piezoelectric film, and it also allows directly contacting the soft layer for the purpose of measuring the MTJ resistance to read the stored bit. This layer is thin enough that it does not impede stress transfer from the piezoelectric to the soft layer significantly.

This type of memory is called a "toggle" memory since every write cycle is preceded by a read cycle. Only when the read bit is not the desired bit, the memory is toggled with voltage generated stress to write the desired bit. To verify that the bit indeed toggled, one can, of course, read the resistance of the MTJ, but there is another way to verify the toggling action. If the magnetic state of the magnetostrictive nanomagnet changed, then it will induce a magnetoelectric voltage in the piezoelectric layer which can be read. If no toggling occurs, this voltage will not be produced [132].

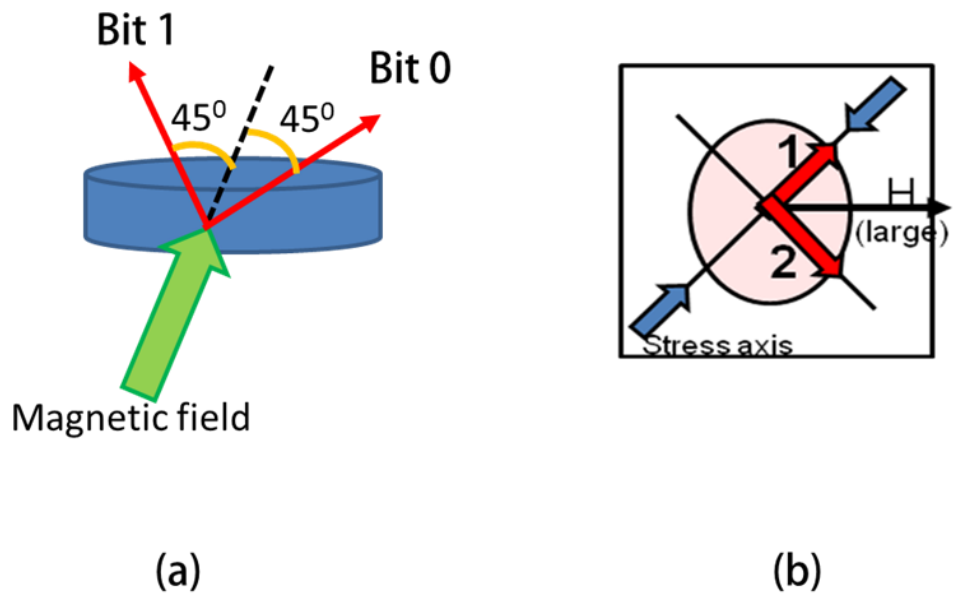

**Fig. 30**: (a) A magnetic field applied in-plane in the direction of the minor axis of an elliptical nanomagnet brings the two stable states out of the major axis into two mutually perpendicular directions in the plane of the nanomagnet; (b) Uniaxial stress is applied along one of these directions. Compressive stress will drive the magnetization to the other stable direction while tensile stress will keep the magnetization aligned along the stress axis, if the magnetostriction is positive. The opposite will happen if the magnetostriction is negative. The illustration and switching scheme is based on [177].

It is also possible to implement a non-toggle memory [177]. If a magnetic field is applied on an elliptical magnetostrictive nanomagnet along the minor axis, it brings the two stable states out of the major axis. By adjusting the field strength, the angle between the two stable states can be adjusted to ~$90^0$ as shown in Fig. 30. These two states encode the bits 0 and 1. Uniaxial stress is applied along one of the stable directions. If the magnetostriction coefficient of the nanomagnet is positive, then compressive stress will drive the magnetization to the other stable direction and tensile stress will keep the magnetization pointing in the direction of the stress. The signs of the stresses will be reversed if the magnetostriction of the nanomagnet is negative.

Clearly, in this strategy, we need to have no prior knowledge of the stored bit if we wish to write either bit 1 or bit 0. Thus, there is no need for a read cycle to precede the write cycle. If we wish to write bit 1, we will simply apply one sign of the stress, and if we wish to write bit 0, then we will apply the opposite sign of the stress. Hence, this is a "non-toggle" memory.

One final issue that needs to be addressed in the context of straintronic memory is the issue of endurance. Since piezoelectric materials suffer from piezoelectric fatigue, a memory cell cannot be cycled through many cycles of program/erase reliably. There are reports of thin piezoelectric films of thickness ~1 μm not experiencing fatigue after $10^5$ cycles [178], which would indicate that endurance comparable to flash memory may be achievable. There are no reports of endurance in thinner films. Better endurance might be possible by optimizing film quality.

# 5. Straintronic Magneto-Tunneling Junction

In the previous two sections, when we talked about straintronic logic and memory, we invoked a magneto-tunneling junction (MTJ) whose soft layer's magnetization is rotated with electrically generated strain. This requires the soft layer to be the magnetostrictive layer of a two-phase (magnetostrictive/piezoelectric) multiferroic, i.e. the soft layer is realized with a magnetostrictive material in elastic contact with an underlying piezoelectric thin film as shown in Fig. 31. Application of a voltage across the piezoelectric will produce strain in the magnetotrictive nanomagnet and rotate its magnetization, thereby changing the resistance of the MTJ.

Straintronic MTJs have been demonstrated by a number of groups [35, 117]. In Ref. [35], a straintronic MTJ was implemented with CoFeB soft layer fabricated on a piezoelectric PMN-PT substrate, a MgO spacer layer and a thicker CoFeB layer acting as the hard layer. Gate pads were delineated around the MTJ to apply voltages that generated strain in the PMN-PT layer, which rotated the magnetization of the soft layer and caused the resistance of the MTJ to change. Repeated toggling of the MTJ resistance by pulsing the voltage applied on the PMN-PT substrate was demonstrated. The tunneling magnetoresistance (TMR) was greater than 100% at room temperature, meaning that the resistance could be changed by more than a factor of 2 with the applied voltage, which generated strain.

In Fig. 31, the schematic of the MTJ fabricated in ref. [35] is shown, along with the simulated strain profile in the substrate upon applying a voltage. The strain profile was calculated with COMSOL Multiphysics package. It also shows the experimentally measured magnetoresistance curves under different voltages applied across the PMN-PT substrate, as well as the variation of the MTJ switching (magnetic) field [squares] and tunneling magnetoresistance ratio TMR [circles] of the MTJ as a function of voltage across the piezoelectric.

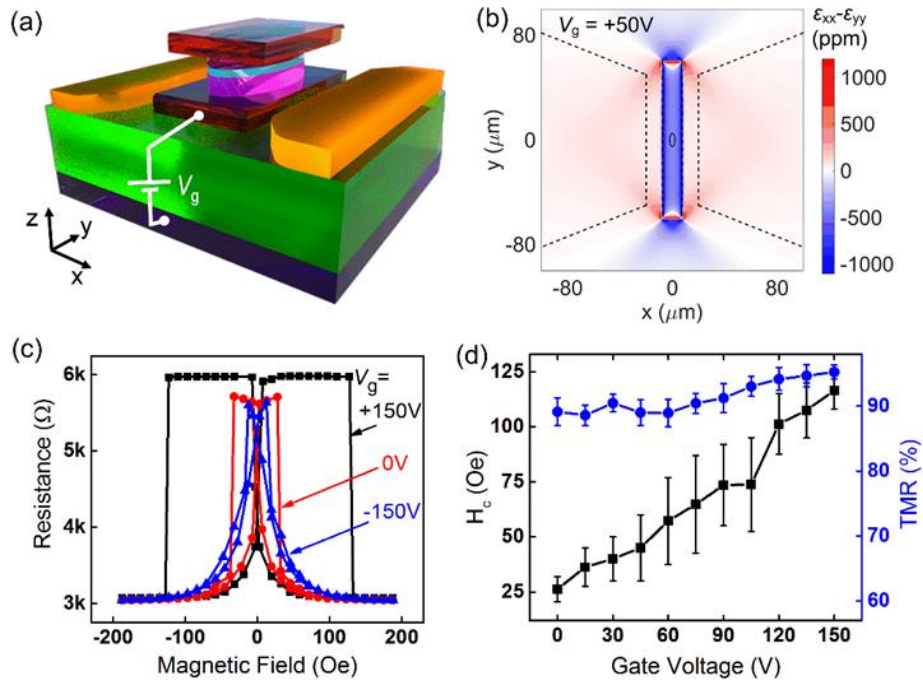

**Fig. 31**: (a) Schematic of the straintronic magnto-tunneling junction. A voltage $V_g$ is applied across the piezoelectric; (b) The in-plane anisotropic strain $\varepsilon_{xx} - \varepsilon_{yy}$ profile generated in the piezoelectric substrate upon application of a gate voltage $V_g = +50\,\text{V}$. The solid line ellipse at the center denotes the MTJ pillar, and the dashed lines denote the positions of electrodes and side gates shown in (a); (c) Magnetoresistance traces measured under different gate voltages $V_g$. (d) Variation of the switching (magnetic) field [squares] and tunneling magnetoresistance ratio (TMR) [circles] of the MTJ as a function of $V_g$. Reprinted from Zhao Z. Y., Jamali M., D'Souza N., Zhang D., Bandyopadhyay S., Atulasimha J. and Wang J. P. 2016 Giant voltage manipulation of MgO-based magnetic tunnel junctions via localized anisotropic strain: A potential pathway to ultra-energy-efficient memory technology *Appl. Phys. Lett.* **109** 092403 with permission of AIP Publishing.

In Fig. 32, micromagnetic simulation results showing the spin texture within the elliptical soft layer at two different voltages -80 V and +80 V applied across the PMN-PT substrate, are presented. Also shown are the magnetoresistance loops for -80 V and +80 V. Finally, when the gate voltage is pulsed between - 80 V and +80 V, the magnetization of the soft layer alternates between the two spin textures shown in this figure, and this then toggles the resistance between a high and a low value that differ by more than a factor of 2. Thus, the resistance of the MTJ can be alternated between two values (which encode the binary bits 0 and 1) with a voltage generating strain in the soft layer. This is the demonstration of a straintronic MTJ.

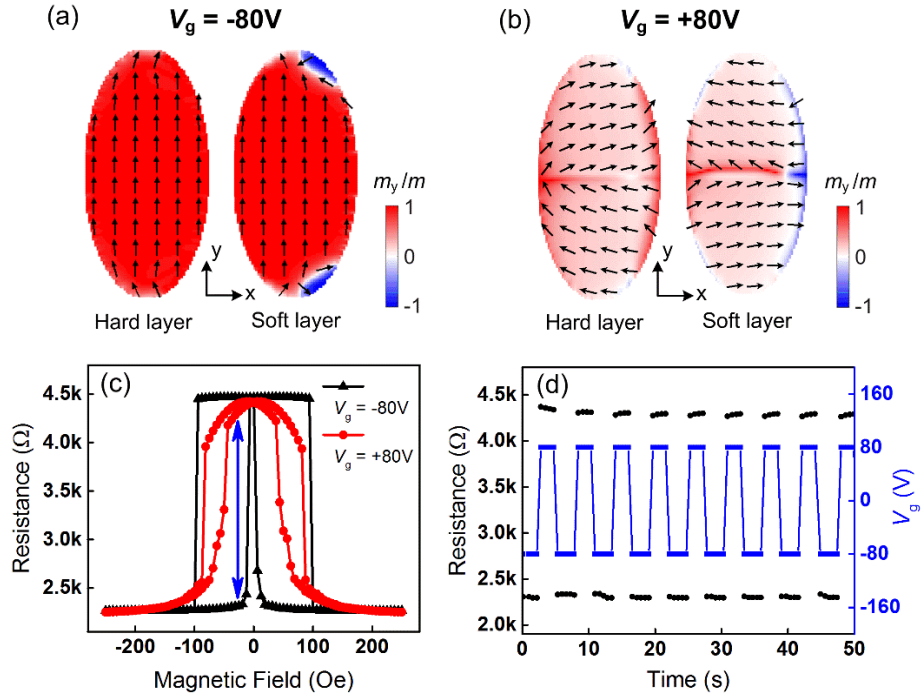

**Fig. 32:** (a)-(b) Micromagnetic simulation results showing the magnetization configurations of the hard and soft CoFeB layers after application of (a) $V_g = -80$ V and (b) $V_g = +80$ V. A small bias field of 30 Oe is applied along the major axis to overcome any effect of dipole interaction. The dimension of the magnet is 3 µm × 6 µm. Black arrows indicate the direction of magnetic moments. (c) Measured magnetoresistance loops for $V_g = -80$ V and $V_g = +80$ V. The blue arrow indicates the switchable high- and low-resistance states. (d) Toggling of the MTJ between high- and low-resistance states with application of ±80 V gate voltage pulsing. A small bias magnetic field of 30 Oe is applied along the +y-axis (refer to Fig. 32a) to overcome the dipole interaction between the two magnetic layers. Reprinted from Zhao Z. Y., Jamali M., D'Souza N., Zhang D., Bandyopadhyay S., Atulasimha J. and Wang J. P. 2016 Giant voltage manipulation of MgO-based magnetic tunnel junctions via localized anisotropic strain: A potential pathway to ultra-energy-efficient memory technology *Appl. Phys. Lett.* **109** 092403 with permission of AIP Publishing.

In Figs. 32a and 32b, it can be seen that the magnetizations of the elliptical hard and soft layers are aligned along their major axes when the voltage across the piezoelectric is negative and along their minor axes when the voltage is positive. This happens because of the following reason. When the applied voltage is negative, it generates compressive strain in the elliptical hard and soft layers in the direction of their major axes and tensile strain in the direction of their minor axes. Since the magnetostriction of CoFeB is negative, such a biaxial strain profile aligns the magnetizations of both layers along their respective major axis. When the sign of the voltage is reversed, the signs of the stresses reverse as well and the new strain profile aligns the magnetizations of both layers along their minor axes, but in *opposite* directions. As a result, the magnetizations of the hard and soft layers are approximately parallel for negative gate voltage and approximately anti-parallel for positive gate voltage. This results in the resistance switching from low to high value, or vice versa, when the gate voltage sign is reversed.

It is interesting to note that when the magnetizations are along the major axes, they are roughly parallel but when they are along the minor axes, they are anti-parallel. This happens because a small bias magnetic field is applied along the major axis to overcome any effect of dipole coupling between the hard and soft layers. This field ensures that when the magnetizations lie along the major axis, they are parallel. When the voltage polarity changes, the new stress profile rotates the magnetizations by $90^0$ to align them along the minor axes. Since there is no magnetic field along the minor axes, dipole coupling is not suppressed, and this time the magnetization of one layer will rotate clockwise and the other anti-clockwise so that the two magnetizations become mutually anti-parallel. The small bias magnetic field along the major axes and dipole coupling are responsible for this behavior. This effect increases the tunneling magnetoresistance ratio (TMR) and hence is beneficial.

In Fig. 32(d), the resistance switching occurs every ~1 second. This should not lead to the inference that the switching takes ~1 second. Commercial resistance meters usually make several measurements of resistance over 1 second and then averages over them to reduce noise. This is the reason that the switching was repeated every second. The actual switching occurs in ~1 nanosecond according to simulations.

The voltage of 80 V was dropped across a 0.5 mm thick piezoelectric substrate. If, instead, we had a 100 nm thick piezoelectric thin film, then the required voltage would have been $80 \times \left(10^{-7} / 5 \times 10^{-4}\right) \times 10 = 160$ mV, even after accounting for a 10-fold reduction in the piezoelectric coefficient going from a substrate to a thin film. The reported relative dielectric constant of PMN-PT is about 1,000 [179]. Hence, if we assume a gate electrode pad area of 100 nm × 100 nm to apply the voltage across the PMN-PT film, then the gate capacitance *C* will be ~ 1fF. The resulting $CV^2$ will be ~25 aJ. This would make it a relatively energy-efficient switch.

# 6. Straintronic Non-Boolean Circuits

Boolean circuits usually need devices to possess two attributes: a large ratio of off-resistance to on-resistance, which makes the circuit error-resilient, and a small dynamic switching error rate, which makes the circuit robust and reliable. These attributes are especially desirable for logic circuits. Unfortunately, magneto-tunneling junction (MTJ) switches, which are the mainstay of nanomagnetic logic circuits, do not possess these attributes. While CMOS transistors could have an off-to-on resistance ratio of ~$10^6$:1, MTJs barely have an off-to-on resistance ratio of 5:1 at room temperature. In fact, as of this writing, the largest MTJ resistance ratio demonstrated at room temperature is slightly higher than 7:1 [180]. However, MTJs have an advantage that make them attractive –non-volatility. These attributes are best utilized in *non-Boolean* circuits that are often relatively forgiving of errors and also accommodating of low off-to-on resistance ratios, but would benefit from the non-volatility. In this section, we will discuss some non-Boolean circuits that can be implemented with straintronic magneto-tunneling junctions and other devices. They will be energy-efficient (as one can potentially implement some non-Boolean functionalities with much fewer MTJs or hybrid MTJ-CMOS devices compared to a purely CMOS implementation) and non-volatile, which is why they are attractive.

## 6.1 Equality bit comparators

An equality bit comparator is a critical element of electronic locks and other cybersecurity hardware. A string of bits, called "reference" bits, acts as the lock's "combination", while another string of bits, called "input" bits, acts as the lock's "key". A multi-bit equality comparator compares the input bit stream with the reference bit stream. If each bit in one stream matches the corresponding bit in the other stream, then the lock opens; otherwise, it does not. Thus, only an authorized user, equipped with the "key" can access the contents protected by the lock.

A transistor-based implementation of a 16-bit equality comparator will require 16 XOR gates and one 16-input AND gate [181]. Each XOR gate will need 9 transistors [182] and the 16-input AND gate will require 17 transistors, making the total transistor count 161.

A straintronic implementation of a 16-bit equality comparator needs only 16 straintronic spin valves and one MTJ [183]. The spintronic version is *non-volatile* unlike any transistor-based rendition and will retain the result of the bit comparison indefinitely in the comparator itself since the elements are magnetic.

Moreover, the programmed (or reference) bit stream can be stored in-situ in the comparator without the need for refresh cycles and/or the need to fetch them from a remote "memory" for the purpose of comparison with the input bit stream. Frequent communication between processor and memory increases delay, energy dissipation and likelihood of faults unnecessarily. Finally, a non-volatile electronic lock is more secure since the reference bit stream is not stored at a different location (memory) but is stored in the comparator itself. To breach security, one would have to compromise the comparator, which is harder to do than to intercept bits in the communication channel between the comparator and memory.

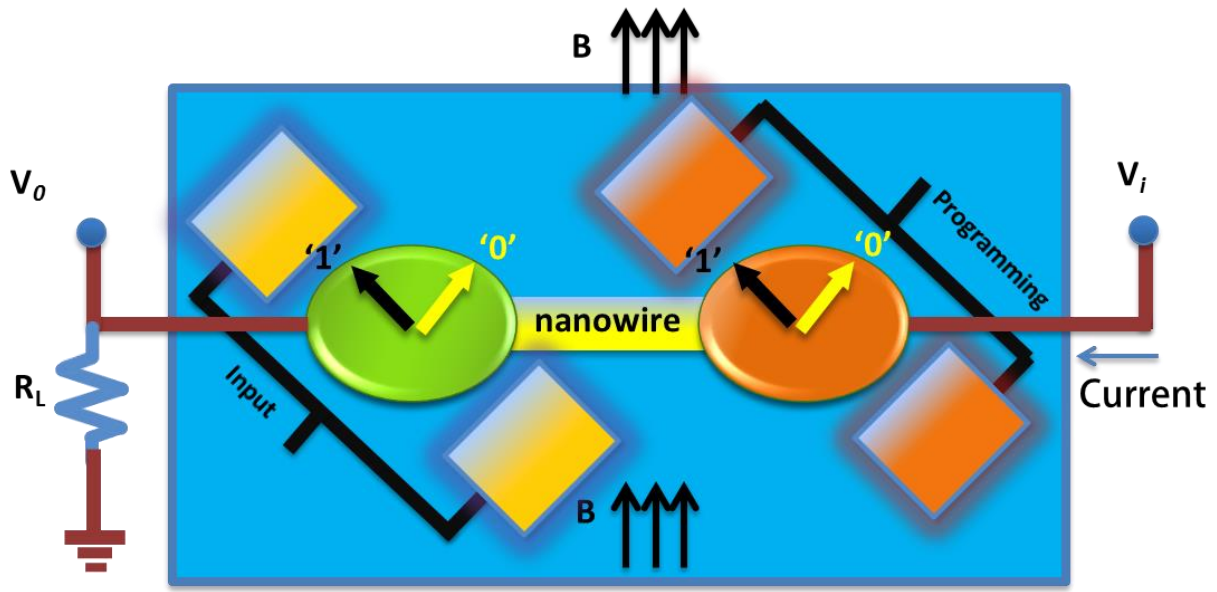

**Fig. 33:** A straintronic single equality bit comparator.

The way a single bit comparator works is illustrated in Fig. 33. It employs a nanowire "spin valve" realized with a nanowire spacer placed between two elliptical magnetostrictive nanomagnetic contacts on a piezoelectric substrate. Gate pads are delineated around each contact as shown in Fig. 33. A global magnetic field B is applied along the minor axes of the elliptical contacts to bring their stable magnetization states out of their major axes and make them lie in the planes of the nanomagnets subtending an angle of ~$90^0$ between them.

When an electrically shorted gate pad pair is activated with a voltage, either compressive or tensile stress is generated in the intervening magnetostrictive nanomagnet (lying between the pair) in the direction joining the centers of the pair, depending on the voltage polarity. Assume that the magnetostriction coefficient of the nanomagnet is positive. In that case, compressive stress will drive the magnetization state to an orientation perpendicular to the stress direction, i.e. to the state encoding bit 0 in Fig. 33, while tensile stress will drive the magnetization to the other state. Thus, the magnetization of either contact will have one of two orientations depending on the polarity of the voltage applied to the shorted gate pair surrounding it.

The input bit and the reference (programming) bit are encoded in the *polarities* of the voltages applied to the two sets of shorted gate pads. The reference bits are pre-programmed into the comparator by applying voltages of the appropriate polarity to the appropriate electrode pads and they determine the magnetization states of the corresponding nanomagnets that store the reference bits. The reference bits can be changed by applying a new set of voltage polarities, whenever desired, making the device reconfigurable.

The input bits, encoded in voltage polarities, are applied to the other set of electrodes. If the two bit streams match (every input bit matches the corresponding reference bit), then the polarities of voltages applied to the two contacts of every spin valve are the same and the magnetizations of the two elliptical nanomagnetic contacts of every spin valve become mutually *parallel* because they both experience the same sign of stress. In that case, the spin valve resistance will be low. On the other hand, if the reference and input bits are different, then the magnetizations of the two elliptical contacts will be mutually *perpendicular*. The spin valve will then have a higher resistance. The spin valve resistance can be measured with the voltage divider circuit arrangement (with a load resistance) shown in Fig. 33. When the bits match, the spin valve resistance is low and a larger fraction of the power supply voltage $V_i$ will be dropped across the load resistor $R_L$, resulting in the output voltage $V_0$ being high. When the bits do not match, a smaller fraction of $V_i$ will be dropped across $R_L$ and $V_0$ will be low. Thus, by measuring $V_0$, we can determine if the bits match or not. This construct was analyzed in ref. [183] and was found to be sufficiently robust against thermal noise. It is also very energy-efficient. A different magnetic bit comparator, that did not employ straintronics, was proposed in ref. [184].

## 6.2   Analog arithmetic operators

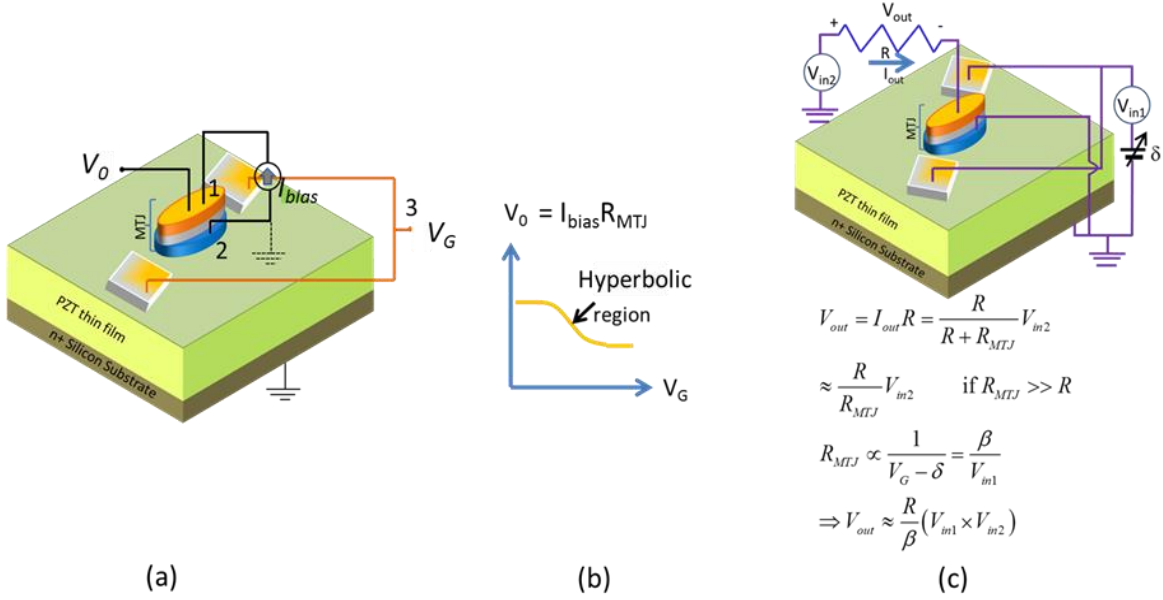

**Fig. 34**: (a) An MTJ configured to produce a hyperbolic region in the transfer characteristic $V_0$ (output voltage) versus $V_G$ (gate voltage). (b) The transfer characteristic showing the hyperbolic region. (c) An analog multiplier implemented with a single MTJ. The two operands are encoded in $V_{in1}$ and $V_{in2}$ and the product of them is encoded in $V_{out}$ or $I_{out}$. The s-MTJ is biased in the hyperbolic region of the transfer characteristic where the s-MTJ resistance is inversely proportional to $(V_G - \delta)$.

Straintronic MTJs can implement arithmetic operators like adders, subtractors, multipliers and dividers, which have applications in analog computing and in probabilistic Bayesian inference engines, as shown in ref. [19]. In Fig. 34(a), we show the schematic of an analog multiplier implemented with a basic straintronic s-MTJ biased by a constant current source $I_{bias}$. Two shorted side gates are biased by a voltage $V_G$ which generates biaxial strain in the piezoelectric thin film underneath the s-MTJ which is transferred to the soft layer of the s-MTJ in contact with the piezoelectric. This rotates the magnetization of the soft layer and changes the resistance of the MTJ, thereby changing the output voltage $V_0$.

In Fig. 34(a), terminal '2' is grounded and hence $V_0 = R_{MTJ} I_{bias}$, where $R_{MTJ}$ is the s-MTJ resistance which depends on the relative orientations of the magnetizations of the soft and hard layers and hence can be altered by the gate voltage $V_G$ generating strain in the soft layer. We have modeled the rotation of the soft layer's magnetization as a function of the gate voltage $V_G$ in the presence of thermal noise using stochastic Landau-Lifshitz-Gilbert simulations. The resulting $V_0$ (or $R_{MTJ}$) versus $V_G$ characteristic is shown qualitatively in Fig. 34(b) where, with proper choice of s-MTJ parameters, we can produce a region in which $V_0 = I_{bias} R_{MTJ} \propto 1/(V_G - \delta)$, i.e. the transfer characteristic $V_0$ versus $V_G$ is roughly *hyperbolic*. When the s-MTJ is biased in that region (by tuning $V_G$), one can perform an analog multiplication of two voltages $V_{in1}$ and $V_{in2}$ with a single s-MTJ as shown in Fig. 34(c) by using a (variable) voltage source $\delta$ such that $V_G = V_{in1} + \delta$. The energy dissipated in a multiplication operation is $C(V_{in1} + \delta)^2 + V_{in2}^2 / (R + R_{MTJ})$ plus any internal dissipation within the soft layer, where $C$ is the capacitance of the gate pads. Simulations show that the energy dissipated to perform this multiplication operation in *optimized* devices at 300 K is ~ 1 aJ [176] while the switching time is less than 1 ns, resulting in a performance figure > *1 Giga-MAC/s/nW*.

## 6.3 Straintronic spin neuron

In artificial neural networks, neurons implemented with CMOS-based operational amplifiers dissipate enormous amounts of energy and consume too much real estate on a chip. However, neurons can also be realized with magneto-tunneling junctions (MTJs) that are switched with a spin-polarized current (representing weighted sum of input currents) which either delivers a spin transfer torque or induces domain wall motion in the soft layer of the MTJ [185-189]. When the spin polarized current passing through the MTJ exceeds a threshold value, the magnetization of the soft layer rotates abruptly, resulting in a sudden change in the MTJ resistance and a concomitant change in the current through or voltage across the MTJ. This implements a "threshold" neuron behavior which fires when the weighted sum of inputs exceeds a threshold value. The MTJ can obviously be also switched with mechanical strain generated in the soft layer with a voltage (representing weighted sum of input voltages) if the soft layer is the magnetostictive component of a magnetostrictive/piezoelectric multiferroic. The latter would be a *straintronic spin-neuron* [190]; it is a voltage driven spin neuron as opposed to a current-driven one in refs. [185-189].

The transfer function of a neuron is usually expressed as

$$O = f\left(\sum_i w_i x_i + b\right),  \quad (3.3)$$

where $f$ is some non-linear function, $w_i$-s are programmable weights of synapses, $x_i$-s are the input signals (representing dendrites) $b$ is a fixed bias and $O$ is the output (representing a neuron's axon). In "threshold neurons", the non-linear function $f$ approximates a unit step (or Heaviside) function whose value is 1 if the argument $\left(\sum_i w_i x_i + b\right)$ exceeds a threshold value and 0 otherwise.

Fig. 35 shows the structure of a *straintronic* spin neuron with programmable synapses implemented with a s-MTJ whose soft layer is magnetostrictive and is in contact with an underlying piezoelectric thin film. The inputs $x_i$-s and the fixed bias $b$ are voltages $V_i$ and $b$; the latter is realized with a constant current source $I \left[ b = I \left( R_1 \| R_2 \| r_1 \| r_2 \| \bullet \bullet \bullet \| r_{N-1} \| r_N \right) \right]$. The voltage appearing at node $P$ is dropped across the piezoelectric layer underneath the (shorted) contact pads A and A'. This voltage is a weighted sum of input voltages and bias, and is given by (using standard superposition principle)

$$V_P = \sum_{i=1}^{N} w_i V_i + b \ , \qquad (3.4)$$

where

$$w_i = \frac{R_1 \| R_2 \| r_1 \| r_2 \| \bullet \bullet \bullet \| r_{i-1} \| r_{i+1} \| \bullet \bullet \bullet \| r_N}{R_1 \| R_2 \| r_1 \| r_2 \| \bullet \bullet \bullet \| r_{i-1} \| r_{i+1} \| \bullet \bullet \bullet \| r_N + r_i} \ . \qquad (3.5)$$

The resistances $R_1$ and $R_2$ are the resistances of the piezoelectric layer underneath the contact pads and $r_i$-s are the series resistances (connected to the input terminals) that implement the programmable weights.

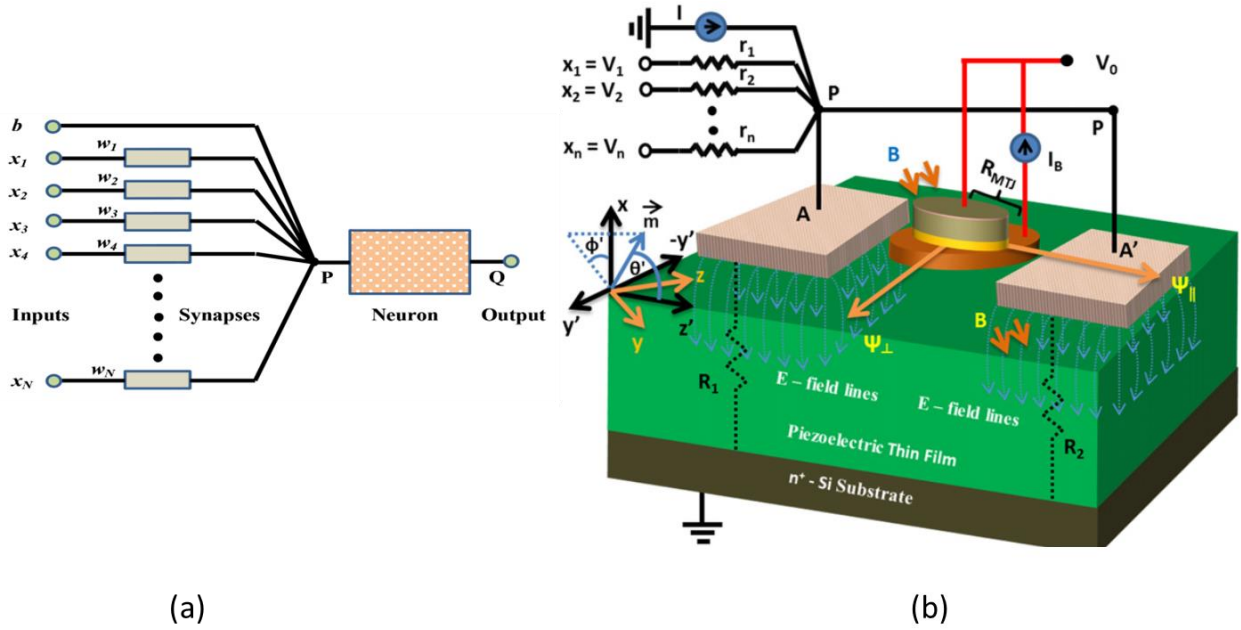

**Fig. 35**: (a) The structure of a threshold neuron with synapses; (b) a "straintronic spin neuron" with synapses realized with a straintronic magneto-tunneling junction, resistors and constant current sources. Reproduced from ref. [190] with permission of the Institute of Physics.

In the s-MTJ, both the hard and the soft layer are shaped like elliptical disks. A fixed magnetic field *B* in the plane of the soft layer directed along its minor axis makes its magnetization bistable, with the two stable directions shown as $\Psi_{\|}$ and $\Psi_{\perp}$ which subtend an angle of ~$90^0$ between them. The hard layer's major axis is made collinear with one of the stable magnetization orientations (say $\Psi_{\|}$) of the soft layer resulting in a "skewed MTJ stack" where the major axes of the two nanomagnets are at an angle. The hard layer is then magnetized permanently in the direction that is anti-parallel to $\Psi_{\|}$. Thus, when the soft layer is in the stable state $\Psi_{\|}$, the magnetizations of the hard and soft layers of the s-MTJ are mutually antiparallel, resulting in high s-MTJ resistance, while when the soft nanomagnet is in the other stable state, the magnetizations of the two layers are roughly perpendicular to each other, resulting in lower s-MTJ resistance.

The electrodes A and A' are placed on the piezoelectric layer such that the line joining their centers is parallel to $\Psi_{\|}$ and hence also to the major axis of the hard layer. The voltage appearing at these two electrodes (which are electrically shorted together) is the voltage at node $P$ given by Equations (3.4) and (3.5). The piezoelectric layer is poled in the vertical direction. Assume that the magnetostriction of the soft layer is positive, which would be the case if it is made of a highly magnetostrictive material like Terfenol-D or Galfenol.

If the voltage $V_P$ (the weighted sum of input voltages and the bias) is low, then there is insufficient stress generated in the soft layer and it will remain magnetized along $\Psi_\parallel$ antiparallel to the magnetization of the hard layer because of dipole coupling between the hard and soft layers. However, if $V_P$ exceeds a threshold value, then sufficient biaxial strain, compressive along the direction of $\Psi_\parallel$ and tensile along the direction of $\Psi_\perp$, will be generated in the soft layer, which will rotate the magnetization by ~90⁰ and place it in an orientation that is perpendicular to the major axis (easy axis) of the hard layer, i.e. collinear with $\Psi_\perp$. This will abruptly reduce the s-MTJ resistance $R_{MTJ}$ (because the hard and soft layers become perpendicular and are no longer antiparallel) and hence the output voltage $V_0$ will drop suddenly since [195]

$$V_0 \approx I_B R_{MTJ} , \qquad (3.6)$$

where $I_B$ is the bias current shown in Fig. 35. Thus, the output voltage has an abrupt dependence on the weighted sum of the input voltages and we can write

$$V_0 = f(V_P) = f\left(\sum_{i=1}^{N} w_i V_i + b\right), \qquad (3.7)$$

which mirrors Equation (3.2). Therefore, it implements a spin neuron.

The current $I_B$ can be made very small and is limited by the requirement that $V_0$ is at least 10 times the thermal fluctuation voltage $\sqrt{kT/C}$ for noise immunity, where $C$ is the capacitance of the s-MTJ. The current source therefore entails little dissipation. A global current source can supply every neuron, reducing the fabrication complexity.

Ref. [190] carried out stochastic Landau-Lifshitz-Gilbert simulations of the neuron firing behavior in the presence of thermal noise and found that the neural behavior is degraded by noise, *but not completely inhibited*. It found that if we allow for 1% broadening of the switching threshold due to noise, then for realistic device parameters, the energy dissipated in the firing action will be ~2.4 fJ, whereas a CMOS-based neuron will dissipate ~0.7 pJ [190]. Curiously, it was also found in [190] that the straintronic spin neuron is orders of magnitude more energy efficient than current driven spin neurons of the type proposed in refs. [185-189].

## 6.4 Ternary content addressable memory implemented with skewed straintronic magneto-tunneling junctions

In ternary content-addressable memory (TCAM), a memory cell is searched based on its content instead of its row and column addresses. The TCAM compares input search data against a table of stored data to return the memory address of entirely or partially matching data. In each TCAM cell search and storage bits have three states: "0," "1," and "X" (don't care). The "don't care" state allows masking, i.e., a match regardless of the storage and/or search data bit. TCAMs are useful for high-speed and parallel data processing and have been applied in network routers, IP filters, virus-detection processors, look-up tables, and many more applications. Key challenges in a large-scale TCAM are to achieve higher cell density, faster search speed, and lower power consumption.

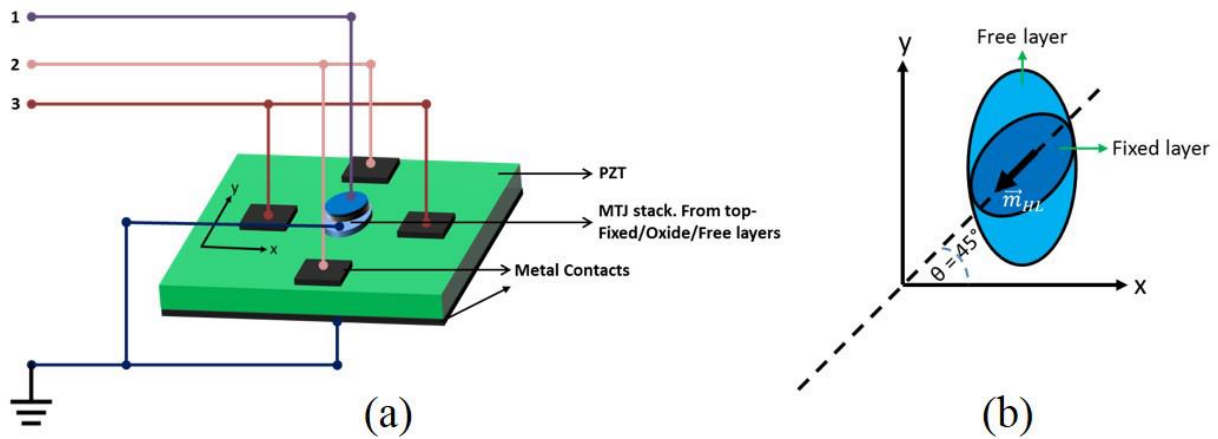

**Fig. 36:** (a) Schematic of a skewed straintronic magneto-tunneling junction (ss-MTJ) with four terminals; (b) top view of the hard (fixed) and soft (free) layers of the ss-MTJ. © 2017 IEEE. Reprinted, with permission, from ref. [20].

TCAM cells can often be implemented better with certain types of magneto-tunneling junctions (MTJs) than CMOS transistors. The use of MTJs reduces device count, energy dissipation and even improves speed. Ref. [20] proposed to implement TCAM cells with *skewed* straintronic MTJs mentioned earlier. The skewed straintronic MTJ (ss-MTJ) is a straintronic MTJ whose hard and soft layers are elliptical but the major axes (easy axes) of these two layers are non-collinear (hence "skewed"). The resistance of the ss-MTJ depends on the angle between the magnetizations of the hard and soft layers. Because of the skewed nature and because of the dipole coupling between the hard and soft layers, this angle $\theta$ is bound by the limits $90^0 \leq \theta \leq 180^0$. The schematic of an ss-MTJ is shown in Fig. 36 where the angle $\theta$ is $135^0$. When the magnetization of the soft layer rotates due to strain or any other influence, the angle θ will go through the value $180^0$ and not $0^0$, because of dipole coupling. That is, in Fig. 36b, the magnetization of the free (soft) layer will rotate clockwise and not counter clockwise. When this rotation occurs, the angle $\theta$ will start out at $135^0$ and go through $180^0$ at which point the resistance of the ss-MTJ assumes its maximum value. As the magnetization rotates further and $\theta$ decreases from $180^0$, the resistance of the ss-MTJ falls. Therefore, it is clear that if we plot the resistance as a function of the voltage applied to stress the piezoelectric (or, equivalently, the angle of rotation of magnetization), then it will exhibit a *non-monotonic* behavior with at first the resistance increasing, reaching a peak, and then decreasing again.

Fig. 36a shows an ss-MTJ with four terminals (including ground). The current flowing through the ss-MTJ ($I_{1G}$) is measured between terminals 1 and ground. Stress is applied to rotate the magnetization of the soft layer by applying a potential ($V_{2G}$) between terminals 2 and ground, while the position of the resistance maximum in the plot of the MTJ resistance or $I_{1G}$ versus $V_{2G}$ can be varied by applying a voltage between terminals 3 and ground.

Fig. 37 shows the angle between the hard and soft layers' magnetizations as a function of the voltage $V_{2G}$ calculated with (stochastic) Landau-Lifshitz-Gilbert simulations at temperatures of 0 K and 300 K (the scatter data points are results for 300 K and the scatter is due to room temperature thermal noise). It also shows the current $I_{1G}$ as a function of $V_{2G}$ for two different values of the voltage $V_{1G}$ applied between terminals 1 and ground. There is a clear notch in the transfer characteristic ($I_{1G}$ versus $V_{2G}$) which occurs when the angle $\theta$ becomes $180^0$. This type of transfer characteristic is unique and is either very difficult or impossible to realize with transistors. This unique characteristic lends itself to convenient implementation of TCAMs.

The "sharpness" of the notch in the transfer characteristic depends on the strength of dipole coupling between the hard and soft layers. The influence of dipole coupling in the soft layer can be represented by an effective magnetic field. In Fig. 38, we show the transfer characteristic as a function of dipole coupling. Weaker dipole coupling makes the notch sharper.

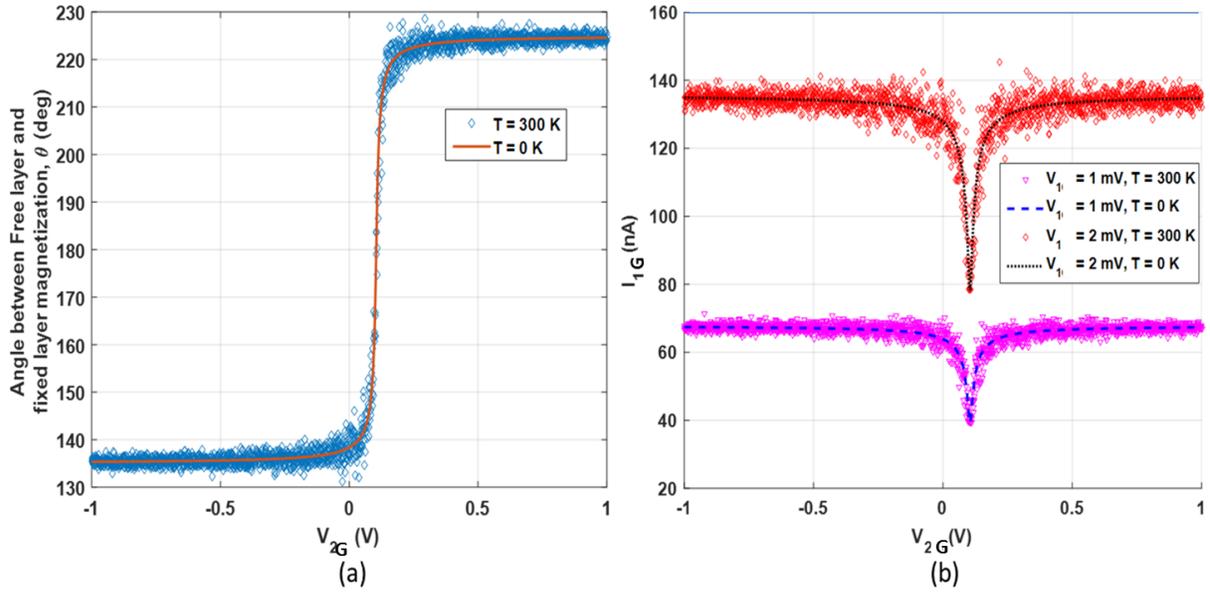

**Fig. 37:** (a) The angle between the magnetizations of the hard and soft layers of an ss-MTJ as a function of $V_{2G}$; (b) the current $I_{1G}$ flowing through the ss-MTJ as a function of $V_{2G}$ when a voltage $V_1$ is applied across the MTJ. © 2017 IEEE. Reprinted, with permission, from ref. [20].

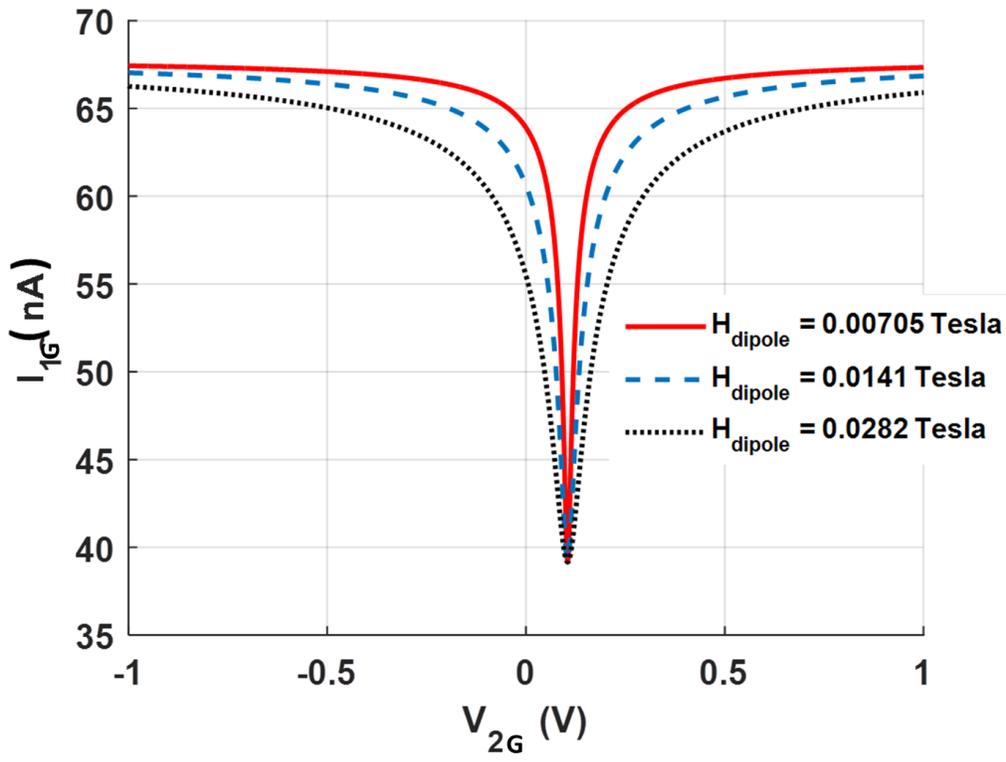

**Fig. 38:** Transfer characteristic of an ss-MTJ as a function of the dipole coupling strength between the hard and soft layers. © 2017 IEEE. Reprinted, with permission, from ref. [20].

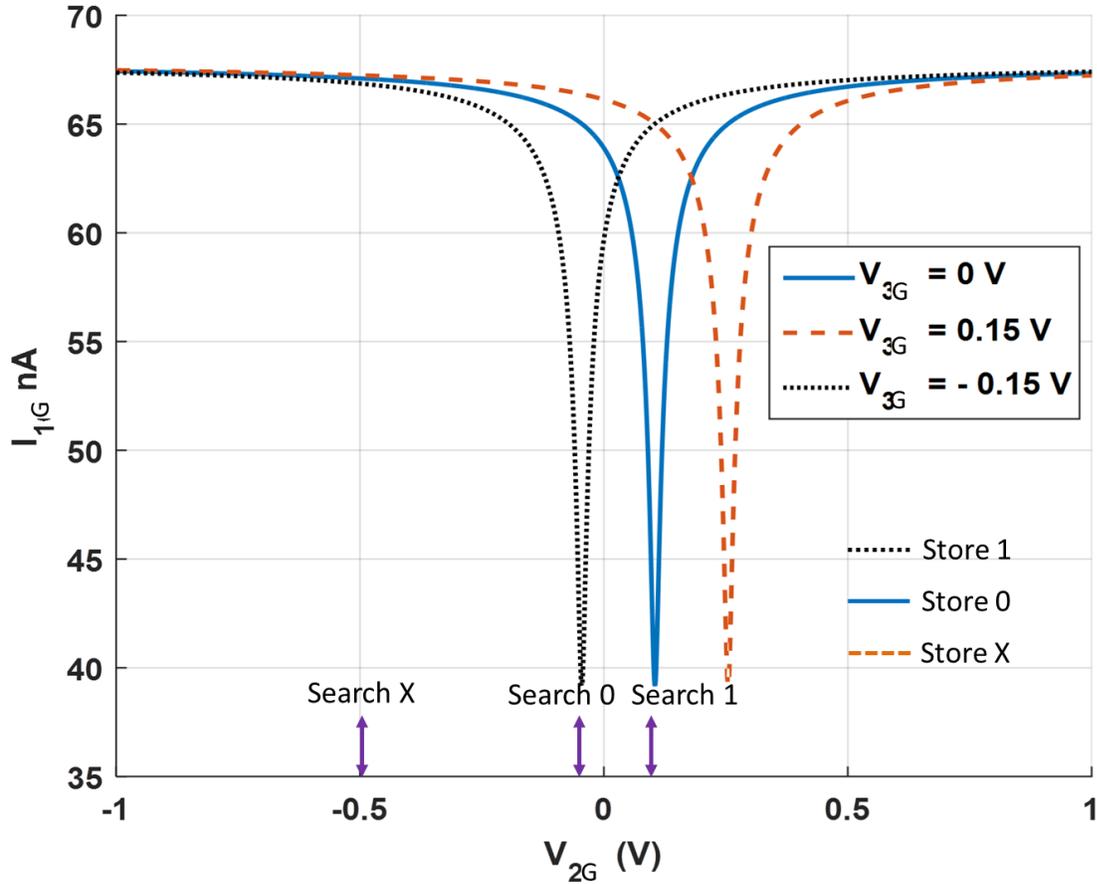

**Fig. 39**: The current $I_{1G}$ through an ss-MTJ as a function of the voltage $V_{2G}$ for three different values of $V_{3G}$.

In an ss-MTJ, the current $I_{1G}$ flowing through the stack can be controlled by the potentials $V_{2G}$ and $V_{3G}$ as shown in Fig. 39. The current $I_{1G}$ is lowest when $V_{3G} = V_{2G} + V_F$, where $V_F$ is a fixed voltage (offset voltage). The current $I_{1G}$ increases steeply when $V_{2G}$ and $V_{3G}$ deviate from this "match" condition.

In the TCAM operation, the search bits are encoded in the potential $V_{2G}$ and the stored bits in the potential $V_{3G}$. Let us say that the search bits X, 0 and 1 are encoded in voltages -0.5, -0.05 and +0.1 V, respectively. The store bits 1, 0 and X, encoded in $V_{3G}$, are such that they place the centers of the notches in the transfer characteristics at -0.05 V, +0.1V and +0.25 V, respectively as shown in Fig. 39. In the encoding scheme, a high current $I_{1G}$ denotes a match between the stored and search bits. When the stored bit is 1 and the search bit is 0, we are located in a notch (the far left notch in Fig. 39) so that the current through the ss-MTJ is low and we have the correct "no-match" result. Similarly, when the search bit is 1 and the stored bit is 0, we are in the center notch and the current is again low indicating no match. When the search and stored bits are the same, we are clearly not in a notch (see Fig. 39), so the current through the ss-MTJ is high, and the match is correctly indicated. Let us now examine what happens with the "don't care" bit. Since the notch for the stored bit X is farthest to the right and exceeds the voltages encoding all search bits, the current $I_{1G}$ remains high for all search bits 0, 1 and X, indicating a match no matter what the search bit is, as long as the stored bit is X. Finally, the search bit X is encoded in a voltage to the left of all notches. Hence, when the search bit is X, no matter what the stored bit is, we are never in a notch and the current is always high, indicating a match. Thus, the correct TCAM operation is realized. The ss-MTJ significantly reduces the complexity of the match operation in a TCAM.

If we had tried to implement a static TCAM cell with CMOS transistors, we will need 16 transistors [191]. However, we can implement the same cell with just a single ss-MTJ, as shown here. This obviously reduces cell footprint and energy dissipation. Ref. [20] has examined a large number of TCAM based circuitry realized with ss-MTJs and found significant reduction in energy as well as increase in speed compared to equivalent circuits realized with CMOS, not to mention the reduction in cell footprint.

## 6.5    Dipole-coupled nanomagnets for logic (DC-NML)

Dipole coupling between two nanomagnets can elicit a variety of functionalities. For example, if two elliptical nanomagnets (each with two stable orientations of magnetization along the major axis) are placed such that the line joining their centers lies along the major axis, then the magnetizations tend to be parallel (ferromagnetic ordering). On the other hand, if this line lies along the minor axis, then the magnetizations tend to be antiparallel (anti-ferromagnetic ordering). This is illustrated in Fig. 40.

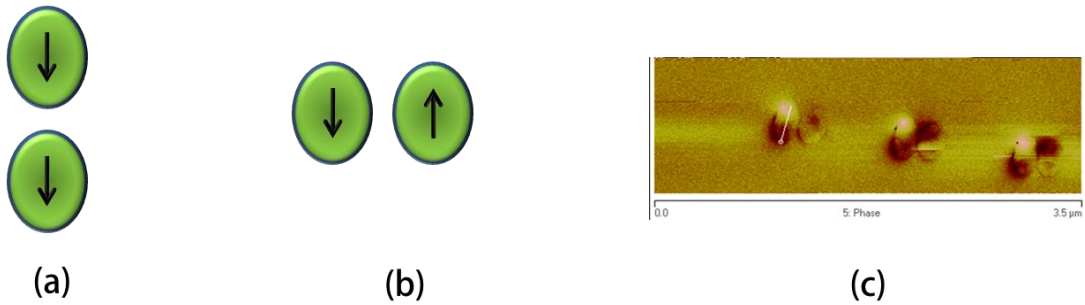

**Fig. 40**: Two elliptical nanomagnets, placed such that the line joining their centers lies along the major axis, assume parallel magnetizations in the ground state; (b) if the line joining the centers lies along the minor axis, then the magnetizations become antiparallel in the ground state; (c) magnetic force micrograph (MFM) images of closely spaced dipole coupled pairs showing the antiparallel configuration.

This behavior can be exploited to build Boolean logic gates. An example of a NAND gate that exploits the anti-ferromagnetic ordering of nearest neighbors due to dipole coupling is shown in Fig. 41. Here, the dipole coupling between neighbors and a weak magnetic field directed along the major axis ensures that the magnetization of the central nanomagnet (encoding the output bit) is the NAND function of the two input bits encoded in the magnetization orientations of the two peripheral nanomagnets.

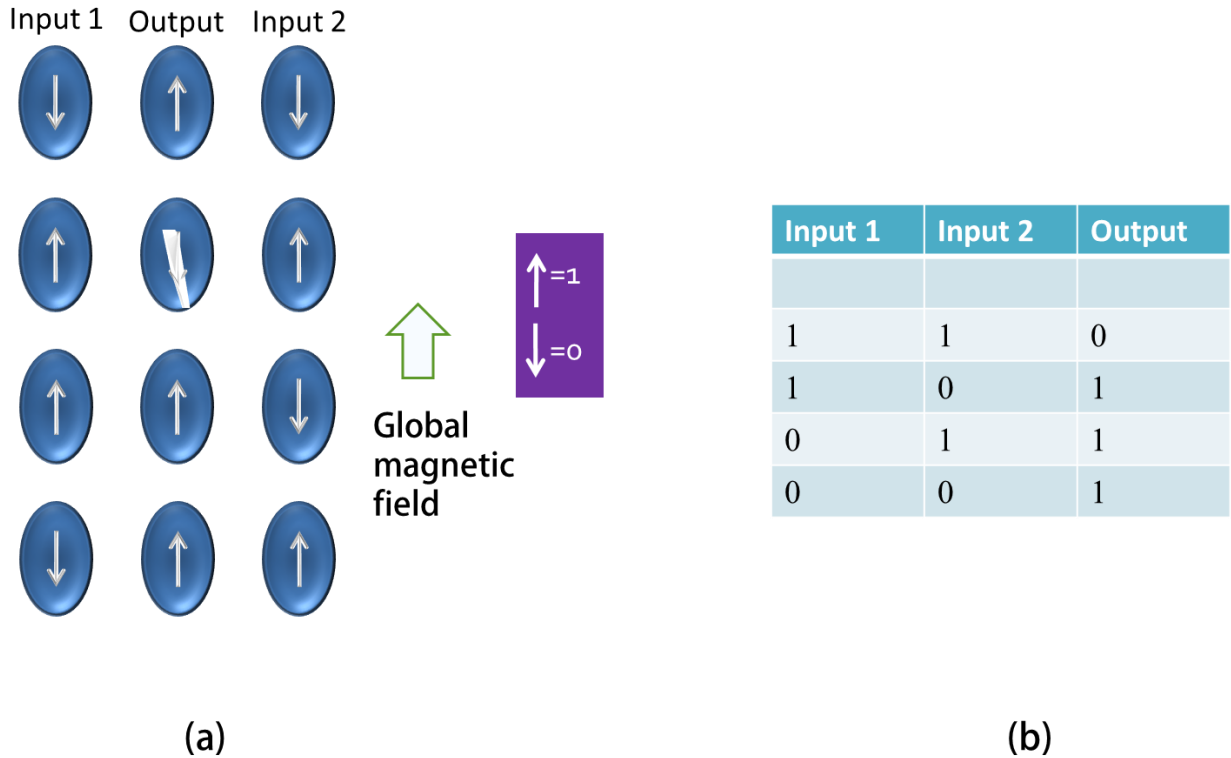

(a) (b)

**Fig. 41**: (a) A NAND gate implemented with three dipole-coupled nanomagnets placed in a weak magnetic field directed along the major axes of the nanomagnets. Magnetization pointing in the direction of the magnetic field encodes the bit 1 and magnetization in the opposite direction encodes the bit 0. (b) The truth table relating the output bit to the input bits, showing that the NAND functionality has been realized.

Logic gate ideas like this are inspired by the *Single Spin Logic* paradigm proposed more than two decades ago [192, 193] where single electron spins were utilized instead of nanomagnets to realize a NAND gate, and exchange coupling between spins, instead of dipole coupling, ensured that neighboring spins prefer to be antiparallel. The difference is that while Single Spin Logic required cryogenic operation, the nanomagnetic version described here is capable of operating at room temperature.

Numerous ideas of DC-NML can be found in the literature and have been known by various names such as "magnetic quantum dot cellular automata" [194]. These ideas aroused some interest, but unfortunately, dipole coupling is not resilient against thermal noise at room temperature. As a result, these paradigms are extremely error-prone [170, 172, 195-200] and hence unsuitable for Boolean logic which is very unforgiving of errors. As mentioned earlier, logic requires the error probability associated with the switching of a logic element to be no more than perhaps $10^{-15}$, while DC-NML gates have error probabilities not less than ~$10^{-9}$ at room temperature (considering only thermal noise and no defects). In the presence of defects that pin magnetization, the switching error probability could become several orders of magnitude larger than $10^{-9}$. Thus, dipole coupled nanomagnetic logic (DC-NML) is not likely to be viable in the short term. On the other hand, dipole coupled *non-Boolean* information processing may have a much better future and an example of that is provided in the next section.

## 6.6 Image processing with dipole coupled strain switched nanomagnets

The idea of collective computation is an old one. Here, the activity of any single device is not vital since the cooperative activity of many devices, working in unison, elicits the computational activity. Consequently, the correct result of the computation emerges even if a substantial fraction of the devices fails.

A well-known example of collective computation is the Ising computer which solves NP-hard optimization problems by representing the solution as the ground state of an Ising Hamiltonian [201]. Hardware for such computers has been implemented with CMOS [202], trapped ions [203] and electromechanical systems [204]. Combinatorial optimization problems have also been solved via simulated annealing [205] and quantum annealing [206] in collective computing systems. Other approaches for solving optimization and/or NP-hard problems have involved cellular neural networks [207], lasers [208], quantum dots [209, 210] and nanomagnets [211-214].

The paradigm described in ref. [214] works as follows: A two-dimensional array of dipole coupled strain-switched nanomagnets can perform a variety of image processing tasks. Each nanomagnet has two stable magnetization states shown in Fig. 42, which encode pixel colors black or white. These two stable states are created by applying a bias magnetic field along the minor axis of the elliptical nanomagnet acting as the soft layer of a skewed straintronic magneto-tunneling junction (ss-MTJ). The hard layer is permanently magnetized in the direction of one of the stable states. An input image containing only black and white pixels is first converted to voltage states (white = positive polarity voltage; black = negative polarity voltage) with photodetectors and these voltages are applied across the ss-MTJ. When the voltage polarity is negative (black pixel), electrons are injected from the hard into the soft layer which drives the soft layer into state 1 shown in Fig. 42 and writes the pixel color "black" into the nanomagnet. When the voltage polarity is positive (white pixel) electrons with spins aligned along the direction of state 1 are extracted from the soft layer and the latter's magnetization switches to the other stable state 2. This is how pixel colors are "written" or mapped into the magnetization states of the soft layers. A pixel color is "read" by measuring the resistance of the ss-MTJ.

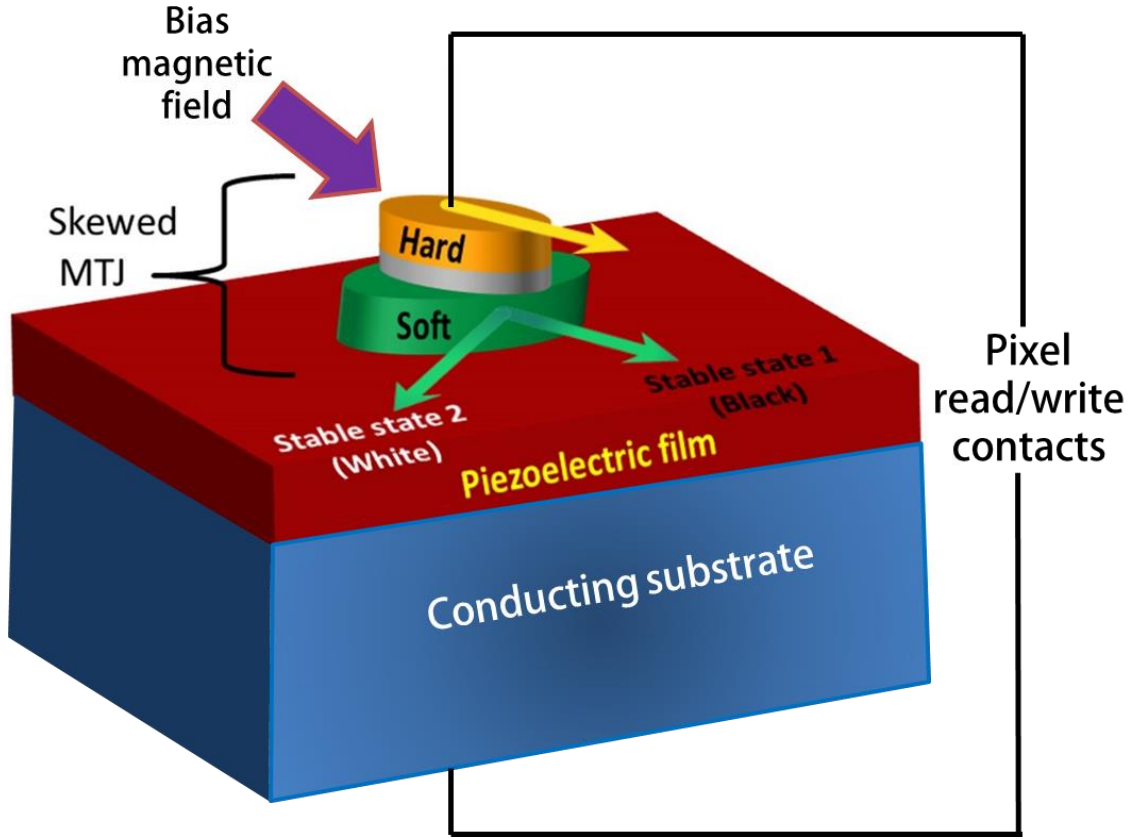

**Fig. 42**: A skewed MTJ whose soft layer has two stable magnetization orientations corresponding to black and white pixels. The two stable orientations subtend an angle of ~$90^0$ between themselves. A black pixel is written by applying a negative voltage between the hard and soft layers while a white pixel is written by applying a voltage of opposite polarity. The pixel color is read by measuring the MTJ resistance. A low resistance corresponds to a black pixel and a high resistance to a white pixel. © 2017 IEEE. Adapted, with permission, from ref. [214].

We consider a two-dimensional array dipole-interacting nanomagnets acting as the soft layers of ss-MTJs. The potential energy of the *i*-th nanomagnet is given by

$$U_i = E^i_{shape} + E^i_{mag} + E^i_{ex} + \sum_{j \neq i} E^{i-j}_{dipole} \qquad (5.1)$$

where $E_{shape}$ is the shape anisotropy energy (due to the elliptical shape), $E_{mag}$ is the magnetostatic energy due to the applied bias magnetic field, $E_{ex}$ is the exchange energy due to exchange interaction between spins and $E_{\text{dipole}}^{i-j}$ is the interaction energy between two different nanomagnets.

The ground state of the array corresponds to $\min \sum_i U_i$. When the input pixels arrive, each nanomagnet's magnetization will be aligned along either the "black" or the "white" orientation depending on the input pixel colors. These alignments will raise the system to an excited state where $\sum_i U_i$ is not at its minimum value. The system will then relax to the ground state by emitting magnons, phonons, etc. in some finite time (provided there are no energy barriers between the excited and ground states that prevent relaxation, or if such energy barriers are temporarily eroded with clock signals) and that will reorient the magnetization states of some or all of the nanomagnets. The reoriented magnetization orientations correspond to the pixel colors of the processed image.

The image processing function requires a nanomagnet to transition to the magnetization orientation corresponding to the system ground state. However, there may be energy barriers in the nanomagnets that will prevent the magnetization from going into the global system ground state, leaving the system stuck in a metastable state. Therefore, an external agent will be needed to erode the energy barrier(s) temporarily and allow the system to migrate to the lowest energy state, thus completing the image processing function. This external agent is strain, which lowers the energy barrier between the stable orientations and allows the migration to occur. The strain is, of course, generated by applying a voltage across the piezoelectric underneath the soft layer with electrodes placed in such a way that biaxial strain is generated in the piezoelectric (compressive along the major axis of the ellipse and tensile along the minor axis, or vice versa, depending on the voltage polarity). This strain will allow the system to transition to the ground state and complete the image processing function. The advantage of this approach over any transistor-centric approach is the lower energy dissipation and the non-volatility.

Ref. [214] presented many examples of image processing with dipole coupled nanomagnets. Two such examples are shown in Figs. 43 and 44.

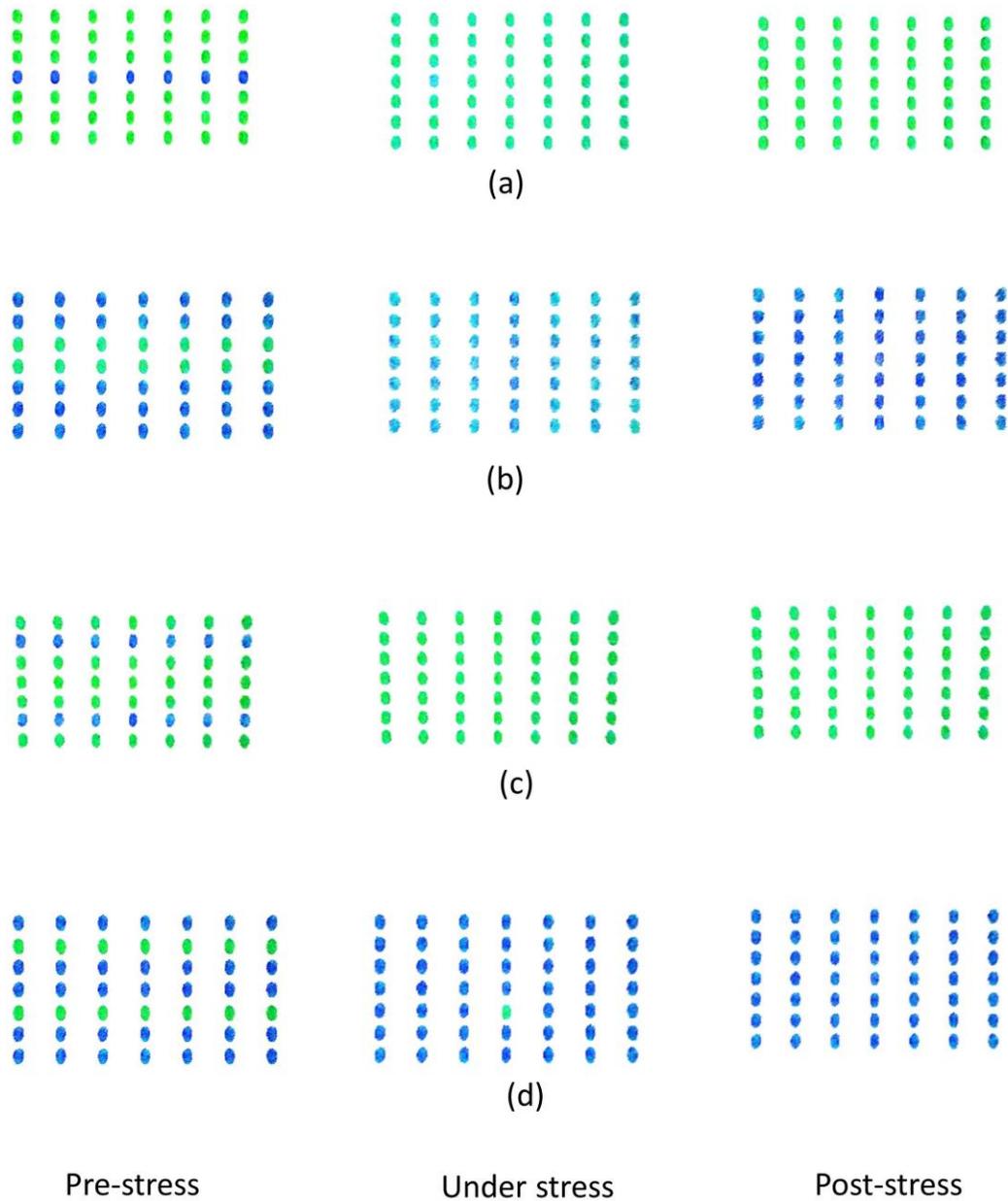

**Fig. 43**: (a) The white segment of an image (7 × 7 pixels) where the entire fourth row has been corrupted and turned black. The nanomagnets are made of Terfenol-D. Green=white; blue=black. Application and subsequent withdrawal of global stress (compressive along the major axes) corrects the corrupted row. (b) Array of black pixels in the black segment of an image where two consecutive rows have been corrupted and turned white. Stress application and withdrawal corrects both corrupted rows. (c) Array of white pixels in the white segment of an image where two nonconsecutive rows have been corrupted and turned black. Stressing corrects both rows. (d) Array of black pixels where two nonconsecutive rows have been corrupted and turned white. Again, stressing corrects both rows. © 2017 IEEE. Reprinted, with permission, from ref. [214].

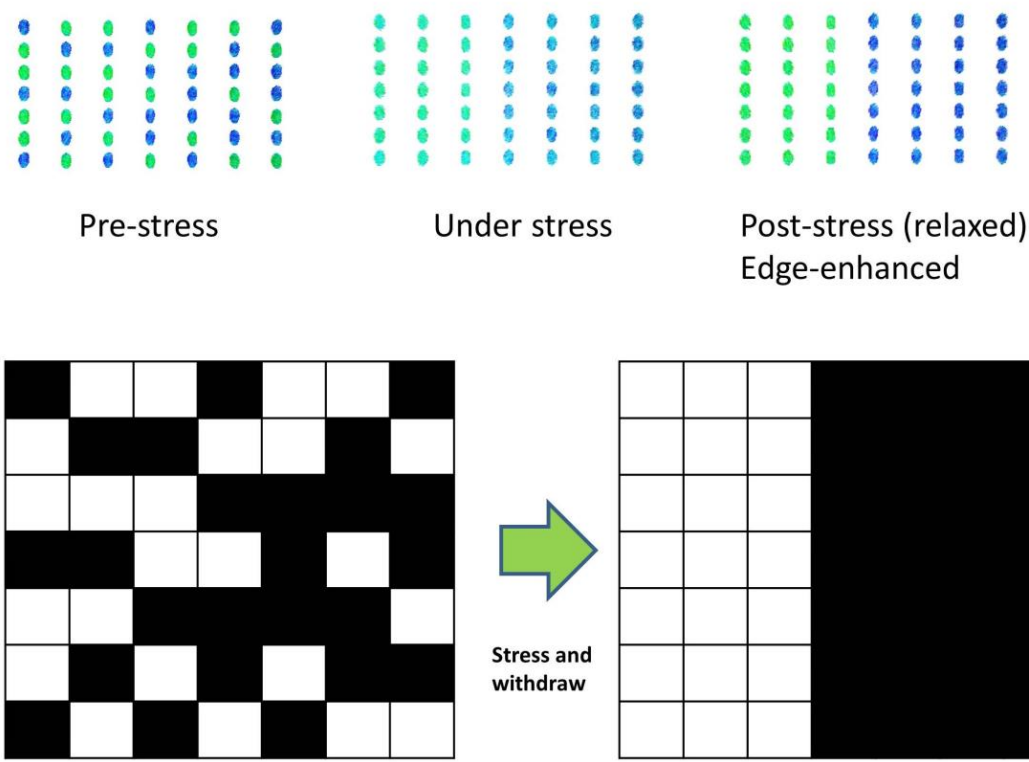

Fig. 44: Edge enhancement detection. The left segment (left three columns) has majority white pixels and the right segment (four right columns) has majority black segments. After application and removal of global stress applied along the major axes of the elliptical nanomagnets, the white-dominant segment turns all-white and the black-dominant segment turns all-black. This enhances the edge contrast between the two segments. © 2017 IEEE. Reprinted, with permission, from ref. [214].

## 7. Conclusions and outlook

In this article, we have provided a broad overview of computing with nanomagnetic switches. These switches have two characteristics that make them appealing: the potential for excellent energy-efficiency (at least comparable to state-of-the-art transistors) if switched with appropriate voltage controlled mechanisms such as straintronics or switched with spin current using appropriately designed spin Hall effect and spin orbit torque based switching paradigms, and the non-volatility that can be exploited for unconventional computing architectures for which transistors (volatile switches) are inefficient. Nanomagnetic switches, such as straintronic magneto-tunneling junctions (s-MTJs), possess these advantages, but like all MTJs, have low resistance off/on ratios and the switching can be error-prone. Therefore, these devices are not suitable for Boolean logic, which demands large resistance off/on ratios and resilience against switching errors. They are however excellent for applications in some types of unconventional computing paradigms such as neuromorphic computing, bit comparison, image processing, ternary content addressable memory, Bayesian inference engines and perhaps Boltzmann machines.

Experimental progress in straintronic switching has been stymied by the lack of magnetostrictive materials with sufficiently large magnetostriction. The strain anisotropy energy density that can be generated with reasonable stress in most common magnetostrictive nanomagnets such as Co or Ni is usually less than the shape anisotropy energy density required for good thermal stability in nanomagnets of lateral dimension less than ~100 nm. That makes switching the magnetization of these nanomagnets with stress difficult and error-prone. Materials such as Terfenol-D or Galfenol, which have much higher magnetostriction than Co or Ni, may be able to overcome this impasse, but these materials are binary or ternary alloys and have multiple phases which introduce additional complexity. These alloys are also likely to have more magnetization pinning sites due to material defects. Thus, there are significant materials challenges in the field of straintronics which will have to be mitigated before the field can mature. Moreover, switching the magnetization repeatedly over many cycles with stress is challenging because of piezoelectric fatigue and also because of the low effective field generated at reasonable stress levels (< 100 MPa) in low-magnetostriction materials like Co or Ni. The allure of straintronics is in low-energy unconventional computing with nanomagnets and that is likely to endure, motivating further progress in the field.

## ACKNOWLEDGEMENT

The straintronics research in the groups of the corresponding authors has been funded by the US National Science Foundation under grants ECCS-1124714, CCF-1253370 (J. A. CAREER grant), CCF-1216614, ECCS-1609303 and CCF-1815033, the Semiconductor Research Corporation under Task 2203.001 and the State of Virginia under grant MF15-006-MS. We acknowledge numerous discussions with our collaborators – Prof. Avik W. Ghosh (University of Virginia), Prof. Amit Ranjan Trivedi (University of Illinois at Chicago), Prof. Csaba Andras Moritz (University of Massachusetts at Amherst), Prof. Jianping Wang (University of Minnesota) and Prof. Greg Carman (University of California at Los Angeles). Other students who had contributed to some of the research described here are Dr. Kuntal Roy and Ms. Justine Drobitch.

# REFERENCES

1. Coombs A. W. M. 1983 The making of Colossus *IEEE Annals of the History of Computing* 5 260.

2. Eckert Jr. J. P. and Mauchly J. W 1964 Electronic numerical integrator and computer *US Patent* 3,120,606.

3. Bardeen J. and Brattain W. 1948 The transistor, a semi-conductor triode *Phys. Rev.* **74** 230.

4. Moore G. E. 1965 Cramming more components onto integrated circuits *Electronics Magazine* page 4.

5. Datta S. Diep V. Q. and Behin-Aein B. 2015 What constitutes a nanoswitch? A perspective *Emerging Nanoelectronic Devices* Eds. Chen A., Hutchby J., Bourianoff G. and Zhirnov V. (John Wiley & Sons, West Sussex, UK) Ch. 2, p. 15.

6. https://ark.intel.com/products/88195/Intel-Core-i7-6700K-Processor-8M-Cache-up-to-4_20-GHz.

7. Tuckerman D. B. and Pease R. F. W. 1981 High performance heat sinking for VLSI *IEEE Elec. Dev. Lett.* **EDL-2** 126.

8. Anton S. R. and Sodano H. A. 2007 A review of power harvesting using piezoelectric materials (2003-2006) *Smart Mater. Struc.,* **16** R1.

9. Sackellares J. C. 2010 Seizure Prediction and Monitoring *Epilepsy Behavior* **18** 106.

10. Zaurin R. and Catbas F. N. 2010 Integration of computer imaging and sensor data for structural health monitoring of bridges *Smart Mater. Struct.* **19** 015019.

11. Bandyopadhyay S. 2007 Power dissipation in spintronic devices: A general perspective *J. Nanosci. Nanotechnol.*, **7** 168.

12. Nikonov D. E. and Young I. A. 2013 Overview of beyond-CMOS devices and a uniform methodology for their benchmarking *Proc. IEEE* **101** 2498.

13. Cowburn R. P., Koltsov D. K., Adeyeye A. O., Welland M. E. and Tricker D. M. 1999 Single domain circular nanomagnets *Phys. Rev. Lett*. **83** 1042.

14. Salahuddin S. and Datta S. 2007 Interacting systems for self-correcting low power switching *Appl. Phys. Lett.* **90**, 093503.

15. Brown W. F. 1963 Thermal fluctuations of a single-domain particle *Phys. Rev.* **130** 1677.
95

16. Gaunt P. 1977 The frequency constant for thermal activation of a ferromagnetic domain wall *J. Appl. Phys.* **48** 3470.

17. Bhati I., Chan, M-T., Chisti Z., Lu S-L. and Jacob B. 2016 DRAM refresh mechanisms, penalties and trade-offs *IEEE Trans. Computers*, **65** 108.

18. Khasanvis S., Li M. Y., Rahman M., Salehi-Fashami M., Biswas A. K., Atulasimha J., Bandyopadhyay S. and Moritz C. A. 2015 Self-similar magneto-electric nanocircuit technology for probabilistic inference engines *IEEE Trans. Nanotechnol*. **14** 980.

19. Khasanvis S., Li M. Y., Rahman M., Biswas A. K., Salehi-Fashami M., Atulasimha J., Bandyopadhyay S. and Moritz C. A. 2015 Architecting for causal intelligence at nanoscale *COMPUTER* **48** 54.

20. Dey Manasi S., Al-Rashid M. M., Atulasimha J., Bandyopadhyay S. and Trivedi A. R. 2017 Straintronic magneto-tunneling junction based ternary content-addressable memory (Parts I and II), *IEEE Trans. Elec. Dev*. **64** 2835.

21. Draaisma H. J. G. and de Jonge W. J. M. 1988 Surface and volume anisotropy from dipole-dipole interactions in ultrathin ferromagnetic films *J. Appl. Phys*. **64** 3610.

22. Johnson M. T., Bloemen P. J. H. den Broeder F. J. A. and de Vries J. J. 1996 Magnetic anisotropy in metallic multilayers *Rep. Prog. Phys*. **59** 1409.

23. Verma S., Kulkarni A. A. and Kaushik B. K. 2016 Spintronics based devices to circuits: Perspectives and challenges *IEEE Nanotechnology Magazine* **10** 13.

24. Salehi-Fashami, M., Atulasimha, J. and Bandyopadhyay, S. 2012 Magnetization dynamics, throughput and energy dissipation in a universal multiferroic nanomagnetic logic gate with fan-in and fan-out *Nanotechnology,* **23,** 105201.

25. Chikazumi, S. *Physics of Magnetism* (Wiley, New York, 1964).

26. Kaushik, B. K., Verma S., Kulkarni A. A. and Prajapati S. *Next Generation Spin Torque Memories* (Springer, New York, 2017).

27. Cui J., Hockel J. L., Nordeen P. K., Pisani D. M., Liang C. Y., Carman G. P. and Lynch C. S. 2013 A method to control magnetism in individual strain-mediated magnetoelectric islands *Appl. Phys. Lett*. **103** 232905.

28. Liang C. Y., Keller S. M., Sepulveda A. E., Sun W. Y., Cui J. Z., Lynch C. S. and Carman G. P. 2014 Electrical control of a single magnetoelastic domain structure on a clamped piezoelectric thin film – analysis *J. Appl. Phys.* **116** 123909.

29. Cui J. Z., Liang C. Y., Paisley E. A., Sepulveda A., Ihlefeld J. F., Carman G. P. and Lynch C. S. 2015 Generation of localized strain in a thin film piezoelectric to control individual magnetoelectric heterostructures *Appl. Phys. Lett*. **107** 092903.

30. Roy K., Bandyopadhyay S. and Atulasimha J. 2013 Binary switching in a 'symmetric' potential landscape *Sci. Rep.* **3** 3038.

31. Atulasimha, J. and Bandyopadhyay, S. 2010 Bennett clocking of nanomagnetic logic using multiferroic single-domain nanomagnets *Appl. Phys. Lett.,* **97**, 173105.

32. Roy K., Bandyopadhyay S. and Atulasimha J. 2012 Energy dissipation and switching delay in stress-induced switching of multiferroic devices in the presence of thermal fluctuations *J. Appl. Phys.* **112** 023914.

33. Ahmad H., Atulasimha J. and Bandyopadhyay S. 2015 Reversible strain induced magnetization switching in FeGa nanomagnets: Pathway to rewritable, non-volatile, non-toggle, extremely low energy straintronic memory *Sci. Rep*. **5** 18264.

34. D'Souza N., Fashami M. S., Bandyopadhyay S. and Atulasimha J. 2016 Experimental clocking of nanomagnets with strain for ultralow power Boolean logic *Nano Lett*. **16** 1069.

35. Zhao Z. Y., Jamali M., D'Souza N., Zhang D., Bandyopadhyay S., Atulasimha J. and Wang J. P. 2016 Giant voltage manipulation of MgO-based magnetic tunnel junctions via localized anisotropic strain: A potential pathway to ultra-energy-efficient memory technology *Appl. Phys. Lett.* **109** 092403.

36. Sampath V., D'Souza N., Atkinson G. M., Bandyopadhyay S. and Atulasimha J. 2016 Experimental demonstration of acoustic wave induced magnetization switching in dipole coupled magnetostrictive nanomagnets for ultralow power computing *Appl. Phys. Lett.* **109** 102403.

37. Sampath V., D'Souza N., Bhattacharya D., Atkinson G. M., Bandyopadhyay S. and Atulasimha J. 2016 Acoustic-wave-induced magnetization switching of magnetostrictive nanomagnets from single-domain to non-volatile vortex states *Nano Lett.* **16** 5681.

38. Slonczewski J. C. 1996 Current-driven excitation of magnetic multilayers. *J. Magn. Magn. Mater.* **159**, L1.

39. Berger L. 1996 Emission of spin waves by a magnetic multilayer traversed by a current *Phys. Rev. B* **54** 9353.

40. Sun J. Z. 2000 Spin-current interaction with a monodomain magnetic body: A model study *Phys. Rev. B* **62** 570.

41. Ralph D. C. and Stiles M. D. 2008 Spin transfer torques *J. Magn. Magn. Mater.* **320** 1190.

42. Sankey J. C., Cui Y-T, Sun J. Z., Slonczewski J. C., Buhrman R. A. and Ralph, D. C. 2008 Measurement of the spin transfer torque vector in magnetic tunnel junctions *Nat. Phys.* **4** 67.

43. Kubota H., Fukushima A., Nagaham T., Shinji Y., Ando K., Maehara H., Nagamine Y., Tsunekawa K., Djayaprawira D. D., Watanabe N. and Suzuki Y. 2008 Quantitative measurement of voltage dependence of spin-transfer torque in MgO-based magnetic tunnel junctions *Nat. Phys.* **4** 37.

44. Camsari K. Y., Gangulay S. and Datta S. 2015 Modular approach to spintronics *Sci. Rep.* **5** 10571.

45. Wang K. L. and Khalili-Amiri P. 2012 Nonvolatile spintronics: Perspectives on instant-on non-volatile nanoelectronic systems *Spin* **2** 1250009.

46. Wang K. L., Alzate J. G. and Khalili-Amiri P. 2013 Low-power non-volatile spintronic memory: STT-RAM and beyond *J. Phys. D: Appl. Phys.* **46** 074003.

47. Li Z. and Zhang S. 2004 Thermally assisted magnetization reversal in the presence of a spin-transfer torque *Phys. Rev. B* **69** 134416.

48. Prejbeanu I. L., Kula W., Ounadjela K., Sousa R. C., Redon O., Dieny B. and Nozieres J. P. 2004 Thermally assisted switching in exchange-biased storage layer magnetic tunnel junctions *IEEE Trans. Magn.* **40** 2625.

49. Prejbeanu I. L., Kerekes M., Sousa R. C., Sibuet H., Redon O., Dieny B. and Nozieres J. P. 2007 Thermally assisted MRAM *J Phys: Condens. Matter.* **19** 165218.

50. Bedau D., Liu H., Sun J. Z., Katine J. A., Fullerton E. E., Mangin S. and Kent A. D. 2010 Spin-transfer pulse switching: From the dynamic to the thermally activated regime *Appl. Phys. Lett.* **97**, 262502.

51. Liu L., Pai C-F., Li Y., Tseng H. W., Ralph D. C. and Buhrman R. A. 2012 Spin-torque switching with the giant spin Hall effect of tantalum *Science* **336** 555.

52. Pai C-F. Liu L., Li Y., Tseng H. W., Ralph D. C. and Buhrman R. A. 2012 Spin transfer torque devices utilizing the giant spin Hall effect of tungsten *Appl. Phys. Lett*. **101** 122404.

53. Niimi Y., Kawanishi Y., Wei D. H., Deranlot C., Yang H. X., Chshiev M., Valet T., Fert A. and Otani Y. 2012 Giant spin Hall effect induced by skew scattering from bismuth impurities inside the thin film CuBi alloys *Phys. Rev. Lett*. **109** 156602.

54. Edelstein V. M. 1990 Spin polarization of conduction electrons induced by electric current in two dimensional asymmetric electron systems *Solid State Commun*. **73** 233.

55. Chernyshov A., Overby M., Liu X., Furdyna J. K., Lyanda-Geller Y. and Rokhinson L. P. 2009 Evidence for reversible control of magnetization in a ferromagnetic material by means of spin-orbit magnetic field *Nat. Phys.* **5** 656.

56. Bychkov Y. A. and Rashba E. I. 1984 Oscillatory effects and the magnetic susceptibility of carriers in inversion layers *J. Phys. C: Solid State Physics* **17** 6039.

57. Bandyopadhyay S. and Cahay M. *Introduction to Spintronics*, 2nd ed. (CRC Press, Boca Raton, 2015).

58. Liu, L., Moriyama, T., Ralph, D. C. and Buhrman, R. A. 2011 Spin-torque ferromagnetic resonance induced by the spin Hall effect *Phys. Rev. Lett*. **106** 036601.

59. Bhowmik D., You L. and Salahuddin S. 2014 Spin Hall effect clocking of nanomagnetic logic without a magnetic field *Nat. Nanotech*. **9** 59.

60. Sayed, S., Diep, V-Q., Camsari, K. Y. and Datta, S. 2016 Spin funneling for enhanced spin injection into ferromagnets *Sci. Rep*. **6** 28868.

61. Hasan M. Z. and Kane C. L. 2010 *Colloquium*: Topological insulators *Rev. Mod. Phys*. **82** 3045.

62. Moore J. E. 2010 The birth of topological insulators *Nature* **464** 194.

63. Mellnik A. R., Lee J. S., Richardella A., Grab J. L., Mintun P. J., Fischer M. H., Vaezi A., Manchon A., Kim E-A, Samarth N. and Ralph D. C. 2014 Spin transfer torque generated by a topological insulator *Nature* **511** 449.

64. King P. D. C. et al. 2011 Large tunable Rashba spin splitting of a two dimensional electron gas in $Bi_2Se_3$ *Phys. Rev. Lett*. **107** 096802.

65. Manchon A. and Zhang S. 2008 Theory of nonequilibrium intrinsic spin torque in a single nanomagnet *Phys. Rev. B* **78** 212405.

66. Manchon A. and Zhang S. 2009 Theory of spin torque due to spin-orbit coupling *Phys. Rev. B* **79** 094422.

67. Pesin D. A. and MacDonal A. H. 2012 Quantum kinetic theory of current-induced torques in Rashba ferromagnets *Phys. Rev. B* **86** 014416.

68. Semenov Y. G., Duan X. and Kim K. W. 2012 Electrically controlled magnetization in ferromagnet-topological insulatoir heterostructures *Phys. Rev. B*. **86** 161406(R).

69. Yamanouchi M., Chiba D., Matsukura F. and Ohno H. 2004 Current-induced domain-wall switching in a ferromagnetic semiconductor structure *Nature* **428** 539.

70. Fukami S., et al. 2009 Low-current perpendicular domain wall motion cell for scalable high-speed MRAM *2009 VLSI Technology Digest of Technical Papers* p.230.

71. Parkin S. S. P., Hayashi M. and Thomas L. 2008 Magnetic domain-wall racetrack memory *Science* **320** 190.

72. Allwood D. A., Xiong G., Faulkner C. C., Atkinson D., Petit D. and Cowburn R. P. 2005 Magnetic domain-wall logic *Science* **309** 1688.

73. Lei N., et al. 2013 Strain-controlled magnetic domain wall propagation in hybrid piezoelectric/ferromagnetic structures *Nat. Commun*. **4** 1378.

74. Edrington W., Singh U., Dominguez Alexander J. R., Nepal R. and Adenwalla S. 2018 SAW assisted domain wall motion in Co/Pt multilayers *Appl. Phys. Lett*. **112** 052402.

75. Mathurin T., Giordano S., Dusch Y., Tiercelin T., Pernod P. and Preobrazhensky V. 2016 Stress-mediated magnetoelectric control of ferromagnetic domain wall position in multiferroic heterostructures *Appl. Phys. Lett*. **108** 082401.

76. Miron I. M., Gaudin G., Auffret S., Rodmacq B., Schuhl A., Pizzini S., Vogel J. and Gambardella P. 2010 Current-driven spin torque induced by the Rashba effect in a ferromagnetic layer *Nat. Mater*. **9** 230.

77. Fan Y., et al. 2014 Magnetization switching through giant spin-orbit torque in a magnetically doped topological insulator heterostructure *Nat. Mater.* **13** 699.

78. Yu G., et al. 2014 Switching of perpendicular magnetization by spin-orbit torques in the absence of external magnetic fields *Nature Nanotechnol.* **9** 548.

79. Ryu K-S., Thomas L. and Parkin S. 2013 Chiral spin torque at magnetic domain walls *Nat. Nanotechnol*. **8** 527.

80. Bhowmik D., Nowakowski M. E., You L., Lee O. J., Keating D., Wong M., Bokor J. and Salahuddin S. 2015 Deterministic domain wall motion orthogonal to current flow due to spin orbit torque *Sci. Rep*. **5** 11823.

81. Manipatruni S., Nikonov D. E. and Young I. A. 2014 Energy delay performance of giant spin Hall effect switching for dense magnetic memory *Appl. Phys. Express* **7** 103001.

82. Shiota Y., et al. 2009 Voltage-assisted magnetization switching in ultrathin $Fe_{80}Co_{20}$ alloy layers *Appl. Phys. Express* **2** 063001.

83. Shiota Y., et al. 2012 Induction of coherent magnetization switching in few atomic layers of FeCo using voltage pulses *Nat. Mater*. **11** 39.

84. Heron J. T., Trassin M., Ashraf K., Gajek M., He Q., Yang S. Y., Nikonov D. E., Chu Y-H, Salahuddin S. and Ramesh R. 2011 Electric-field-induced magnetization reversal in a ferromagnet-multiferroic heterostructure *Phys. Rev. Lett.* **107** 217202.

85. Dzyaloshinskii I. E. 1957 Thermodynamic theory of "weak" ferromagnetism in antiferromagnetic substances *Sov. Phys. JETP* **5** 1259.

86. Moriya T. 1960 Anisotropic superexchange interaction and weak ferromagnetism *Phys. Rev*. **120** 91.

87. Heron J. T. et al. 2014 Deterministic switching of ferromagnetism at room temperature using an electric field *Nature* **516** 370.

88. Kyuno K., Ha J-G., Yamamoto R. and Asano S. 1996 First-principles calculation of the magnetic anisotropy energies of Ag/Fe(001) and Au/Fe(001) multilayers *J. Phys. Soc. Jpn*., **65** 1334.

89. Maruyama T., et al. 2009 Large voltage-induced magnetic anisotropy change in a few atomic layers of iron *Nat. Nanotechnol*. **4** 158.

90. Shiota Y., Maruyama T., Nozaki T., Shinjo T., Shiraishi M. and Suzuki Y. 2009 Voltage assisted magnetization switching in ultrathin $Fe_{80}Co_{20}$ alloy layers *Appl. Phys. Express* **2** 063001.

91. Pertsev N. A. 2013 Origin of easy magnetization switching in magnetic tunnel junctions with voltage-controlled interfacial anisotropy *Sci. Rep*. **3** 2757.

92. Duan C-G, Velev J. P., Sabirianov R. F., Zhu Z., Chu J., Jaswal S. S. and Tsymbal E. Y. 2008 Surface magnetoelectric effect in ferromagnetic metal films *Phys. Rev. Lett.* **101** 137201.

93. Niranjan M. K., Duan C-G, Jaswal S. S. and Tsymbal E. Y. 2010 Electric field effect on the magnetization at the Fe/MgO(001) interface *Appl. Phys. Lett*. **96** 222504.

94. Nakamura K., Shimabukuro R., Fujiwara Y., Tomonori I. and Freeman A. J. 2009 Giant modification of the magnetocrystalline anisotropy in transition metal monolayers by an electric field *Phys. Rev. Lett.* **102** 187201.

95. Miwa S., Matsuda K., Tanaka K., Kotani Y., Goto M., Nakamura T. and Suzuki Y. 2016 Voltage controlled magnetic anisotropy in Fe/MgO tunnel junctions studied by x-ray absorption spectroscopy *Appl. Phys. Lett*. **107** 162402.

96. Endo M., Kanai S., Ikeda S., Matsukura F. and Ohno H. 2010 Electric field effects on thickness dependent magnetic anisotropy of sputtered MgO/Co$_4$OFe$_4$OB$_2$O/Ta structures *Appl. Phys. Lett.* **96** 212503.

97. Wang W-G., Li M., Hageman S. and Chien C. L. 2012 Electric-field-assisted switching in magnetic tunnel junctions *Nat. Mater*. **11** 64.

98. Rajanikanth A., Hauet T., Montaigne F., Mangin S. and Andrieu S. 2013 Magnetic anisotropy modified by electric field in V/Fe/MgO(001)/Fe epitaxial magnetic tunnel junction *Appl. Phys. Lett*. **103** 062402.

99. Nozaki T., Shiota Y., Shiraishi M., Shinjo T. and Suzuki Y. 2010 Voltage-induced perpendicular magnetic anisotropy change in magnetic tunnel junctions *Appl. Phys. Lett*. **96** 022506.

100. Kanai S., Yamanouchi M., Ikeda S., Nakatani Y., Matsukura F. and Ohno H. 2012 Electric field induced magnetization reversal in a perpendicular-anisotropy CoFeB-MgO magnetic tunnel junction *Appl. Phys. Lett.* **101** 122403.

101. Shiota Y., Nozaki T., Bonell F., Murakami S., Shinjo T. and Suzuki Y. 2012 Induction of coherent magnetization switching in few atomic layers of FeCo using voltage pulses *Nat. Mater*. **11** 39.

102. Amiri P. K. and Wang K. L. 2012 Voltage-controlled magnetic anisotropy in spintronic devices *Spin* **2** 1240002.

103. Amiri P. K., et al. 2015 Electric-field-controlled magnetoelectric RAM: Progress, challenges and scaling *IEEE Trans. Magn*. **51** 3401507.

104. Ong P. V. Kioussis N., Khalili Amiri P. and Wang K. L. 2016 Electric-field-driven magnetization switching and nonlinear magnetoelasticity in Au/FeCo/MgO heterostructures *Sci. Rep*. **6** 29815.

105. Grezes C., Ebrahimi F., Alzate J. G., Cai X., Katine J. A., Langer J., Ocker B., Khalili Amiri P. and Wang K. L. 2016 Ultra-low switching energy and scaling in electric-field-controlled nanoscale magnetic tunneling junctions with high resistance-area product *Appl. Phys. Lett*. **108** 012403.

106. Stöhr J., Siegmann H. C., Kashuba A. and Gamble S. J. 2009 Magnetization switching without charge or spin currents *Appl. Phys. Lett*. **94** 072504.

107. Grezes C., Rijas Rozas A., Ebrahimi F., Alzate J. G., Cai X., Katine J. A., Langer J., Ocker B., Khalili Amiri P. and Wang K. L. 2016 In-plane magnetic field effect on switching voltage and thermal stability in in electric-field-controlled perpendicular magnetic tunnel junctions *AIP Advances* **6** 075014.

108. Drobitch J. L., Abeed M. A. and Bandyopadhyay S. 2017 Precessional switching of a perpendicular-anisotropy-magneto-tunneling-junction without a magnetic field *Jpn. J. Appl. Phys.* **56** 100309.

109. Bhattacharya, D., Al-Rashid, M.; Atulasimha, J. 2016 Voltage Controlled Core Reversal of Fixed Magnetic Skyrmions without a Magnetic Field. *Sci. Rep.* **6**, 31272.

110. R Nakatani, Y., Hayashi, M., Kanai, S., Fukami, S., Ohno, H. 2016 Electric Field Control of Skyrmions in Magnetic Nanodisks. *Appl. Phys. Lett.* **108**, 152403.

111. Bhattacharya, D., Atulasimha, J. 2018 Skyrmion mediated voltage controlled switching of ferromagnets for reliable and energy efficient 2-terminal memory, *ACS Appl. Mat. Int*. **10**, 17455.

112. Bhattacharya, D., Al-Rashid, M., Atulasimha, J. 2017 Energy efficient and fast reversal of a fixed skyrmion 2-terminal memory with spin current assisted by voltage controlled magnetic anisotropy. *Nanotechnology*, **28,** 425201.

113. Ong P V, Kioussis N, Odkhuu D, Amiri P K, Wang K L and Carman G P 2015 Giant voltage modulation of magnetic anisotropy in strained heavy metal/magnet/insulator heterostructures Phys. Rev. B **92**, 020407; Hibino Y, Koyama T, Obinata A, Hirai T, Ota S, Miwa K, Ono S, Matsukura F, Ohno H and Chiba D 2016 Peculiar temperature dependence of electric-field effect on magnetic anisotropy in Co/Pd/MgO system *Appl. Phys. Lett.* **109**, 082403.

114. Bauer U., Emori S. and Beach G. S. D. 2012 Voltage-gated modulation of domain wall creep dynamics in an ultrathin metallic ferromagnet *Appl. Phys. Lett.* **101** 172403.

115. Bernard-Mantel A., Herrera-Diez L., Ranno L., Pizzini S., Vogel J., Givord D., Auffret S., Boulle O., Miron I. M. and Gaudin G. 2013 Electric-field control of domain wall nucleation and pinning in a metallic ferromagnet *Appl. Phys. Lett*. **102** 122406.

116. Upadhyaya P., Dusad, R., Hoffman S., Tserkovnyak Y., Alzate J. G., Khalili Amiri P. and Wang K. L. 2013 Electric field induced domain-wall dynamics: Depinning and chirality switching *Phys. Rev. B*. **88** 224422.

117. Li P., Chen A., Li D., Zhao Y., Zhang S., Yang L., Liu Y., Zhu M., Zhang H. and Han X. 2014 Electric field manipulation of magnetization rotation and tunneling magnetoresistance of magnetic tunnel junctions at room temperature *Adv. Mater.* **26** 4320.

118. Roy K., Bandyopadhyay S. and Atulasimha J. 2011 Hybrid spintronics and straintronics: A magnetic technology for ultra-low energy computing and signal processing *Appl. Phys. Lett.* **99** 063108.

119. Eerenstein W., Mathur N. D. and Scott J. F. 2016 Multiferroic and magnetoelectric materials *Nature* **442**, 759.

120. Brintlinger T., Lim S., Baloch K. H., Alexander P., Qi Y., Barry J., Melngailis J., Salamanca-Riba L, Takeuchi I. and Cumings J. **2010**, In Situ Observation of Reversible Nanomagnetic Switching Induced by Electric Fields *Nano Lett.* **10** (4), 1219.

121. Ghidini M., Pellicelli R., Prieto J. L., Moya X., Soussi J., Briscoe J., Dunn S. and Mathur N.D. 2013 Non-volatile electrically-driven repeatable magnetization reversal with no applied magnetic field *Nature Commun.s* **4**, 1453.

122. Gopman D B, Dennis C L, Chen P J, Iunin Y L, Finkel P, Staruch M and Shull R D 2016 Strain-assisted magnetization reversal in Co/Ni multilayers with perpendicular magnetic anisotropy *Sci. Rep.* **6** 27774.

123. Chopdekar R V., Heidler J, Piamonteze C, Takamura Y, Scholl A, Rusponi S, Brune H, Heyderman L J and Nolting F 2013 Strain-dependent magnetic configurations in manganite-titanate heterostructures probed with soft X-ray techniques *Eur. Phys. J. B* **86**, 241.

124. Heidler J, Piamonteze C, Chopdekar R V., Uribe-Laverde M A, Alberca A, Buzzi M, Uldry A, Delley B, Bernhard C and Nolting F 2015 Manipulating magnetism in $La_{0.7}Sr_{0.3}MnO_3$ via piezostrain *Phys. Rev. B - Condens. Matter Mater. Phys.* **91** 024406.

125. Venkataiah G, Shirahata Y, Suzuki I, Itoh M and Taniyama T 2012 Strain-induced reversible and irreversible magnetization switching in $Fe/BaTiO_3$ heterostructures *J. Appl. Phys.* **111**, 033921.

126. Dusch Y, Tiercelin N, Klimov A, Giordano S, Preobrazhensky V and Pernod P 2013 Stress-mediated magnetoelectric memory effect with uni-axial $TbCo_2$/FeCo multilayer on 011-cut PMN-PT ferroelectric relaxor *J. Appl. Phys.* **113**, 17C719.

127. Rementer C R, Fitzell K, Xu Q, Nordeen P, Carman G P, Wang Y E and Chang J P 2017 Tuning static and dynamic properties of FeGa/NiFe heterostructures *Appl. Phys. Lett.* **110**, 242403.

128. Dai G, Zhan Q, Liu Y, Yang H, Zhang X, Chen B and Li R-W 2012 Mechanically tunable magnetic properties of $Fe_{81}Ga_{19}$ films grown on flexible substrates *Appl. Phys. Lett.* **100**, 122407.

129. Corredor E C, Coffey D, Ciria M, Arnaudas J I, Aisa J and Ross C A 2013 Strain-induced magnetization reorientation in epitaxial Cu/Ni/Cu rings *Phys. Rev. B - Condens. Matter Mater. Phys.* **88,** 054418.

130. Sohn H, Nowakowski M E, Liang C Y, Hockel J L, Wetzlar K, Keller S, McLellan B M, Marcus M A, Doran A, Young A, Kläui M, Carman G P, Bokor J and Candler R N 2015 Electrically driven magnetic domain wall rotation in multiferroic heterostructures to manipulate suspended on-chip magnetic particles *ACS Nano* **9**, 4814.

131. Finizio S, Foerster M, Buzzi M, Krüger B, Jourdan M, Vaz C A F, Hockel J, Miyawaki T, Tkach A, Valencia S, Kronast F, Carman G P, Nolting F and Kläui M 2014 Magnetic anisotropy engineering in thin film Ni nanostructures by magnetoelastic coupling *Phys. Rev. Appl.* **1** 1.

132. Klimov A, Tiercelin N, Dusch Y, Giordano S, Mathurin T, Pernod P, Preobrazhensky V, Churbanov A and Nikitov S 2017 Magnetoelectric write and read operations in a stress-mediated multiferroic memory cell *Appl. Phys. Lett.* **110** 222401.

133. Chung T K, Keller S and Carman G P 2009 Electric-field-induced reversible magnetic single-domain evolution in a magnetoelectric thin film *Appl. Phys. Lett.* **94**, 132501.

134. Sohn H, Liang C yen, Nowakowski M E, Hwang Y, Han S, Bokor J, Carman G P and Candler R N 2017 Deterministic multi-step rotation of magnetic single-domain state in Nickel nanodisks using multiferroic magnetoelastic coupling *J. Magn. Magn. Mater.* **439** 196.

135. Buzzi M, Chopdekar R V., Hockel J L, Bur A, Wu T, Pilet N, Warnicke P, Carman G P, Heyderman L J and Nolting F 2013 Single domain spin manipulation by electric fields in strain coupled artificial multiferroic nanostructures *Phys. Rev. Lett.* **111** 1–5.

136. Roy K., 2014 Critical analysis and remedy of switching failures in straintronic logic using Bennett clocking in the presence of thermal fluctuations *Appl. Phys. Lett.* **104** 013103.

137. Novosad V., Otani Y., Ohsawa A., Kim S. G., Fukamichi K., Koike J. and Maruyama K. 2000 Novel magnetostrictive memory device *J. Appl. Phys*. **87** 6400.

138. Biswas A. K., Bandyopadhyay S. and Atulasimha J. 2014 Complete magnetization reversal in a magnetostrictive nanomagnet with voltage-generated stress: A reliable energy-efficient non-volatile magneto-elastic memory *Appl. Phys. Lett*. **105** 072408.

139. Biswas A. K., Ahmad H., Atulasimha J. and Bandyopadhyay S. 2017 Experimental demonstration of complete $180^0$ reversal of magnetization in isolated Co nanomagnets on a PMN-PT substrate with voltage generated strain *Nano Letters,* **17,** 3478.

140. Ahmad, H., Atulasimha, J. and Bandyopadhyay, S. 2015 Electric field control of magnetic states in isolated and dipole coupled FeGa nanomagnets delineated on a PMN-PT substrate *Nanotechnology* **26** 401001.

141. Ding J., Fan L., Zhang S-y., Zhang H. and Yu W-w. 2018 Simultaneous realization of slow and fast acoustic waves using a fractal structure of Koch curve. *Sci. Rep.* **8** 1481.

142. Roy, K., Bandyopadhyay, S. and Atulasimha, J. 2010 Energy efficient mixed mode switching of a multiferroic nanomagnet for logic and memory *arXiv:1012.0819*.

143. Biswas, A. K., Bandyopadhyay, S. and Atulasimha, J. 2013 Acoustically assisted spin-transfer-torque switching of nanomagnets: An energy-efficient hybrid writing scheme for non-volatile memory *Appl. Phys. Lett*., **103**, 232401.

144. Davis, S., Baruth, A. and Adenwalla, S. 2010 Magnetization Dynamics Triggered by Surface Acoustic Waves *Appl. Phys. Lett*. **97**, 232507.

145. Li W, Buford B, Jander A and Dhagat P 2014 Acoustically assisted magnetic recording: A new paradigm in magnetic data Storage *IEEE Trans. Magn.* **50**, 2285018.

146. Li W, Buford B, Jander A and Dhagat P 2014 Writing magnetic patterns with surface acoustic waves *J. Appl. Phys.* **115** 17E307.

147. Kovalenko O, Pezeril T and Temnov V V. 2013 New concept for magnetization switching by ultrafast acoustic pulses *Phys. Rev. Lett.* **110** 266602.

148. Foerster M, Macià F, Statuto N, Finizio S, Hernández-Mínguez A, Lendínez S, Santos P V., Fontcuberta J, Hernàndez J M, Kläui M and Aballe L 2017 Direct imaging of delayed magneto-dynamic modes induced by surface acoustic waves *Nat. Commun.* **8** 407.

149. Bombeck, M., Salasyuk, A. S., Glavin, B. A., Scherbakov, A. V., Bruggemann, C., Yakovlev, D. R., Sapega, V. F., Liu, X., Furdyna, J. K., Akimov, A. V. et al. 2012 Excitation of Spin Waves in Ferromagnetic (Ga,Mn)As Layers by Picosecond Strain Pulses *Phys. Rev. B*, **85,** 195324.

150. Scherbakov A V., Salasyuk A S, Akimov A V., Liu X, Bombeck M, Brüggemann C, Yakovlev D R, Sapega V F, Furdyna J K and Bayer M 2010 Coherent magnetization precession in ferromagnetic (Ga,Mn)As induced by picosecond acoustic pulses *Phys. Rev. Lett.* **105** 1–4.

151. Thevenard L, Peronne E, Gourdon C, Testelin C, Cubukcu M, Charron E, Vincent S, Lemaître A and Perrin B 2010 Effect of picosecond strain pulses on thin layers of the ferromagnetic semiconductor (Ga,Mn)(As,P) *Phys. Rev. B* **82** 104422.

152. Weiler, M., Dreher, L., Heeg, C., Huebl, H., Gross, R., Brandt, M. S. and Goennenwein, S. T. B. 2011 Elastically Driven Ferromagnetic Resonance in Nickel Thin Films *Phys. Rev. Lett*. **106,** 117601.

153. Janusonis, J., Chang, C. L., van Loosdrecht, P. H. M. and Tobey, R. I. 2015 Frequency Tunable Surface Magneto Elastic Waves *Appl. Phys. Lett*. **106**, 181601.

154. Singh U and Adenwalla S 2015 Spatial mapping of focused surface acoustic waves in the investigation of high frequency strain induced changes *Nanotechnology* **26**, 25.

155. Thevenard, L., Duquesne, J.-Y., Peronne, E., von Bardeleben, H., Jaffres, H., Ruttala, S., George, J.-M., Lemaître, A. and Gourdon, C. 2013 Irreversible Magnetization Switching Using Surface Acoustic Waves *Phys. Rev. B*, **87**, 144402.

156. Chudnovsky E M and Jaafar R 2016 Manipulating the Magnetization of a Nanomagnet with Surface Acoustic Waves: Spin-Rotation Mechanism *Phys. Rev. Appl.* **5** 031002.

157. Tejada, J.; et al. 2017 Switching of magnetic moments of nanoparticles by surface acoustic waves *Europhys. Lett.,* **118**, 37005.

158. Salehi-Fashami M, Al-Rashid M, Sun W-Y, Nordeen P, Bandyopadhyay S, Chavez A C, Carman G P and Atulasimha J 2016 Binary information propagation in circular magnetic nanodot arrays using strain induced magnetic anisotropy *Nanotechnology* **27** 43LT01.

159. Chavez A. C., Sun W., Atulasimha J, Wang K. L., and Carman G. P., 2017 Voltage induced artificial ferromagnetic-antiferromagnetic ordering in synthetic multiferroics, *J. Appl. Phys.* **122**, 224102.

160. Carlton D., Lambson B, Scholl A., Young A., Ashby P., Dhuey S., and Bokor J. 2012 Investigation of Defects and Errors in Nanomagnetic Logic Circuits , *IEEE Trans. Nanotech.* **11**, 4, 2196445.

161. Abeed M., Atulasimha J., Bandyopadhyay S., 2018 Magneto-elastic switching of magnetostrictive nanomagnets with in-plane shape anisotropy arXiv preprint arXiv:1803.05869.

162. Panduranga K. P., Lee T., Chavez A., Prikhodko S. V., and Carman G. P. 2018 Polycrystalline Terfenol-D thin films grown at CMOS compatible temperature *AIP Advances* **8**, 056404.

163. Cowburn, R. P. and Welland, M. E. 2000 Room temperature magnetic quantum cellular automata *Science* **287**, 1466.

164. Ney, A., Pampuch, C., Koch, R. and Ploog, K. H. 2003 Programmable computing with a single magnetoresistive element. *Nature* **425**, 485.

165. Lee, S., Choa, S., Lee, S. and Shin, H. 2007 Magneto-logic device based on a single-layer magnetic tunnel junction. *IEEE Trans. Elec. Dev.* **54** 2040.

166. Wang, J., Meng, H. and Wang J-P. 2005 Programmable spintronics logic device based on a magnetic tunnel junction element. *J. Appl. Phys.* **97**, 10D509.

167. Behin-Aein, B., Datta, D., Salahuddin, S. & Datta S. 2010 Proposal for an all-spin logic device with built-in memeory. *Nature Nanotech.* **5**, 266.

168. Srinivasan, S., Sarkar, A., Behin-Aein, B. & Datta, S. 2011 All-spin logic device with in-built non-reciprocity. *IEEE Trans. Magn.* **47** 4026.

169. Waser, R. (ed.) *Nanoelectronics and Information Technology*, Ch. III (Wiley-VCH, Weinheim, Germany, 2003).

170. Spedalieri, F. M., Jacob, A. P., Nikonov D. E. and Roychowdhury, V. P. 2011 Performance of magnetic quantum cellular automata and limitations due to thermal noise. *IEEE Trans. Nanotech.* **10**, 537.

171. Salehi-Fashami, M., Munira, K., Bandyopadhyay, S., Ghosh, A. W. and Atulasimha, J. 2013 Switching of dipole coupled multiferroic nanomagnets in the presence of thermal noise: reliability analysis of hybrid spintronic-straintronic nanomagnetic logic. *IEEE Trans. Nanotech.* **12**, 1206.

172. Carlton, D. et al. 2012 Investigation of defects and errors in nanomagnetic logic circuits. *IEEE Trans. Nanotech.* **11**, 760.

173. Alam, M. T. et al. 2010 On-chip clocking for nanomagnetic logic devices. *IEEE Trans. Nanotech.* **9**, 348.

174. Sharad, M., Yogendra, K., Gaud, A., Kwon, K. W. and Roy, K. 2013 Ultra-high density, high performance and energy-efficient all spin logic. *arXiv:1308:2280 [cond-mat:mes-hall]*.

175. Behin-Aein, B., Sarkar, A., Srinivasan, S. and Datta, S. 2011 Switching energy-delay of all spin logic devices. *Appl. Phys. Lett.* **98**, 123510.

176. Biswas, A. K., Atulasimha, J. and Bandyopadhyay, S. 2014 An error-resilient non-volatile magneto-elastic universal logic gate with ultralow energy-delay product *Sci. Rep*. **4**, 7553.

177. Tiercelin, N., Dusch, Y., Preobrazhensky, V. and Pernod, P. 2011 Magnetoelectric memory using orthogonal magnetization states and magnetoelastic switching *J. Appl. Phys.* **109** 07D726.

178. Yaseen, M., Ren, W., Chen, X., Feng, Y., Shi, P. and Wu, X. 2018 Effects of thickness, pulse duration, and size of strip electrode on ferroelectric electron emission of lead zirconate titanate films *J. Elec. Mater*. **57**, 1183.

179. Sun, E., Zhang, S., Luo, J., Shrout, T. R. and Cao, W. 2010 Elastic, dielectric and piezoelectric constants of $Pb(In_{1/2}Nb_{1/2}O_3)$-$Pb(Mn_{1/3}Nb_{2/3}O_3)$-$PbTiO_3$ single crystal poled along $[011]_c$ *Appl. Phys. Lett.* **97**, 032902.

180. Ikeda, S., Hayakawa, J., Ashizawa, Y., Lee, Y. M., Miura, K., Hasegawa, H., Tsunoda, M., Matsukura, F. and Ohno, H. 2008 Tunnel magnetoresistance of 604% at 300 K by suppression of Ta diffusion in CoFeB/MgO/CoFeB pseudo spin valves annealed at high temperature *Appl. Phys. Lett*. **93**, 082508.

181. http://www.learnabout-electronics.org/Digital/dig43.php.

182. Jacob Millman, *Micro-electronics: Digital and Analog Circuits and Systems*, (Mc-Graw Hill, New York, 1979).

183. Biswas A. K., Atulasimha, J. and Bandyopadhyay, S. 2017 Energy-efficient hybrid spintronic-straintronic nonvolatile reconfigurable equality bit comparator *SPIN* **7** 1750004.

184. Datta, S., Salahuddin, S. and Behin-Aein, B. 2012 Non-volatile spin switch for Boolean and non-Boolean logic, *Appl. Phys. Lett.*, **101**, 252411.

185. Sharad, M., Augustine, C., Panagopoulos, G. and Roy, K. 2012 Spin-based neuron model with domain-wall magnets as synapse *IEEE Trans. Nanotech*. **11** 843.

186. Sharad, M., Fan, D. and Roy, K. 2013 Spin-neurons: A possible path to energy-efficient neuromorphic computers *J. Appl. Phys*. **114**, 234906.

187. Sengupta, A., Choday, S. H., Kim, Y. and Roy, K., 2015 Spin orbit torque based electronic neuron *Appl. Phys. Lett*. **106** 143701.

188. Sengupta, A. Parsa, M., Han, B. and Roy, K. 2016 Probabilistic deep spiking neural systems enabled by magnetic tunnel junction *IEEE Trans. Elec. Dev*. **63** 2963.

189. Sengupta, A., Shim, Y. and Roy, K. 2016 Proposal for an all-spin artificial neural network: emulating neural and synaptic functionalities through domain wall motion in ferromagnets *IEEE Trans. Biomed Circuits Syst*. **99** 1.

190. Biswas, A. K., Atulasimha, J. and Bandyopadhyay, S. 2015 The straintronic spin neuron *Nanotechnology* **26** 285201.

191. Pagiamtzis, K. and Sheikholeslami A. 2006 Content addressable memory (CAM) circuits and arhcitectures: A tutorial and survey *IEEE J. Solid State Circuits* **41** 712.

192. Bandyopadhyay, S., Das, B. and Miller A. E. 1994 Supercomputing with spin polarized single electrons in a quantum coupled architecture *Nanotechnology* **5** 113.

193. Agarwal, H., Pramanik, S. and Bandyopadhyay, S. 2008 Single spin universal Boolean logic gate *New J. Phys*., **10** 015001.

194. Imre, A., Csaba, G., Ji, L., Orlov, A., Bernstein, G. H. and Porod, W. 2006 Majority logic gate for magnetic quantum dot cellular automata *Science* **311** 205.

195. Al-Rashid, M. M., Bandyopadhyay, S. and Atulasimha, J. 2016 Dynamic error in strain-induced magnetization reversal of nanomagnets due to incoherent switching and formation of metastable states: A size dependent study *IEEE Trans. Elec. Dev*. **63** 3307.

196. Al-Rashid, M. M., Bhattacharya, D., Bandyopadhyay, S. and Atulasimha, J. 2015 Effect of nanomagnet geometry on reliability, energy dissipation and clock speed in strain clocked DC-NML *IEEE Trans. Elec. Dev*. **62** 2978.

197. Munira, K., Xie, Y., Nadri, S., Forgues, M. B., Salehi-Fashami, M., Atulasimha, J., Bandyopadhyay, S. and Ghosh, A. W. Reducing error rates in straintronic multiferroic nanomagnetic logic by pulse shaping *Nanotechnology* **26** 245202.

198. Salehi-Fashami, M., Atulasimha, J. and Bandyopadhyay, S. 2013 Energy dissipation and error probability in fault-tolerant binary switching *Sci. Rep*. **3** 3204.

199. Salehi-Fashami, M., Munira, K., Bandyopadhyay, S., Ghosh, A. W. and Atulasimha, J. 2013 Switching of dipole coupled multiferroic nanomagnets in the presence of thermal noise: Reliability of nanomagnetic logic *IEEE Trans. Nanotechnol*. **12** 1206.

200. Carlton D. B., Emley, N. C., Tuchfeld E. and Bokor, J. 2008 Simulation Studies of Nanomagnet Based Logic Architecture *Nano Letters,* **8**, 4173.

201. Barahona, F. 1982 On the computational complexity of Ising spin glass models. *J. Phys. Math. Gen*. **15** 3241.

202. Yamaoka, M., Yoshimura, C., Hayashi, M., Okuyama, T., Aoki, H. and Mizuno, H. 2016 A 20 k-spin Ising chip to solve combinatorial optimization problems with CMOS annealing *IEEE J. Solid State Circuits* **51** 303.

203. Kim, K., Chnag, M. S., Korenblit, S., Islam, R., Edwards, E. E., Freericks, J. K., Lin, G. D., Duan, L. M. and Monroe, C. 2010 Quantum simulation of frustrated Ising spins with trapped ions *Nature* **465** 590.

204. Mahboob, I., Okamoto, H. and Yamaguchi, H. 2016 An electromechanical Ising Hamiltonian *Sci. Adv.* **2** e1600236.

205. Kirkpatrick, S., Gelatt, C. D. and Vecchi, M. P. 1983 Optimization by simulated annealing *Science* **220** 671.

206. Johnson, M. W. et al. 2011 Quantum annealing with manufactured spins *Nature* **473** 194.

207. Chua L. O. and Yang L. 1988 Cellular neural networks: applications *IEEE Trans. Circuits. Syst*. **35** 1273.

208. Utsunomiya, S. Takata, K. and Yamamoto, Y. 2011 Mapping of Ising models into injection-locked laser systems *Opt. Express* **19** 18091.

209. Roychowdhury, V. P., Janes, D. B., Bandyopadhyay, S. and Wang, X. 1996 Collective computational activity in self-assembled arrays of quantum dots: A novel neuromorphic architecture for nanoelectronics, *IEEE Trans. Elec. Dev*. **43** 1688.

210. Karahaliloglu, K., Balkir, S., Pramanik, S. and Bandyopadhyay, S. 2003 A quantum dot image processor *IEEE Trans. Elec. Dev*. **50** 1610.

211. Bhanja, S., Karunaratna, D. K., Panchumarthy, R., Rajaram, S. and Sarkar, S. 2015 Non-Boolean computing with nanomagnets for computer vision applications *Nature Nanotechnol*. **11** 177.

212. D'Souza, N., Atulasimha, J. and Bandyopadhyay, S. 2012 An ultrafast image recovery and recognition system implemented with nanomagnets possessing biaxial magnetocrystalline anisotropy I*EEE Trans. Nanotechnol*. **11** 896.

213. Sutton, B., Camsari, K. Y., Behin-Aein, B. and Datta, S. 2016 Intrinsic optimization using stochastic nanomagnets *Sci. Rep*. **7** 44370.

214. Abeed, M. A., Biswas, A. K., Al-Rashid, M. M., Atulasimha, J. and Bandyopadhyay, S. 2017 Image processing with dipole-coupled nanomagnets: Noise suppression and edge enhancement detection *IEEE Trans. Elec. Dev*. **64** 2417.